\documentclass[final,3p,times,pdflatex]{elsarticle}
\usepackage{array}
\usepackage{booktabs}
\usepackage{changes}
\usepackage{amsmath}
\usepackage{amssymb}
\usepackage{pifont}
\usepackage{caption}
\usepackage{morefloats}
\usepackage{fancyhdr}
\usepackage{multirow}
\usepackage{graphicx}
\usepackage{hhline}
\usepackage{float}
\usepackage{subfig}
\newcommand{\as}{\alpha_s}

\newcommand{\dif}{\mathrm{d}}             % finite-dimensional differential
\newcommand{\ui}{\mathrm{i}}             % imaginary unit
\newcommand{\To}{\Rightarrow}
 
\journal{Computer Physics Communications}

\begin{document}

\begin{frontmatter}

\begin{flushright}
DAMTP-2018-25 
\end{flushright}

\title{{\tt reSolve} - A Transverse Momentum Resummation Tool}

\author[damtp]{F.~Coradeschi}
\cortext[cor1]{Corresponding author}
\author[damtp,kavli]{T.~Cridge}
\ead{t.cridge@damtp.cam.ac.uk}
\address[damtp]{DAMTP, CMS, University of Cambridge, Wilberforce road,
  Cambridge, CB3  0WA, United Kingdom}
\address[kavli]{KITP, University of California, Santa Barbara, CA 93106, USA}

\begin{abstract}

In this note, we introduce the new tool reSolve, a Monte Carlo differential cross-section and parton-level event
generator whose main purpose is to add transverse momentum resummation to a general class of inclusive processes
at hadron colliders, namely all those which do not involve hadrons or jets in the measured final state. This documen-
tation refers to the first main version release, which will form the basis for continued developments,
consequently it only implements the key features of those we plan to ultimately include. This article acts
as a manual for the program; describing in detail its use, structure, validation and results; whilst also highlighting
key aspects of the resummation formalism applied. It details the two classes of processes so far included; these are
diphoton production and Drell-Yan production.

A main concept behind the development of the tool is that it is a hands-on white box for the user: significant
effort has been made to give the program a modular structure, making the various parts which comprise it independent
of each other as much as possible and ensuring they are transparently documented, customizable and, in principle, replaceable with something that may better serve the user’s needs.

reSolve is a new C++ program, based on an evolution of the private Fortran code 2gres, previously used for
the calculations in refs. [1] and [2]; it is also influenced by the DYRes Fortran code of refs. [3] and [4].
This initial version calculates the low transverse momentum contribution to the fully differential cross-section for two
main categories of processes; the inclusive production of two photons, and inclusive Drell-Yan production. In all cases
resummation up to Next-to-Next-to-Leading Logarithm (NNLL) is included. We aim to extend the program to several
more processes in the near future. The program is publicly available on Github.
\end{abstract}

\normalsize 

\end{frontmatter}

\section{Program Summary}
\noindent{\em Program title:} {\tt reSolve} \\
{\em Program obtainable   from:} {\tt https://github.com/fkhorad/reSolve/} \\
{\em Distribution format:}\/ tar.gz \\
{\em Programming language:} {\tt C++}, {\tt fortran} \\
{\em Computer:}\/ Personal computer. \\
{\em Operating system:}\/ Tested on Linux 4.13.0-38-generic, Linux 4.4.0-119-generic and on MAC-OS 10.11.6.
\\
{\em Word size:}\/ 64 bits. \\
{\em External routines:}\/ minuit and Cuba \\
{\em Typical running time:}\/ 0.1-1 seconds per Monte Carlo point per core used. \\
{\em Nature of problem:}\/ Calculating the transverse momentum spectrum, including resummation, for a general process at hadron colliders.\\
{\em Solution method:}\/ Monte Carlo generation of the phase space points and phase space integration to allow the production of differential distributions, each phase space point itself has to be inverse Fourier transformed and double inverse Mellin transformed to allow the resummation, following the usual transverse momentum resummation impact parameter space formalism. {\tt reSolve} operates up to Next-to-Next-to-Leading Order including Next-to-Next-to-Leading Logarithm resummation (NNLO+NNLL). \\
{\em Restrictions:}\/ So far only diphoton production in the Standard Model (background, not including Higgs) and Drell-Yan production are included, nonetheless the program is designed to allow further extensions to additional processes, including by the user. The limitations on the processes possible to implement are that they must be hadron-hadron collisions producing a non-strongly interacting measured final state system. This first main implementation of {\tt reSolve} calculates only the resummed part of the differential cross-section, which is dominant at low transverse momentum, this has to be matched with the usual finite contributions to obtain the spectrum over the whole transverse momentum range. \\

\section{Introduction}

Since the early days of the Large Hadron Collider, a large part of the experimental effort has been focused on direct searches for signals of New Physics at ever-increasing energy scales. However, given the lack of any clear evidence for new phenomena so far, the focus has been shifting, both theoretically and experimentally, to alternative strategies. Amongst these is what can be called the ``precision physics at colliders'' program; this involves an effort to extend both our experimental measurements and theoretical predictions of as many observables as possible, in the context of the Standard Model (SM) and beyond, to an ever increasing precision. This will allow us both to extend our knowledge and understanding of the properties of the model, and hopefully to eventually detect small deviations from the SM which will signal the presence of the as yet undiscovered New Physics which is expected to exist at higher scales.

An interesting observable for this precision strategy is the transverse momentum spectrum of generic final states. 
This is theoretically interesting to check our understanding of perturbative QCD and also important for the precise determination of observables such as the $W$ mass~\cite{Aaboud:2017svj} and many others, and will be a focus of ongoing and future measurements at $13$TeV at Run 2 of the LHC and beyond.

The theoretical calculation of the transverse momentum ($q_T$) spectrum for generic systems of invariant mass $M$ is challenging towards the low end of the spectrum, where $q_T \ll M$, due to the well-known presence of logarithmically-enhanced terms which appear to all orders in the perturbative expansion. In particular, once $log^{-1}(M/q_T)$ is of order of $\as(M)$ (with $\as$ the strong coupling constant) the convergence of the perturbative series is completely ruined, and it is necessary to resum such large logarithmic contributions to all orders theoretically if we are to produce meaningful theoretical predictions at low transverse momenta. Indeed without this resummation, transverse momentum spectra diverge as $q_T \rightarrow 0$ (for instance see figure~1 in \cite{Cieri:2015rqa} for an explicit example of this divergence in the context of diphoton production); rendering both the total cross-section and differential transverse momentum distribution predictions without such resummation inaccurate at best, and often useless. The $q_T$ resummation is an involved but well studied problem; in the present article, and in the program {\tt reSolve}, we will follow the impact parameter space formalism of references~\cite{Dokshitzer:1978hw,Parisi:1979se,Curci:1979bg,Collins:1981uk,Kodaira:1981nh,Collins:1984kg,Catani:1988vd,deFlorian:2000pr,Catani:2010pd,Catani:2013tia}.

The aim of this work is to outline the development of a general tool, the {\tt reSolve} program, which allows the implementation of $q_T$ resummation to a large class of inclusive processes, of the generic form $h_1 h_2 \to F + X$, where $h_1$ and $h_2$ are hadrons, $F$ is the measured final state, which can include an arbitrary number of $non strongly-interacting$ particles (no jets or hadrons are allowed in $F$), and $X$ denotes unresolved extra radiation. This therefore allows event generation and predictions for arbitrary observables at the fully differential level for such processes. {\tt reSolve} is a new program, written completely from scratch in C++, based around an evolution of a previous private code {\tt 2gres}, which was used for the calculations in refs.~\cite{Cieri:2015rqa} and~\cite{Cieri:2017kpq}, and which is in turn related to the resummation codes DYres~\cite{Bozzi:2010xn,Catani:2015vma} and Hres~\cite{deFlorian:2012mx}. In this first main release, two processes are so far included, namely diphoton production ($F = \gamma\gamma$) and Drell-Yan production ($F = W,$ $Z$, $Z/\gamma^*$); in both Next-to-Next-to-Leading-Logarithmic (NNLL) precision can be reached in the resummation. The diphoton process is interesting as it provides the SM background for Higgs production and decay in one of the experimentally cleanest channels. Hopes of understanding the nature of Electroweak Symmetry Breaking at a deeper level, and more generally of improving our knowledge of Higgs boson physics, make the diphoton channel one of significant experimental focus, in both signal and background. The Drell-Yan process is similarly intriguing, used experimentally to measure the mass of the W boson, to constrain PDFs, and for many other purposes, it could also be extended to Beyond Standard Model $W'$ and $Z'$ models. In future iterations of {\tt reSolve} further processes will be added, with Higgs production and decay (and signal-background interference) being amongst the easiest potential additions of the possible interesting candidates. The program has been designed in such a way so as to make the addition of new processes as straightforward as possible, taking full advantage of the substantial process-independence of the formalism, in order to also allow perspective users to add processes of their own interest. Indeed a description of how to add a process to {\tt reSolve} is outlined in Section~\ref{AddProcess}.

The outline of the paper is as follows; we begin in section~\ref{usingprog} with the details of how to use the program -- its input and output files, options for running with one or more PDF fits, parallelisation and using the incorporated histogramming program to convert the generated events into histogram data for the differential cross-section spectra. This section serves as a guide to users of the program. Following this, in section~\ref{programstruc}, we go into detail about the modular structure of the program, designed to make the addition of new processes straightforward, including providing an outline of how the code actually works. In section~\ref{theory}, a review of the theoretical impact parameter space resummation formalism is included, with several pointers towards how the formulae were practically implemented in the code, before section~\ref{validation} provides further information on the validation of the {\tt reSolve} program and results that can be generated by the program for both the diphoton and Drell-Yan processes. In addition its speed is briefly discussed. Finally, we end by signposting some of the future developments we intend to complete following this first version in section~\ref{future}, before concluding in section~\ref{conclusions}. \ref{app:usage} provides a summary of how to use {\tt reSolve} along with the sample input files included with the program, all included here for ease of reference; meanwhile \ref{app:coeffs} contains further information on the details of the theoretical formalism applied for the resummation.

\section{Using the {\tt reSolve} Program} \label{usingprog}

We are presenting the new Monte Carlo tool {\tt reSolve}. The objective of the program is to calculate differential cross-sections and generate events at parton-level in generic processes of the class $h_1 h_2 \to F + X$, where $h_1$ and $h_2$ are hadrons and $F$ is a final state containing an arbitrary number of particles, but no strongly interacting ones, that is, no jets and no hadrons, including transverse momentum resummation; whilst $X$ represents unresolved extra radiation. As of the current version, the processes implemented are $F=\gamma\gamma$ and $F=W^+$, $W^-$, $Z$, $Z/\gamma$. The structure of the program is however general, and most of the instructions that will follow will apply equally well to other processes and future versions. This paper will therefore serve as the main manual for users of the {\tt reSolve} program.

\subsection{\textbf{Getting it running}}

\begin{itemize}
\item In order to run the {\tt reSolve} program, first download the zipped tarball {\tt reSolve.tar.gz}, this is available both on Github at {\tt https://github.com/fkhorad/reSolve}, and with the published paper. Extract this to the desired working directory.
\item Next enter the makefile, found in {\tt code/src} and adapt any compilation flags as appropriate to your machine, note if you wish to use the {\tt Cuba} integrator \cite{Hahn:2004fe}, rather than just the built-in {\tt k\_vegas} integrator provided with the code, you will need to tell the program where to find it. Interface codes for both {\tt k\_vegas} (our integrator) and {\tt Cuba} are provided within the program in {\tt code/src/main} and the relevant interface is automatically used once the integrator chosen is given in the input file, see section~\ref{inputfile}.
\item Finally run {\tt ``make''} to compile the program and produce the {\tt ./reSolve.out} executable.
\item Running the program then involves simply entering in the terminal \newline {\tt ./reSolve.out \{path to input file\}}, e.g. {\tt ./reSolve.out input/Diphoton\_NNLO\_test\_1.dat }.
\item Note that the output working directory to which the output is directed is included in the input file. In order to avoid overwriting of data or corruption of events, {\tt reSolve} will not run if there are already events (or a file {\tt ``reSolve\_main\_out.dat''}) in the specified folder. Therefore to re-run into the same folder, first either move or delete the event files so the working directory for the output (specified in {\tt ``workdir/''} in the input file) is empty before running.
\end{itemize}

\subsection{\textbf{Input File}} \label{inputfile}

\begin{figure}[ht!]
\centerline{\includegraphics[height = 15cm, width = 16cm]{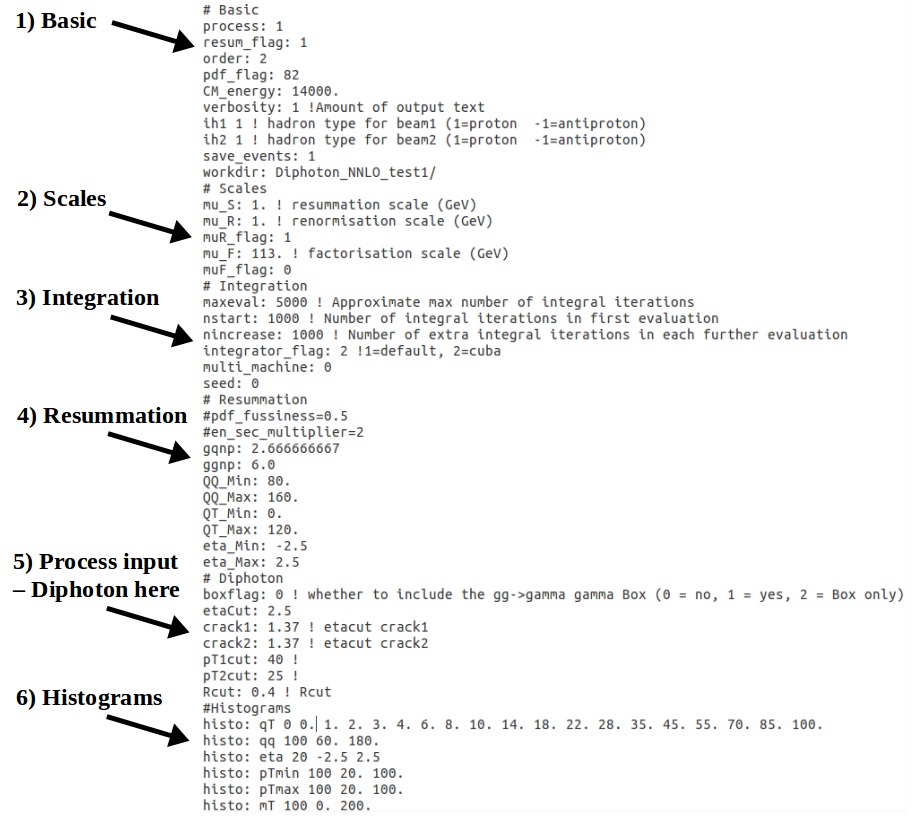} }
\caption{Input file for {\tt reSolve}, this file shown is the example file found in {\tt input/Diphoton\_NNLO\_test\_1.dat}. It is split up into appropriate sections - Basic, Scales, Integration, Resummation, Process-specific input (Diphoton in this case), and Histograms.} \label{input1} 
\end{figure}

Sample input files are provided with the program, including all those used for figures and results in this paper, these are summarised in \ref{inputsincluded}. These are found in the directory labelled {\tt input}, figure~\ref{input1} shows the input file {\tt Diphoton\_NNLO\_test\_1.dat}. There are a number of different options within the input file and these are split into sections:
\begin{enumerate}
\item Basic - The first section, labelled ``Basic'' at the top of the input files includes general input. These are the {\tt process}, which is $1$ for diphoton resummation or $2$ for Drell-Yan resummation - the only processes so far incorporated. If Drell-Yan is selected then one must also chose the specific Drell-Yan process via the {\tt DYprocess} flag; $1 = W^+$, $2 = W^-$, $3 = W^{\pm}$, $4 = Z$ only, $5 = Z/\gamma^*$. In addition, the $DYnarrowwidthapprox$ flag allows the calculation of the on-shell only cross-section if it is set to $1$. There is the {\tt order} flag to indicate the order of the calculation (leading order (0), next-to-leading order (1) or next-to-next-to-leading order (2)) and the {\tt resum\_flag} to turn the resummation on(1) or off(0), this is useful as when adding a process to the {\tt reSolve} program the main addition required is a new Born cross-section, this can then be tested by setting {\tt order:0} and {\tt resum\_flag:0} to ensure it recovers the well-known Born total cross-section for the given process. {\tt pdf\_flag} allows for the PDFs input into the program to be changed, currently it is set to 82 indicating the {\tt MSTW} (Martin-Stirling-Thorne-Watt) 2008 set \cite{Martin:2009iq} at NNLO - this is the only set naturally incorporated into the program, the program is nonetheless setup to make this easy to change between PDF sets. 80 offers the LO MSTW set and 81 the NLO MSTW PDF set. The MSTW PDF sets are read from {\tt ``Grids/''} in the main program directory. {\tt CM\_energy} indicates the collision energy of the protons/anti-protons in the centre of mass frame in $GeV$, {\tt verbosity} sets the amount of information to be output to terminal, in general we recommend this be kept set to 1 and higher values only used for debugging; {\tt ih1/2} indicate whether beam 1 and beam 2 are proton or anti-proton beams; {\tt save\_events} is set to $0$ if only the total cross-section is required, however to produce differential cross-sections the events must be saved and therefore {\tt save\_events} should be set to $1$ to indicate the events will be saved in ``easy'' form or alternatively to $2$ which is a ``pseudo-lhe'' form. Both allow determination of the overall transverse momentum spectrum, as well as other differential observables, rather than just the overall cross-section which is naturally output; finally {\tt workdir} sets the working directory the events will be output to.
\hfill \break
\item Scales - This section sets the three scales involved in the resummation formalism. First there is the usual factorisation scale {\tt mu\_F} ($\mu_F$) encapsulating scale dependence as a result of the factorisation of the input and output processes of the collision, this is the scale to which the PDFs are evolved. Second there is the usual renormalisation scale {\tt mu\_R} ($\mu_R$) dependence arising from the scale at which the $\alpha_s$ coupling and the partonic differential cross-section are evaluated. Finally there is also the scale {\tt mu\_S} ($\mu_S$), which arises as a result of the truncation of the resummed expression at a given perturbative order, parametrising the ambiguity stemming from the precise definition of the logarithmic terms which are being resummed. The setting of the scales here is however somewhat complicated, should you wish to set $\mu_F$, $\mu_R$, $\mu_S$ directly to fixed values throughout the resummation this is done here, in that case one must ensure that also the flags {\tt muR\_flag} and {\tt muF\_flag} are set to 0. However, rather than fixed scales one can set the values of $\mu_F^2$, $\mu_R^2$, $\mu_S^2$ to fixed fractions of the $qq^2$ invariant mass of each generated event, to do this set {\tt muR\_flag} and {\tt muF\_flag} to 1 and $\mu_R$, $\mu_F$ to the desired fraction of $qq$; the resummation scale $\mu_S$ with be set to half the renormalisation scale $\mu_R$ in this case, as is the convention. In the input file in figure~\ref{input1}, the flags {\tt muR\_flag} and {\tt muF\_flag} are set to $1$ and $0$ respectively, therefore $\mu_R = qq$ and $\mu_S = \frac{qq}{2}$ whilst $\mu_F$ is fixed at $\mu_F = 113GeV$. Here one may also specify the parameter {\tt mu\_min}, which is the starting minimum scale from which the PDF fit factorisation scales are  calculated, see Section~\ref{Multipddfits} for more information. 
\hfill \break
\item Integration - This section deals with the inputs specific to the Monte Carlo phase space integration. {\tt maxeval}, {\tt nstart} and {\tt nincrease} correspond to the approximate maximum number of Monte Carlo evaluations to perform total across all iterations\footnote{This is not true in the {\tt k\_vegas} parallelised case, as described in Section~\ref{parallelisation}, here the number of total iterations is set in this case via the number of iterations desired per core.}, the number of evaluations to perform in the first iteration, and the number of additional evaluations in each subsequent iteration. Therefore the number of evaluations in the $nth$ iteration is given by $n_{eval} = n_{start} + (n-1)n_{increase}$ and the total number of evaluations across $N$ iterations is $n_{tot} = Nn_{start} + \frac{1}{2}N(N-1)n_{increase}$. The program will stop after a whole number of iterations once this number of evaluations $n_{tot}$ exceeds {\tt maxeval}. The {\tt integrator\_flag} determines which integrator algorithm is used. For the moment there are two possibilities, our own internal {\tt k\_vegas} Monte Carlo implementation is used for {\tt integrator\_flag}$ = 1$ or an external {\tt Cuba} Monte Carlo implementation \cite{Hahn:2004fe} is used for {\tt integrator\_flag}$ = 2$. The internal {\tt k\_vegas} implementation is based on the original Lepage Vegas algorithm \cite{1978JCoPh..27..192L}. The interface and calling of {\tt Cuba} will then be done automatically by the program once the directory to the extra {\tt Cuba} integrator is provided in the {\tt reSolve} makefile. Instead one may set {\tt integrator\_flag}$ = 0$, events will then be read in from an event file with name given by the input variable ``randoms\_file'', this is useful for debugging or comparison with other programs. Note {\tt Cuba} will automatically parallelise over the number of cores in the computer used whilst {\tt k\_vegas} will not. {\tt multi\_machine} sets whether you wish to use parallelisation with {\tt k\_vegas}, with $0$ indicating not and $1$ indicating parallelisation, it allows you to run different batches on different cores/computers and combine them all each iteration. Again, more information on how to perform parallelisation with the {\tt k\_vegas} integrator is given in Section~\ref{parallelisation}. {\tt seed} is used to set the seed for the randoms used for the Monte Carlo phase space integration by {\tt k\_vegas} or {\tt Cuba}. This can be used to set the seeds for the randoms for the Monte Carlo integration based on time ($-1$), a random repeatable set of uniformly distributed seeds ($0$) or, if one is using parallelisation with {\tt k\_vegas}, one can set $-2$ to ensure each batch has a different seed, here the seed is set based on the machine tag but is deterministic and repeatable. This is as opposed to setting the seed flag to $-1$ to set the seed based on time, which will also ensure different randoms in each machine batch, but of course in a non-repeatable manner.
\hfill \break
\item Resummation - Here various settings for the general resummation implementation are set, these are process-independent. The maximum and minimum values of the invariant mass squared $qq^2$, transverse momentum squared $q_T^2$ and rapidity $\eta$, are all set here via {\tt QQ\_Max}, {\tt QQ\_Min}, {\tt QT\_Max}, {\tt QT\_Min}, {\tt eta\_Max} and {\tt eta\_Min}. {\tt gqnp} and {\tt ggnp} are factors to account for non-perturbative corrections to the sudakovs, they factor in uncertainty from very low $q_T \sim \Lambda_{QCD}$ and are given in equation~\ref{NonPTfactors} in section~\ref{sec:FourierAndMellin}. In addition, here one may set the variables {\tt en\_sec\_multiplier} and {\tt PDF\_fussiness} which are related to the PDF fit files used. More detail is given in Section~\ref{Multipddfits}. Meanwhile further PDF fit options are available here; {\tt PDF\_fitonly} can be set to 1 to allow one to run the code just to obtain the PDF fit file. This can be useful if running parallel batches, if you start without the PDF fit file here all batches will attempt to evaluate it and this will, at best, increase the time taken to run the code (as it will wait for the slowest core to complete the fit). At worst it could lead to inconsistencies in the fit used. 
\hfill \break
\item Process Inputs - Penultimately, there are the process-specific inputs, those specific to the diphoton/Drell-Yan process.
\begin{enumerate}
\item Diphoton - We first describe the diphoton process inputs. These include {\tt boxflag} which allows the user to include ($1$) or not ($0$) the $gg \rightarrow \gamma \gamma$ box diagram (see figure~\ref{ggbox}), or even to only have this contribution to the process ({\tt boxflag}$ = 2$). There are then the diphoton cuts, these are: {\tt crack1/2} which indicate if a crack in the rapidity sensitivity of the detector is present (often 1.37 to 1.56 for the LHC); {\tt etaCut} which should be less than or equal to $Min(|${\tt eta\_Min}$|$,$|${\tt eta\_Max}$|)$; and $pT1cut$ and $pT2cut$, which cut based on the $q_T$ of the photons. It is required that the larger of the two photons' transverse momenta is larger than $pT1cut$, i.e. $Max(q_{T}^{\gamma_1},q_{T}^{\gamma_2})>${\tt pT1cut}, and that the smaller of the two photons' transverse momenta is larger than $pT2cut$, i.e. $Min(q_{T}^{\gamma_1},q_{T}^{\gamma_2})>${\tt pT2cut}. Finally {\tt Rcut} is a cut placed on the opening angle of the two photons produced as two highly collimated photons may not be resolved experimentally, we require: \newline $\Delta_R = \sqrt{(\eta_1-\eta_2)^2 + (\phi_1-\phi_2)^2} >${\tt Rcut}.
\item Drell-Yan - These process specific inputs similarly detail possible cuts required. First there are the usual general cuts described for the diphoton section - {\tt crack1/2} and {\tt pT1/2cut}. Then there are specific cuts for different Drell-Yan processes, {\tt eta1cut} and {\tt eta2cut} cut on the rapidity of the produced Drell-Yan leptons for neutral current Drell-Yan; in the case of charged current Drell-Yan the observables are different as the (anti-)neutrino is not observed, therefore the standard cuts are {\tt etaecut} - the rapidity of the produced charged lepton, {\tt pTecut} - the transverse momentum of the charged lepton produced, {\tt pTmisscut} - the missing transverse momentum of the event (assumed to be from the (anti-)neutrino), and {\tt tmasscut} the transverse mass of the event as defined by $m_T = 2(|p_{T 1}||p_{T 2}|-p_{T 1}.p_{T 2})$. The cut requirements are $|\eta_e|< ${\tt etaecut}, $p_{T}^{\overset{\scriptscriptstyle(-)}{\nu}} > ${\tt pTmisscut}, $p_{T}^{e^{\pm}} >${\tt pTecut} and $m_{T} > ${\tt tmasscut}.
\end{enumerate}
\hfill \break
\item Histogram - Finally, the input file may be supplemented with inputs to the histogrammer, which is now incorporated within {\tt reSolve} and can determine the histogram files directly. For each differential cross-section desired, one line must be added of the form ``histo: \{variable\} \{additional info\}'' where the variables available are ``qT'', ``qq'', ``eta'', ``mT'', ``pTmin'' or ``pTmax''; the additional information is either the number of bins followed by the lower bound of the first bin and the upper bound of the final bin, or the endpoints of each bin (to allow non-uniformly distributed bins). In addition, the histogrammer can also be used alone, without re-running the whole resummation, provided the user has the event files (from a previous run). In order to run the histogrammer alone include the line {\tt hist\_only:1}, set the usual {\tt workdir} input line to the working directory of the event files to use for the histogrammer, and include the {\tt histo} lines relevant for the histograms you wish to calculate. More information is given in Section~\ref{histogrammer}.

\end{enumerate}

\begin{figure}[ht!]
\centerline{\includegraphics[height = 3cm, width = 3cm]{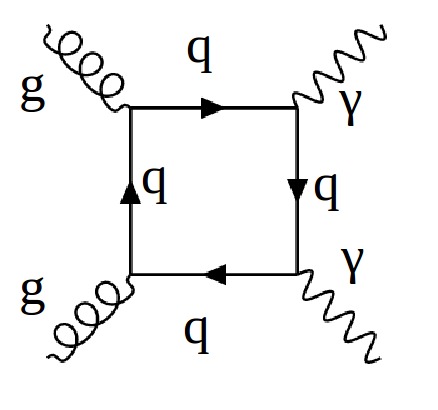}} 
\caption{gluon gluon box diagram contribution to diphoton production} \label{ggbox} 
\end{figure}

\subsection{Pdf Fits} \label{Multipddfits}

In order to implement the b-space resummation formalism of \cite{Catani:2013tia} (and references therein), the PDFs must be defined in Mellin space at generic complex values, which requires that we perform an analytic fit for the PDFs (as a function of $x$, at fixed scale). This PDF fit is done at the very start of the program once and for all, before the integration. If a fixed scale or a narrow range of invariant masses are used, a PDF fit at a single scale is satisfactory, indeed this is what is traditionally done in such resummation programs. In this case, the single fit is done at the input fixed factorisation scale $\mu_F$ at a momentum fraction set via $(QQ\_Max /CM\_energy)^2$. However, in cases where the invariant mass squared region considered is broad and one has a dynamical factorisation scale, then one may wish to improve precision by running with multiple PDF fits at various scales between $QQ\_Min$ and $QQ\_Max$. In order to do this set $muF\_flag$ and $mu\_F$ to $1$ and set the variable {\tt en\_sec\_multiplier} accordingly. It should be noted however that the first version of the program is currently significantly slower for multiple PDF fits, therefore we only recommend using this option in particular cases where very wide $QQ$ regions are used or if very accurate predictions are required. This will be optimised in future versions. The energy scale at which the PDF fits are performed is determined by the invariant mass range and the {\tt en\_sec\_multiplier} variable. The starting scale is taken as $QQ\_Min$, but this can be reset by setting {\tt mu\_min} in the input file in the resummation part of the file, the program starts with the scale {\tt Q\_start}$= min(${\tt mu\_min}$, $ {\tt QQ\_min}$)$. By default the value of {\tt mu\_min} is $20GeV$. The setting of the scales for the multiple PDF fits used is as follows, we denote {\tt en\_sec\_multiplier} by $E_n$:
\begin{enumerate}
\item To determine the first value of the factorisation scale at which a PDF fit is performed, {\tt reSolve} calculates $QQ\_temp = Q\_start \times E_n$, provided this is less than the $QQ\_Max$ then the scale the first PDF fit is performed at is then $QQ\_Min \times \sqrt{E_n}$, i.e. the geometric mean of $QQ\_Min$ and the new scale $QQ\_temp$. The program will then go on to perform another PDF fit, see step~\ref{nextfit}. \label{firstpdffit}
\item If however $QQ\_temp > QQ\_Max$ then {\tt reSolve} performs just one fit at the geometric mean of the endpoints of the invariant mass range, $\mu_F = \sqrt{QQ\_Min \times QQ\_Max}$. \sloppy
\item If in the previous steps, $QQ\_temp < QQ\_Max$ then {\tt reSolve} will perform a PDF fit at a further scale. Once more, the program will take the scale of the previous fit and multiply it by $E_n$ giving a new $QQ\_tempnew = QQ\_tempprev \times E_n$ and compare this new scale (which in the second fit case would now be $QQ\_Min \times E_{n}^2$) with $QQ\_Max$. Again if the new scale $QQ\_tempnew < QQ\_Max$ then the PDF fit will be performed at the new $QQ\_tempnew$ scale, i.e. at $\mu_F = \sqrt{QQ\_tempprev \times QQ\_tempnew}$. $QQ\_tempnew$ then becomes $QQ\_tempprev$ and we repeat this step~\ref{nextfit} until $QQ\_tempnew > QQ\_Max$. Once $QQ\_tempnew > QQ\_Max$ then the fit is the last fit and is performed at $\mu_F = \sqrt{QQ\_tempprev \times QQ\_Max}$. \label{nextfit}
\end{enumerate}

For example, with $QQ\_Min = 80$GeV, $QQ\_Max = 200$GeV and $E_n = 1.5$, the first fit would be performed at $\sqrt{80 \times 120} = 97.98$GeV, $120GeV < QQ\_Max$ so a second fit is performed. For this second fit, $QQ\_temp = 120 \times 1.5 = 180$GeV, which is less than $QQ\_Max$ still so a third fit will be performed and the second fit is performed at $\sqrt{120 \times 180} = 146.97$GeV.  For the third fit, $QQ\_tempnew = 180 \times 1.5 = 270 > QQ\_Max$, therefore this is the final fit and the fit is now performed at $\sqrt{QQ\_tempprev \times QQ\_Max} = \sqrt{180 \times 200} = 189.73$GeV.

{\tt PDF\_fussiness} allows nearby PDF fits to be used, for example setting it to 0.02 will ensure that PDF fits made at a scale within 2\% of the desired $\mu_F$ are used rather than a timely, completely new fit being performed. By default {\tt PDF\_fussiness} will be set to 0.01 if no input is provided.

For validation plots for running the {\tt reSolve} program with multiple PDF fits please see Section~\ref{multipdffits_validation}.

\hfill \break
\subsection{Parallelisation} \label{parallelisation}
With the input {\tt integrator\_flag}, the option to use either the built-in {\tt k\_vegas} Monte Carlo integration (1) or the external {\tt Cuba} Monte Carlo integration (2) are possible. Whilst the {\tt Cuba} implementation will by default parallelise over the number of cores of the machine used, the {\tt k\_vegas} implementation will automatically only use one core. However, the {\tt reSolve} program has been designed to allow multiple batches of events to be run on different cores, and indeed on different machines, and then combined after each iteration before again being distributed across the machines and cores available. This can therefore be used to parallelise across all the cores in the machine used, like {\tt Cuba}, or even to run across multiple machines. This can hasten the process of producing the events required for differential cross-section spectra greatly, depending upon the computer resources available. 

In order to run batches in parallel across multiple machines/cores, one must first turn the {\tt multi\_machine} input to 1, as described in Section~\ref{inputfile}. In addition, in order to avoid producing the same events across multiple cores, one must set {\tt seed} to -2, to allow the randoms' seeds to be set by converting the machine tag into a numerical seed different for each batch (in a deterministic and repeatable manner), or to -1, to set the randoms' seed based on time - which will be marginally different for the batch sent to each machine. For parallel running using {\tt k\_vegas}, unlike the standard running or {\tt Cuba} running, the maximum number of iterations is not set by {\tt maxeval} in the input file, rather it is set at the start of the parallelisation script in the variable {\tt max\_iter}, this is the number of iterations to run per core. The number of evaluations (phase space points) per iteration per core are set as usual via {\tt nstart} and {\tt nincrease} in the input file. The number of cores per machine is set via {\tt max\_cores} at the start of the paralellisation script. In order to parallelise across all the cores of just one machine, use the built in file {\tt single\_machine\_parallel}. To run this, type into the terminal the call shown in figure~\ref{singleparallel}.

\begin{figure}[ht!]
\begin{verbatim}
single_machine_parallel {path to input file}
\end{verbatim}
\caption{Terminal call to the {\tt single\_machine\_parallel} file used to parallelise the in-built {\tt k\_vegas} vegas Monte Carlo implementation across all the cores of a computer, if this call does not work the machine used may require the user to prefix the call with {\tt bash -x} in order to allow the terminal to determine it's a bash script. To run this, {\tt integrator\_flag} and {\tt multi\_machine} must both be set to 1 in the input file. Alternatively, the integrator {\tt Cuba} will automatically parallelise across the cores of the machine, this requires {\tt integrator\_flag} = 2.} \label{singleparallel} 
\end{figure}

\sloppy
It is important to note that for parallel runs, in the case where one uses the {\tt k\_vegas} integrator, whether across the cores of one machine or across many machines, the numbers of integration evaluations at the start and the increase in the number of evaluations from one iteration to the next; {\tt nstart} and {\tt nincrease}; are then the numbers per core. Therefore each core used in parallel will, in total, undertake $n\_\{tot\_per\_core\}$ evaluations, for a total of $n\_tot$ phase space points across all cores, given below in equations~\ref{ntotpercore} and \ref{ntot}. In running the {\tt single\_machine\_parallel} parallelisation script, the working directory for the input file used will be filled with event files of the form {\tt events\_lhe\_\{iter\_number\}.lhe}, each containing all the events from all the cores concatenated for that given iteration ({\tt lhe} here indicates the type of event output selected - {\tt save\_events} set to 2 in the input file). In addition, there are {\tt reSolve\_main\_out\_\_\{core\_number\}.dat} files giving the final iteration result and accumulated results for the total cross-section for the specified core. Finally, the {\tt reSolve\_main\_out\_END\_ITER.dat} file lists the total accumulated results across all cores for the total cross-section. Meanwhile, in the overall directory of the {\tt reSolve} program, {\tt nohup} files are produced, the files {\tt nohup\_\_.\{core\_number\}.out} contain the inputs and results for each core, the {\tt nohup\_iter\_\{iter\_number\}.out} files contain the Monte Carlo grids produced from each iteration.

\begin{equation} \label{ntotpercore}
n\_\{tot\_per\_core\} = [{\tt max\_iter}*{\tt nstart}+ 0.5*{\tt n\_increase}*{\tt max\_iter}*({\tt max\_iter}-1)]
\end{equation}
\begin{equation} \label{ntot}
n\_tot = {\tt max\_cores}*[{\tt max\_iter}*{\tt nstart}+ 0.5*{\tt n\_increase}*{\tt max\_iter}*({\tt max\_iter}-1)].
\end{equation}

\begin{figure}
\centerline{\includegraphics[height = 21cm, width = 16.5cm]{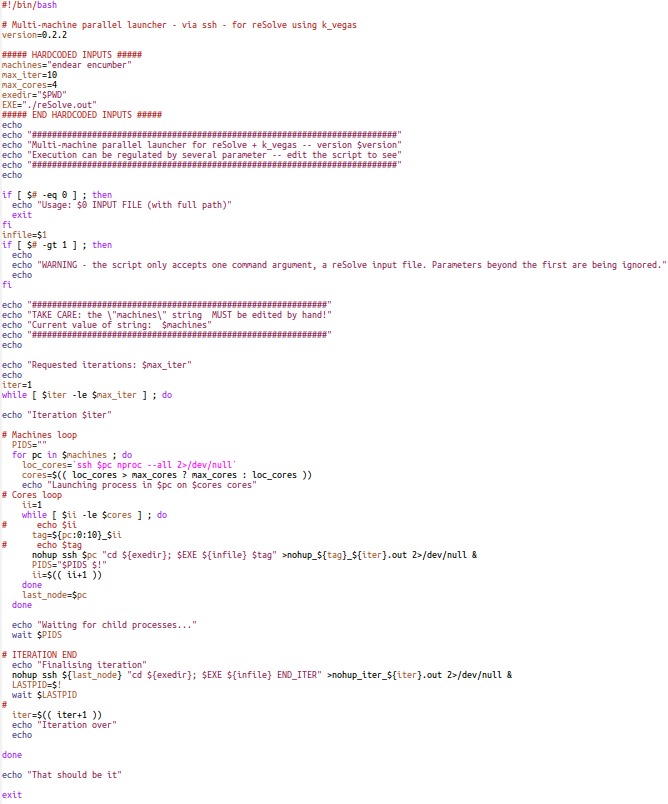}} 
\caption{The parallelisation program {\tt multi\_machine\_parallel\_local} for running across multiple machines, it is included with the program. To use this both {\tt integrator\_flag} and {\tt multi\_machine} must be set to 1 in the input file.} \label{multimachineparallel} 
\end{figure}

\begin{figure}
\centerline{\includegraphics[height = 3cm, width = 8cm]{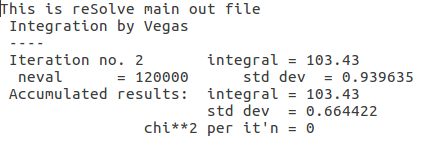}} 
\caption{The accumulated results outputted into the file ``{\tt reSolve\_main\_out\_END\_ITER.dat}'', here the results are after two iterations on each of 3 cores across 2 different machines. These results were for the input file {\tt yZ\_Born\_Tevatron\_parallel.dat} used with the {\tt multi\_machine\_parallel\_local} script, this input file is the $Z/\gamma*$ setup listed later in table~\ref{DYtestinputstable}, with {\tt multi\_machine} and {\tt integrator\_flag} set to 1 to allow the {\tt k\_vegas} multiple machine parallelisation and with the iteration numbers in the input file altered to {\tt nstart}$=5000$ and {\tt nincrease}$=15000$.} \label{parallelfinalanswer} 
\end{figure}

\begin{figure}
\centerline{\includegraphics[height = 21cm, width = 17cm]{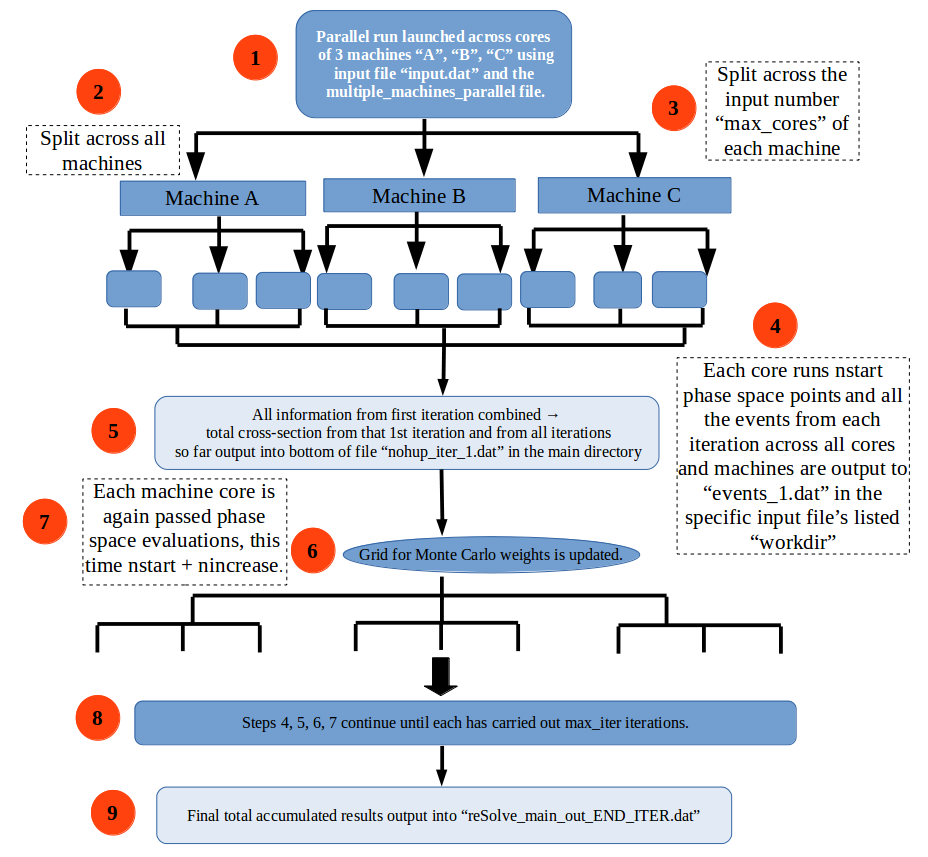}} 
\caption{Flowchart demonstrating the running of a parallel run across the cores of many machines, the script {\tt multi\_machine\_parallel\_local} may be used to perform this.} \label{parallelschematic} 
\end{figure}

\sloppy
If instead of running across all the cores of one machine, one instead wants to reduce the runtime even further by parallelising across many machines, one must use the integrator option {\tt k\_vegas} - setting {\tt integrator\_flag} to 1 in the input file. Again {\tt multi\_machine} must also be 1. With these settings, one may then use the script {\tt multi\_machine\_parallel\_local} to undertake parallel runs across many computers, this file is shown in figure~\ref{multimachineparallel}. This parallelisation script allows the use of multiple machines all of which are accessible via ssh on systems with or without a shared home directory in which to access input files and to output results to. The user must change the {\tt exedir} line to the directory in which the {\tt reSolve} program is installed on your machine, unless you are running the script from the {\tt reSolve} working directory. To enter which machines you wish to run on, enter the machine names into {\tt machines} at the top of the program. After saving the script, simply typing ``{\tt multi\_machine\_parallel\_local \{path to input file\}}'' into the terminal will set off a parallel run across the specified number of cores of all named machines, some machines may require this to be prefixed by {\tt bash -x} to indicate the script is a bash script. The information is then combined at the end of each iteration to update the grid, before using all machine cores for the subsequent iteration, this continues until all iterations are complete. The maximum number of iterations to be performed is set at the top of the file, as is the number of cores to use per machine, in {\tt max\_iter} and {\tt max\_cores} respectively, these were also set at the top of the {\tt single\_machine\_parallel} script. In the same way as was the case for the single machine parallelisation, in the multiple machine parallelisation the {\tt max\_iter} variable is used to set the number of iterations, with the {\tt maxeval} input used for single core {\tt k\_vegas} running or {\tt Cuba} running not relevant for {\tt k\_vegas} parallelisation. In the working directory for the input file, a file of the form {\tt reSolve\_main\_out\_\{machine\_name\}\_\{core\_number\}.dat} is created for each core on each machine used, listing the overall total cross-section for that machine core for both the final iteration and the accumulated results across all iterations, meanwhile {\tt reSolve\_main\_out\_END\_ITER.dat} lists the combined total cross-section across all machines, all cores for the final iteration and then the accumulated result across all machines, all cores and all iterations, an example of this file is given in figure~\ref{parallelfinalanswer}. The output event files are also output into the input file working directory. In addition, {\tt nohup\_\{machine\_name\}\_\{iteration\_number\}.out} files are created in the overall code working directory (one level above that of the input file's working directory), which contain all the usual terminal output for each core on each machine used. A schematic of a parallel run across many machines is given in figure~\ref{parallelschematic}. In parallel, histogram data files will be automatically generated by the {\tt reSolve} program as usual, nonetheless should the user wish to redo them, they can be re-binned or completely redone using the ``{\tt hist\_only}'' mode in exactly the same way as with running on one core or with {\tt Cuba} across one machine, see Section~\ref{histogrammer}. A description of the time taken to run in parallel compared with on one core is given in Section~\ref{speed}.

Sample input files which work with parallelisation, either across many cores of one machine, or across many ssh-accessibe cores, are included with the {\tt reSolve} program. These sample parallelisable input files are in the input directory and called  {\tt Diphoton\_Born\_parallel\_LHC.dat}, {\tt yZ\_Born\_Tevatron\_parallel.dat}, {\tt Diphoton\_NNLO\_test\_1\_\_parallel.dat} and {\tt yZ\_NNLO\_Tevatron\_parallel.dat}. These are the same setups of the {\tt Diphoton\_Born\_LHC.dat}, {\tt yZ\_Born\_Tevatron.dat}, {\tt Diphoton\_NNLO\_test\_1.dat} and {\tt yZ\_NNLO\_Tevatron.dat} files (used in the validation of the {\tt reSolve} program in Sections~\ref{BornCheckDiph}, \ref{BornCheckDY}, \ref{resurevampcomp} and \ref{DYresurevampcomp} respectively) except adapted for {\tt k\_vegas} parallelisation. 

In general, the parallelisation needed depends on the structure of the user's computer network, this varies significantly from one user to another therefore the user may have to make small changes to the scripts as appropriate for their computer resources.

\newpage

\subsection{\textbf{Output}}

\begin{figure}[ht!]
\centerline{\includegraphics[height = 10cm, width = 12cm]{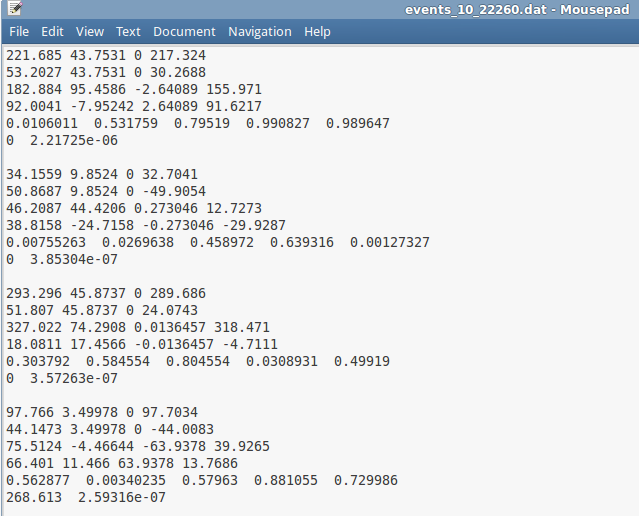}} 
\caption{Output file of {\tt reSolve}, this file is in the ``easy'' form. In this form each event is represented by six rows of information and followed by a blank row. The first two rows are the 4-momenta of the incoming partons, the next two rows are the outgoing photon four-momenta, the fifth row is then the set of 5 random values between 0 and 1 used to generate this event and its momenta. These randoms set the invariant mass squared  - $qq^2$, the transverse momentum squared - $q_T^2$, the rapidity - $\eta$, and the $\theta$ and $\phi$ opening angles of the two photons in the diphoton centre of mass frame. Finally the last row gives the value of the cross-section here (in $GeV$) (0 if cut) and the weight of the event in the Monte Carlo.} \label{outputeasy} 
\end{figure}

The events produced by {\tt reSolve} will be saved provided the flag ``{\tt save\_events}'' is set to 1 for the ``easy'' form or 2 for the ``pseudo-lhe'' form, contrastingly, if it is set to 0 only the total cross-section is output and is in the {\tt reSolve\_main\_out.dat} file. Events are required if differential cross-sections are desired rather than just the total cross section, by default events are output and will be output in the ``easy'' form into your {\tt workdir}, split into a different file for each iteration of the program. The events will all be automatically read by the histogrammer to determine the histograms specified in the input file. A sample output event file  (in the ``easy'' form) is shown in Figure~\ref{outputeasy}. Each event details the 4-momenta of the incoming partons, the 2 outgoing particle 4-momenta, the random values used to define the phase space point by the Monte Carlo and finally the event cross-section (in $GeV$) and event weight.

As for the output to terminal, with verbosity set to 1 the program will only output the result after each iteration. Higher verbosity settings provide more information but the primary use of such information is debugging and will slow down the program as many values may be output to terminal each iteration. Figure~\ref{iterationoutput} gives the form of the output after each iteration.

\begin{figure}
\begin{verbatim}
Iteration 2:  30000 integrand evaluations so far
[1] 3.90839 +- 0.612087  	chisq 0.0447161 (1 df)
\end{verbatim} 
\caption{Terminal output of {\tt reSolve} after its second iteration. It lists the iteration number, the total number of integrand evaluations carried out, the cross-section in $GeV$ with an error, and the chi-squared with number of degrees of freedom. If parallelisation and {\tt k\_vegas} is used such information appears in output files after each iteration, rather than in the terminal.} \label{iterationoutput} 
\end{figure}

The cross-section estimate given by the Monte Carlo iteration on iteration is $\Sigma_i \frac{1}{N} w_i f_i$, where $w_i$ is the weight of the event, $N$ is the total number of events and $f_i$ is the cross-section estimate for that given phase space point. The error estimate meanwhile is given by the square root of the standard variance estimate in \eqref{variance}. The factor of $N-1$ ensures we obtain the variance of the mean rather than of the $f_i$ evaluations.

\begin{equation} \label{variance}
Var = \frac{1}{N-1}\Sigma_i [(f_i^2  w_i^2) - (f_i w_i)^2]
\end{equation}

The Monte Carlo cross-section iterations and their error estimates are then combined by weighted average, where the weights are the inverse of the variance estimates.

The exact same process is carried out in the histogrammer included with the program, except the weighted averaging is now done bin by bin in order to produce a fully differential cross-section, in whichever observables are required, rather than just the total cross-section. The error given on the cross-section is consequently only approximate and should only be used as a judgement after a couple of iterations. After each iteration the grid used to weight the Monte Carlo events is updated, the chi-squared gives an indication of how well the grid approximates the integral, this is estimated via the difference between the weighted events and the cross-section estimate at that iteration (the mean), weighted according to the variance. This chi-squared should be divided by the number of degrees of freedom (which is equal to the iteration number minus 1) to understand how good the estimate is. This is produced as described in \cite{1978JCoPh..27..192L} and in the {\tt Cuba} package \cite{Hahn:2004fe}. The chi squared divided by the number of degrees in freedom (shown in brackets) should therefore decline iteration on iteration, again after the first few iterations are passed. 

The overall accumulated results across all iterations are also output into the file {\tt ``reSolve\_main\_out.dat''}, or the file {\tt ``reSolve\_main\_out\_END\_ITER.dat''} if ran across many machines.

\hfill \break

\subsection{Histogrammer} \label{histogrammer}

The program package also includes a histogrammer package, this can be used to read in the events produced by {\tt reSolve} to produce the necessary differential cross-section invariant mass, transverse momentum or other differential cross-section histogram data.

The histogrammer has now been incorporated into the program in order to simplify the use of {\tt reSolve}. With the histogrammer incorporated into the {\tt reSolve} program, the user must simply include a few lines in the usual {\tt reSolve} input file detailing the desired histograms and the binning. In the section ``Histograms'', add a line of the form given in Figure~\ref{histograminput}, begin by indicating this is histogrammer input with {\tt histo} and then follow by the variable for the histogrammer, the number of bins required and the start bin lower bound and final bin upper bound. This will then calculate the events in each bin for the number given of evenly spaced bins across the range specified. The option {\tt \{variable\}} can be {\tt qT}, {\tt qq}, {\tt eta}, {\tt mT}, {\tt pTmin} or {\tt pTmax} for the transverse momentum spectrum, invariant mass spectrum, rapidity distribution, transverse mass distribution and distribution of the minimum/maximum transverse momentum of the (two) outgoing particles respectively. If, rather than evenly spaced bins, the user requires unevenly distributed bins, enter a ``0.'' where the number of bins is input, and instead proceed by entering the endpoints of every bin, this is useful in allowing finer bin spacings at the lower end of the transverse momentum spectrum, where resummation is crucial. Histogram information will only be calculated for each variable specified, therefore if no lines of the form given in Figure~\ref{histograminput} are included in the input file, the {\tt reSolve} program will produce the events and total cross-section only. Of course, as had to be done in the previous beta version of {\tt reSolve}, these events can then be used to determine the differential cross-sections required via a standalone histogram calculation, the histogrammer can be used without performing the whole resummation calculation again provided there are the required event files from previous runs of {\tt reSolve}. In order to run just the histogrammer, for example should the user wish to re-bin the events or determine further histograms not initially specified in the input file, one must include the {\tt hist\_only} flag set to 1 (if it is not present it is set to 0 by default). In this case the resummation calculation and event generation will all be skipped and instead {\tt reSolve} will read the event files specified in {\tt workdir} and determine the histograms specified by the ``{\tt histo: \{variable\} \{binning info\} }'' input lines. Figure \ref{histonlyinput} provides a sample input file where only the histogramming is done. It is important to note that the histogrammer produces the cross-section in each bin normalised by the bin width (rather than the total cross-section in the bin), this ensures the amplitude is independent of the binning used.

\begin{figure} [!htbp]
\begin{center}
\begin{verbatim}
histo: {variable} {no. of bins} {start bin lower end} {final bin upper end}
\end{verbatim}
\end{center}
\caption{Line required for the histogrammer in the {\tt reSolve} input file.} \label{histograminput} 
\end{figure}

\begin{figure} [!htbp]
\centerline{\includegraphics[height = 3cm, width = 5cm]{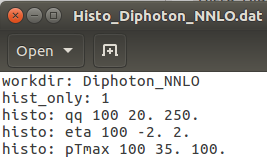}} 
\caption{Sample input file to run {\tt reSolve} so that it only does the histogram calculation, reading in previously generated events, here from the working directory {\tt Diphoton\_NNLO/} and determing the invariant mass ($qq$), rapidity ($eta$) and maximum transverse momentum ($pTmax$) differential cross-section histogram data.} \label{histonlyinput} 
\end{figure}

\begin{figure} [!htbp]
\centerline{\includegraphics[height = 6cm, width = 5cm]{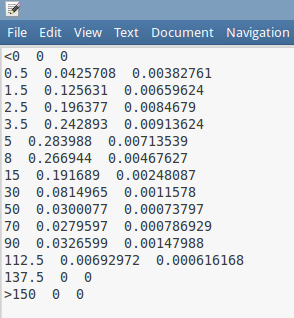}} 
\caption{Output file for $q_T$ from the histogrammer included with the program.} \label{histoout} 
\end{figure}

Once the program has run, it will then produce all the events as before (provided the {\tt save\_events} flag is set to 1 or 2 not 0, and that {\tt hist\_only} mode is not being run) in addition to output files detailing the histogram plotting information for each of the desired differential distributions. With all of the above variables specified, the {\tt reSolve} program will produce output files ``{\tt histo\_0\_qT.dat}'', ``{\tt histo\_1\_qq.dat}'', ``{\tt histo\_2\_eta.dat}'', ``{\tt histo\_3\_mT.dat}'', ``{\tt histo\_4\_pTmin.dat}'', ``{\tt histo\_5\_pTmax.dat}''. Each of these files lists the centre-points of each bin and the corresponding cross-section in that bin followed by a ``0'' column (the error in the position of the bin) and then an estimate of the error in that bin. A sample output file for the qT differential cross-section is given in figure~\ref{histoout}. In addition, for all of the validation and results figures in this paper in Section~\ref{validation}, the histogram data used is provided with the paper.

\newpage

\section{\textbf{Program Structure}} \label{programstruc}

\begin{figure}[ht!]
\centerline{\includegraphics[height = 3cm, width = 12cm]{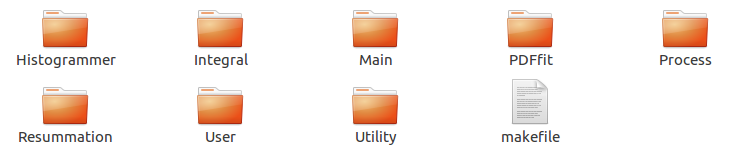}} 
\caption{The folder structure of the {\tt reSolve} code. It is modular, splitting up the different parts of the calculation.} \label{folders} 
\end{figure}

The program structure is intended to be modular, with the different aspects of the calculation divided into different folders, see Figure~\ref{folders}. In principle several of these - Histogrammer, Integral, PDFfit, Resummation and Utility - can be used independently of the main code, this is important for the straightforward extension of the program to additional processes. The details of the calculations in each folder are detailed below in Section~\ref{folderstructure} and following this, an explanation of how the program works is given in Section~\ref{programoutline}. This is intended to allow users to understand how the program functions in order to both simplify its use and enable users to add their own processes should they wish.

\subsection{Modular Structure} \label{folderstructure}

\begin{itemize}

\item Histogrammer - This contains the files required to read in the event files generated by {\tt reSolve} in the {\tt workdir} and calculate the cross-section per bin width in the desired differential cross-sections. The histogrammer can also be used on previously generated events, without re-running the whole resummation calculation, by using the {\tt hist\_only} flag.
\item Integral - This contains the files necessary to perform the Monte Carlo phase space integration, either using the in-house {\tt k\_vegas} integrator or the external {\tt Cuba} integrator, it also contains the Fortran file which determines the 5 random values for the Monte Carlo event generation. The file {\tt events\_readin.cc} can be used to read in events from other programs, its use is largely in debugging and can be ignored with the program running normally. 
\item Main - This contains all the general files used to read in the input file, perform pre-processing, post-processing and interface with the Monte Carlo integration routines.  
\item PDFfit - This contains the routines used to evaluate the fit parameters for the PDF fits and then to link to the Fortran routine used to evaluate the PDF fits, as well as to output them in a form useful for the rest of the program. 
\item Process - This folder contains any process specific code. So far the program only calculates the resummed part of the diphoton or Drell-Yan differential transverse momentum spectra. Each process folder includes a file routine to read the process specific parts of the input file. The file {\tt \{process\}\_ps.cc} uses the randoms generated to determine the phase space point for each event - i.e. the invariant mass squared, transverse momentum squared and rapidity of the diphoton/Drell-Yan system as well as the $\theta$ and $\phi$ angles in the centre of mass frame. These set the particle four-momenta. These variables then determine the $\theta$ angle of the diphoton/Drell-Yan system with respect to the incoming partons as well as the phase space Jacobian and the relevant $s$, $t$, $u$ Mandelstam variables. A cut file is also included, called to determine if a point generated by the Monte Carlo passes the cuts or not and therefore whether it is evaluated. The process dependent parts of the hard function calculation are calculated in the {\tt \{process\}\_hard.cc} file, including the Born cross-section and $H_q^{\gamma\gamma (1)}$ and $H_q^{\gamma\gamma (2)}$ coefficients in Drell-Yan scheme, see section~\ref{rescoeffs} equations~\ref{H1yy} and \ref{H2yy}. Finally there is an integrand file {\tt \{process\}\_integrand.cc} which organises these aspects of the calculation and calls the generic resummation code in order to determine the cross-section for each event. The output ``dumper'' routines are then called to output the event information. 
\item Resummation - This folder contains the resummation routines which are process independent, and therefore will also be called for other processes. This includes the inverse Fourier and inverse Mellin transforms and the process independent parts of the hard function calculation, as well as the Sudakov factors. This aspect of the calculation is all directed by calling the ``resummed'' routine in {\tt inv\_fourier.cc}. More details on how the resummation calculation is performed are given in section~\ref{programoutline} and section~\ref{theory}. The hard functions calculated within the program for now include only those required for the diphoton background and Drell-Yan processes calculations. The hard factors for gluon-gluon initiated processes are only included at leading order currently. This is all that is required for NNLL resummation for the diphoton process and even this is not required for Drell-Yan. For diphoton background calculations, gluon-gluon initiated processes begin at two orders beyond the qq initiated processes (which are evaluated here at next-to-next-to-leading logarithm). For other processes, for example Higgs signals, the gluon gluon contributions will need to be added in up to NNLL, this is left to future versions of {\tt reSolve}.
\item User - This is a folder where the user can call additional routines they may write for pre-processing, Monte Carlo or post-processing, as well as to read process-specific input. Currently, it contains routines to allow parallelisation of the {\tt k\_vegas} integrator across multiple cores and multiple machines, as well as the switch between the interfaces to either read events into the integrator or evaluate the phase space integral using {\tt k\_vegas} or {\tt Cuba}.
\item Utility - This folder contains auxiliary functions necessary for the program; including the {\tt alphaS.f} Fortran routine for $\alpha_s$ evolution, the output ``dumper'' routines to output the events used in ``easy'' or ``pseudo-lhe'' form, the routines used for the Lorentz algebra, and initialisation routines associated with the PDFs.
\end{itemize}

This program structure is designed to modularise the program, this enables the straightforward extension of the program into other processes, indeed we intend to perform this further ourselves in the future. In order to add a new process one must simply replace the process sub-folder with equivalent files for the new process. Depending on the process, additional hard factors may need to be added to the hard function calculation in the resummation and potentially also the different orders included for the qq and gg processes as their relative order contributions depend upon the process. Currently, qq hard factors are included up to NNLL, whilst gg hard factors are only up to LL, this will be resolved in future versions as we extend the program to additional processes. A detailed guide on how to add a process to the {\tt reSolve} program is given in Section~\ref{AddProcess}.

In addition to the code structure given here, the {\tt input} folder in the main {\tt reSolve} directory contains all the input files used for validations and results described later in this paper in Section~\ref{validation}, in addition to a folder for the events produced in running each of these input files. These setups and the results generated are described in detail in Section~\ref{validation}.

\pagebreak

\subsection{Program Flow} \label{programoutline}

\begin{figure}
\centerline{\includegraphics[height = 23cm, width = 16cm]{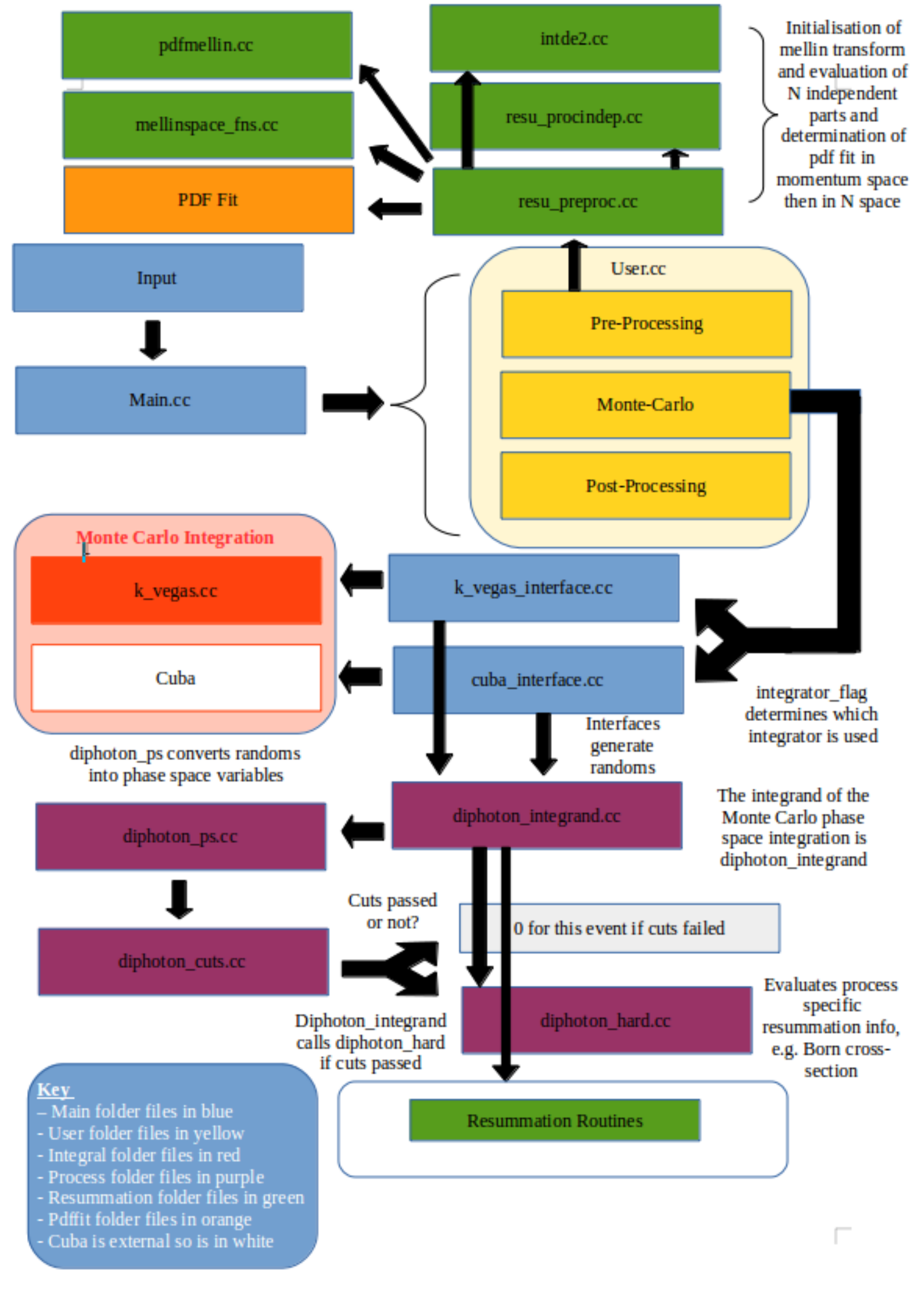}} 
\caption{A flowchart demonstrating the different aspects included in the program and what is called when in the calculations. The different aspects of the program are coloured differently to indicate where they sit in the program folder structure. A zoom in of the resummation routines at the bottom of the flowchart is given in figure~\ref{resummationflowchart}. The program functions analogously in the case of Drell-Yan processes.} \label{programflowchart} 
\end{figure}

\begin{figure}
\centerline{\includegraphics[trim= 0 200 0 40, clip, height = 18cm, width = 16cm]{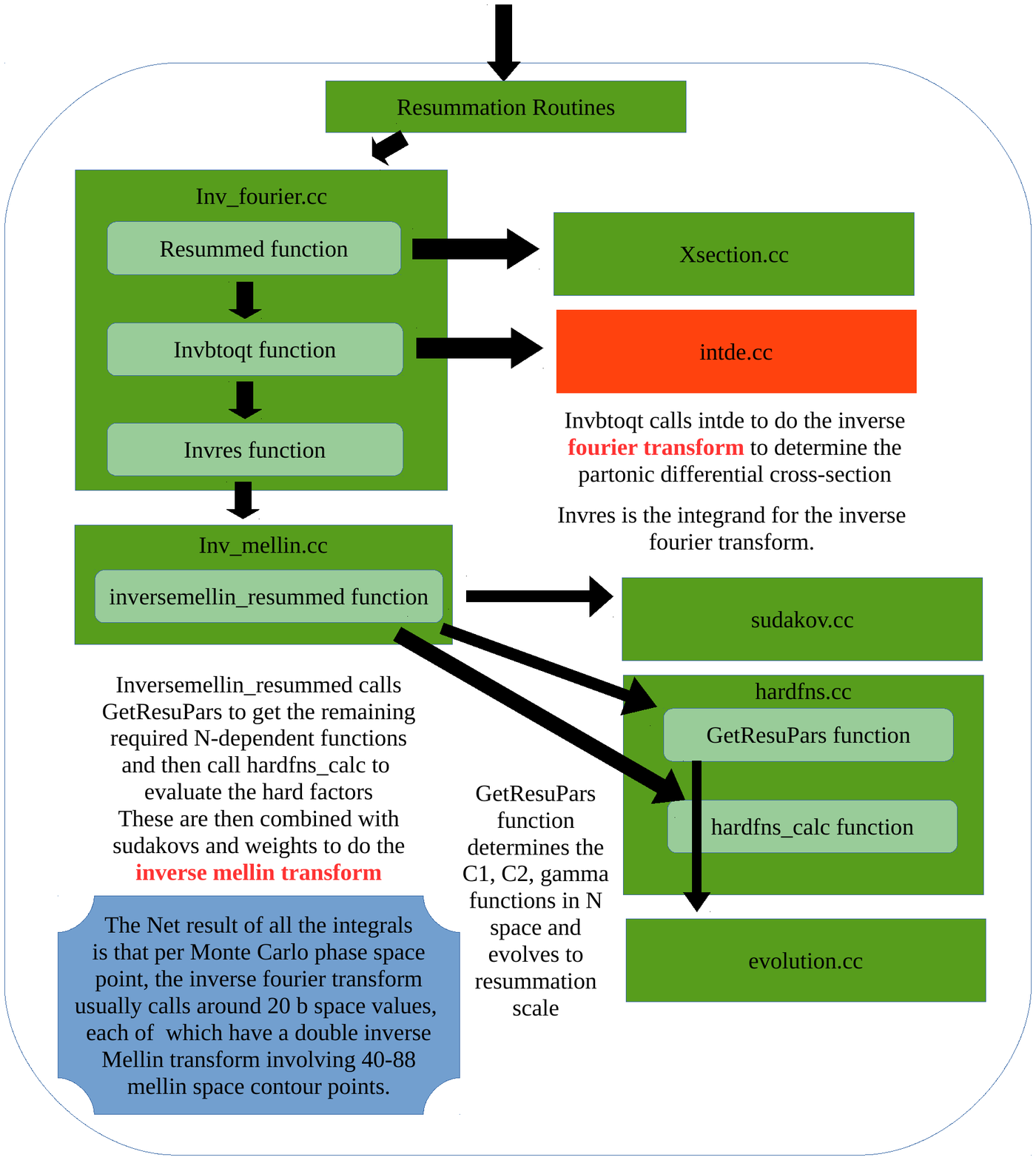}} 
\caption{A flowchart providing more detail on the resummation aspect of the program, which is the main part of the calculation. This highlights how both the inverse Fourier and double inverse Mellin transforms are performed. This part of the program is process independent.} \label{resummationflowchart} 
\end{figure}

As detailed in section~\ref{theory}, the resummation formalism of \cite{Bozzi:2005wk} requires the determination of the following integrals; phase space integral - evaluated via Monte Carlo, inverse Fourier transform from impact parameter ($b$) space to transverse momentum ($q_T$) space, and an inverse Mellin transform from Mellin space ($N_1$, $N_2$) to momentum fraction space ($z_1$, $z_2$) for each of the incoming partons. Figure~\ref{programflowchart} shows in detail the flow of the calculations performed, starting at {\tt main.cc}, figure~\ref{resummationflowchart} then illustrates in more detail the resummation aspects of the calculation. The colours of the boxes indicate where in the program structure the various files and routines lie, with the key also given in figure~\ref{programflowchart}.

\hfill \break

Our program {\tt reSolve} works as follows, this description and the figures are for the diphoton process, but the exact same sequence occurs for the Drell-Yan processes with the appropriate process-specific files in the Process sub folder interchanged:
\begin{enumerate}
\item The calculation begins in {\tt main.cc}, this calls {\tt InputPars.cc} to read in the various inputs provided in the input file and described in Section~\ref{inputfile}. From here {\tt user.cc} is called to carry out required processing before the Monte Carlo integration.
\item This pre-processing includes calling {\tt resu\_preproc.cc}, this carries out various initialisations including those of the inverse Fourier transform, N-independent resummation parameters (via {\tt resu\_procindep.cc}), inverse Mellin transform contour and N-dependent resummation parameters (via {\tt mellinspace\_fns.cc}), and others. It also calls the PDF fitting routines and calls {\tt pdfmellin.cc} to convert the PDF fits into Mellin space.
\item {\tt User.cc} then moves onto the Monte Carlo aspect of the program, calling \\ {\tt k\_vegas\_interface.cc} or {\tt cuba\_interface.cc} if the input {\tt integrator\_flag} is $1$ or $2$ respectively. These programs call the random generator and pass any points evaluated to the Monte Carlo evaluation programs - either the in-house {\tt k\_vegas} or the external {\tt Cuba}.
\item The Monte Carlo interface programs themselves then call the process specific files - first {\tt diphoton\_integrand.cc}, this is the integrand of the Monte Carlo integral. It calls {\tt diphoton\_ps} to convert the randoms generated into a phase space point; determining $qq^2$, $q_T^2$, $\eta$, $\theta_{CM}$ and $\phi_{CM}$, whilst also evaluating the Jacobian for the phase space integral.
\item {\tt diphoton\_integrand.cc} next calls {\tt diphoton\_cuts.cc}, this has the task of reading in the cut information from the input file and determining whether the phase space point being considered passes the cuts or not. Provided the cuts are passed, {\tt diphoton\_hard.cc} is called, this evaluates the Born cross-section and other process-specific resummation variables; such as the $H^{1}_q$, $H^{2}_q$, $H^{1}_g$ and $H^{2}_g$ hard factors in the Drell-Yan scheme.
\item {\tt reSolve} now moves onto the general resummation routines as {\tt diphoton\_integrand.cc} calls resummed in {\tt inv\_fourier.cc}. This section of the calculation is shown in more detail in figure~\ref{resummationflowchart}.
\item First in the resummation routines, {\tt inv\_fourier.cc} determines the correction factor required to account for the fact that the PDFs are fit, this is done by evaluating the LO cross-section directly with standard PDFs and then with the PDF fit and using the ratio as a correction factor. This is all determined in {\tt xsection.cc}.
\item {\tt invbtoqt} is now called. The role of this routine is to call {\tt intde2.cc} to perform the inverse Fourier transform back from $b$ space to $q_T$ space. The integrand of this inverse Fourier transform is the routine {\tt invres}.
\item {\tt invbtoqt} calls invres for a number of different points in impact parameter space, usually of the order of 20, depending on the precise details of the convergence. For each $b$ value, invres must evaluate the double inverse Mellin transform used to perform the resummation. This is done via the routine {\tt inversemellin\_resummed}.
\item {\tt inversemellin\_resummed} in {\tt inv\_mellin.cc} organises the double inverse Mellin transform calculation, this calculation is built directly into the code. Variables $b_*$ ({\tt bstar} in the code) and $b_{log} (${\tt blog} in the code) are introduced to regulate singularities seen at large and small $b$ respectively, see section~\ref{sec:FourierAndMellin}:
\begin{equation*}
b_* = \dfrac{b}{\sqrt{1+b^2/b_{lim}^2}}, \quad b_{log} = \log\left[{1+\dfrac{q^2 b_{*}^{2}}{{b_{0}^{'}}^2}}\right], \ \text{ with} \quad b_{lim} = \dfrac{b_{0}^{'}}{q}\exp\left(\dfrac{1}{2\alpha_{s}\beta_0}\right) \ \text{ and} \quad b_{0}^{'}=2\exp\left[\gamma \dfrac{q}{\mu_{S}}\right]
\end{equation*} 
(notice that the $b_{log}$ variable is just $\tilde{L}$ as defined in eq.~\eqref{tildeL} with the replacement $b \to b_*$)

\item First, the Sudakov form factors for soft gluon and soft quark emission are calculated by calling {\tt sudakov.cc}. Then {\tt GetResuPars} determines the $C_1$, $C_2$, anomalous dimensions and other N-dependent basis functions in Mellin space and evolves them to the resummation scale $\mu_{S}$.
\item {\tt hard\_fns.cc} next determines the hard factors, incorporating the virtual diagram contributions into the resummation.
\item The sudakovs, hard factors and appropriate weights are used at each of $40-88$ points along the contour in Mellin space, with the number of points depending on the rapidities of the two photons, this is done for each inverse Mellin transform. Note this contour is in the complex plane in order to allow for faster convergence.
\item The contributions at each point are then summed along the contour to calculate the double inverse Mellin transform.
\item Putting all this information together gives the inverse Mellin transformations, if these are called for each of around 20 $b$ values this allows the determination of the inverse Fourier transform for each phase space point. Repeating the process for $\mathcal{O}(10^5-10^7)$ randomly distributed phase space points and including the effects of the Jacobian transformation between the randoms space volume and the phase space volume, {\tt reSolve} thereby determines the overall cross-section. This information is printed out after each iteration; meanwhile all the events, their individual cross-sections and the randoms associated with their phase space points are output into output files (one per iteration), this information can be used to re-create the phase space variables and so produce histograms of the differential cross-section in invariant mass ($qq$), transverse momentum ($q_T$), rapidity ($\eta$), transverse mass ($m_T$), or maximum and minimum transverse momenta for the two photons ($p_T^{\{max/min\}}$).

\end{enumerate}

The Monte Carlo phase space integration works by selectively refining the grid from which it draws the randoms iteration on iteration so as to maximise the sampling where the integrand is largest, it does this by importance sampling and stratified sampling \cite{Hahn:2004fe} \cite{1978JCoPh..27..192L}. The result is each successive iteration should produce a more accurate estimate; in addition, the number of evaluations per iteration typically increases iteration on iteration (set by {\tt nincrease}) and therefore the statistical fluctuations will also reduce.

\subsection{Adding a Process to {\tt reSolve}} \label{AddProcess}

\begin{figure}
\centerline{\includegraphics[height = 1.5cm, width = 5cm]{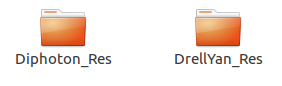}} 
\caption{The Process subfolder, which contains (nearly) all the process-dependent aspects of the resummation program. A new subfolder within it will be created for each additional process. See Figure~\ref{processfiles} for the routines included} \label{processsubfolder} 
\end{figure}

\begin{figure}
\centerline{\includegraphics[height = 4cm, width = 8cm]{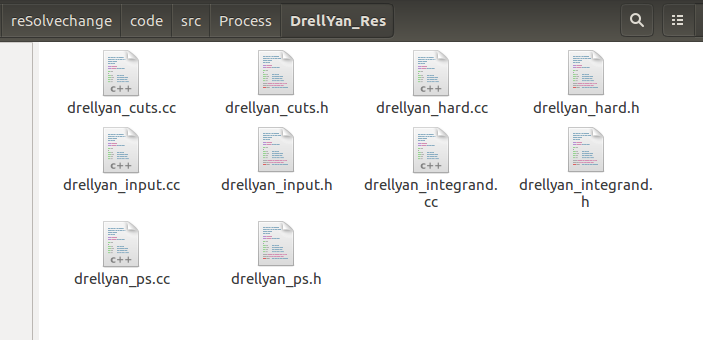}} 
\caption{The {\tt DrellYan\_Res} subfolder (contained in the Process subfolder), which contains the Drell-Yan specific routines and links with the generic resummation parts of the {\tt reSolve} program. An analogous subfolder and routines therein exists for the diphoton process and should be recreated for any processes added.} \label{processfiles} 
\end{figure}

Following the description of the structure of the {\tt reSolve} program, and how it works, in the previous two sections (Sections \ref{folderstructure} and \ref{programoutline} respectively), we now outline how to add a new process to the program. In order to produce the full differential cross-sections, in, for example, transverse momentum, {\tt reSolve} adds $q_T$ resummation to a generic process so as to accurately produce the low $q_T$ part of the spectrum, which is often the technically most-challenging piece. The user must then add this transverse momentum resummed spectrum, produced by {\tt reSolve}, to the usual finite piece and match them appropriately to obtain the complete spectrum over all $q_T$. The resummation part of the {\tt reSolve} program is completely process-independent, therefore to add a new process one must simply undertake the following steps; where {\tt \{process\}} appears in a routine, file or folder name below it is to be replaced by a suitable name for the added process.
\begin{itemize}
\item Set a process number for the added process, given $1$ indicates diphoton and $2$ indicates Drell-Yan, for the first added process take $3$. In {\tt User.cc} extend the else if sequence to include the added process if the process flag is set to 3, to do so copy the code section for either diphoton or Drell-Yan to produce the new calls for the new process. This includes calling a new routine {\tt \{process\}\_setup} which will be defined in the file ``\{process\}\_input.cc'' which we will create later in the new sub folder ``Process/\{process\}\_Res/''. 
\item Enter the Process sub folder (see Figure~\ref{processsubfolder}), this is the sub folder which contains the process dependent routines. There are currently two sub folders within it, one for diphoton ({\tt Diphoton\_Res}) and one for Drell-Yan ({\tt DrellYan\_Res}), each contains files and routines {\tt \{process\}\_cuts.cc}, {\tt \{process\}\_hard.cc}, {\tt \{process\}\_input.cc}, {\tt \{process\}\_integrand.cc},  {\tt \{process\}\_ps.cc}, and corresponding header files, see Figure~\ref{processfiles}. The goal is to produce appropriate corresponding files for the new process. First create the new process sub folder {\tt \{process\}\_Res}. Next, we will begin with the aforementioned process-specific input file  input routine {\tt ReadInput\_\{process\}} in {\tt \{process\}\_input.cc}, this reads the process-specific input from the input file ``\{process\}\_input.cc''. This contains two parts, first there is the {\tt \{process\}\_setup} routine which organises the program pre-resummation for the new process.  To create this, the form of these files for diphoton or Drell-Yan can be copied, with appropriate changes in the number of random variables required for the Monte Carlo ({\tt ndim}) and other variables such as the number of particles ({\tt npart}). This setup routine will call the routine {\tt \{process\}\_ReadInput}, again the basic form of this will be analogous to the diphoton and Drell-Yan cases, with the new relevant cuts for the process under consideration; for example diphoton has an ``R cut'' based on the angular separation of the two photons, whilst Drell-Yan may have a transverse mass cut and ``pTmiss'' cut. Create also the corresponding header file, including the class {\tt \{process\}\_input} used to pass this input information to the cut-checking routine later.
\item Next create the {\tt \{process\}\_integrand.cc} file and corresponding routine. This routine is called from {\tt User.cc} via the {\tt k\_vegas\_call} or {\tt cuba\_call}, and coordinates the main calculations for this process. Once more, the general form of these files for the diphoton or Drell-Yan case may be copied. First, the routine calls a phase-space generating routine {\tt \{process\}\_ps} in order to generate the randoms and phase space for each process event, this is contained in the process specific file {\tt \{process\}\_ps.cc}. Next the cuts relevant to the process phase space are checked by calling the {\tt \{process\}\_cuts} routine in the {\tt \{process\}\_cuts.cc} file, if the cuts are passed (``false'' means not cut here, i.e. cuts passed) then the Born-level cross-section is determined via the routine {\tt sigmaij\{process\}calc} in the file {\tt \{process\}\_hard.cc}. The relevant information is then loaded into the resuminfo object containing all the necessary information needed for the process-independent resummation part of the calculation, which is implemented by calling the ``resummed'' routine in inverse Fourier. This determines the overall cross-section for each event, including resummation up to NNLL, finally the events are then output in whichever form is indicated in the input file.
\item This {\tt \{process\}\_integrand} routine directed the resummation calculation by calling the relevant phase space generating, cut-checking and Born cross-section determining process-specific routines; these are {\tt \{process\}\_ps}, {\tt \{process\}\_cuts} and {\tt \{process\}\_hard} respectively and must be created for the new added process. First consider the phase space generation; this routine reads in randoms generated in the rest of the code and uses them to set the relevant parameters for the process phase space, these are parameters such as the invariant mass squared ($q^2$), transverse momentum squared (${q_T}^2$), $\eta$ of the diphoton/Drell-Yan system and $\theta_{CM}$ and $\phi_{CM}$ angles describing the plane in which the diphotons/Drell-Yan system are emitted within the centre of mass frame of the colliding partons. The Jacobian for the transformation between these random variables, whose values are between 0 and 1, and the phase space variables, is given by ``randsjacob''. A kinematics routine is then called to determine the 4-momenta and angular separation of the relevant particles (e.g. the two photons for diphoton, two leptons for Z Drell-Yan, etc) in the lab frame in order to allow later application of the cuts. Various other variables such as the factorisation, resummation and renormalisation scales are also set here; this will be repeated for the new process and so can be copied; all this information on phase space variables, kinematics and other variables is then passed into the resuvars object to be passed to the relevant sections of the resummation code as and when necessary.
\item The {\tt \{process\}\_cuts.cc} file and routine's overall structure may again be duplicated from the diphoton and Drell-Yan examples. There are two broad types of cuts contained, first generic kinematical constraints are applied to cut if either $|\eta|$ or $q_T$ become too large that the resummation formalism is no longer valid. These are the ``gencuts'' and should be kept for all processes. Secondly, there are the process specific, phase space cuts, ``pscuts'', which are checked via the {\tt PS\{process\}cuts\_1} routine. This routine will have to be written anew for each added process, nonetheless it is straightforward. The process-specific cut information read in from the input file (via the {\tt ReadInput\_\{process\}} routine discussed earlier) is passed in via a {\tt \{process\}\_input} object, whilst the phase space and event information is passed via a ``PSpoint'' object. With this information, the relevant phase space kinematic parameters can be determined for each event and tested against the cuts.
\item Finally, the {\tt \{process\}\_hard.cc} file is where most potential difficulty lies in ensuring consistency of factors. The {\tt sigmaij\{process\}calc} routine uses the input process information and event phase space point parameters to determine the Born-level cross-section for the added process. This is then loaded into the sigmaij vector array to be used elsewhere in the code, for example in computation of the hard factors in the resummation code. Process-specific hard factors are also calculated within this file (although not in the Drell-Yan case as the Drell-Yan scheme has been used - see Section~\ref{sec:resscheme} for further information).
\item It is also necessary to add a new section in the file {\tt hardfns.cc} in {\tt Resummation/}, there are sections of code here which detail the contributions from gluon-gluon initiation, quark-gluon initiation, quark-quark initiation to the hard factors. In the interests of the speed of evaluation of the program, only the non-zero contributions for each process are explicitly summed, for example for diphoton or $Z/\gamma$ Drell-Yan only $q \bar{q}$ is summed over as the final quarks at the stage of the diphoton production must be the same, whilst for $W^{\pm}$ Drell-Yan the contributions are $q \bar{q'}$ and, as a result of the CKM matrix, can occur with $q$ and $q'$ of different generations, such as $u \bar{s}$. Aside from determining which contributions are non-zero and summing these appropriately, one may copy the diphoton and Drell-Yan code here. It is worth noting that the structure of the theoretical formalism here is process-independent, one must just sum over all contributions (including zero contributions) for each process, the small process dependence introduced in {\tt reSolve} is purely a pragmatic one, to avoid wasting time summing many zero contributions in the time-critical part of the program.
\end{itemize}

All of the remainder of the program should remain exactly as it is, the calculation of the hard factors, sudakovs, determination of the inverse Mellin transforms, the inverse Fourier transform from impact parameter space, the Monte Carlo phase space integration and everything else required will be calculated automatically by the program. In this way {\tt reSolve} takes advantage of the generality of the b-space Mellin-space resummation formalism of \cite{Dokshitzer:1978hw,Parisi:1979se,Curci:1979bg,Collins:1981uk,Kodaira:1981nh,Collins:1984kg,Catani:1988vd,deFlorian:2000pr,Catani:2010pd,Catani:2013tia}, which is described in Section~\ref{theory} and in \ref{app:coeffs} of this paper. However, in the current version only the hard factors for qq initiated processes are included up to NNLL, those for gg initiated processes are only included at LL, therefore processes requiring these hard factors beyond LL will need the additional higher order hard factors to be added. We intend to resolve this in the immediate future, indeed it is necessary for the adding of the Higgs diphoton signal process and should not be of great difficulty.

\pagebreak

\section{Transverse momentum resummation theory and formalism} \label{theory}

The central feature of {\tt reSolve} is the addition of transverse momentum ($q_T$) resummation to a rather general class of inclusive hadronic processes, of which the processes included in the current version, $h_1 h_2 \to \gamma\gamma$ and $h_1 h_2 \to W^{\pm}/Z$ (with the $h_i$ here being either a $p$ or a $\bar p$) are just an example. We thus devote this section to a brief review of the specific $q_T$ resummation formalism used in the program.

The physical motivation for $q_T$ resummation lies in the well-known fact that, when measuring transverse momenta $q_T$ which are much softer than the typical hard scale of a process $\mu$, the appearance of large logarithmic terms (of the generic form $\as^n (\log(\mu / q_T))^m$, with $m\leqslant2n$, where $\as$ is the strong coupling constant) spoils the convergence of the perturbative series. In these cases, contributions to all orders in $\as$ must be resummed in order to recover reliable predictions for observable quantities. There are different approaches which can be followed to achieve this resummation; the formalism used in {\tt reSolve} is the impact parameter ($b$) space formalism of refs.~\cite{Dokshitzer:1978hw,Parisi:1979se,Curci:1979bg,Collins:1981uk,Kodaira:1981nh,Collins:1984kg,Catani:1988vd,deFlorian:2000pr,Catani:2010pd,Catani:2013tia}. In the following we will not give any derivation of the final formulae used in the code, but just state them and comment on how they are actually implemented in {\tt reSolve}. We will make an effort to distinguish between what is inherent in the formalism and what is a choice related to our specific implementation. We mainly follow the notation used in ref.~\cite{Catani:2013tia}.

The class of processes we will consider are those of the form $h_1 h_2 \to F + X $, where $h_1$ and $h_2$ are hadrons and the observed final state $F$ is made up of an arbitrary number of particles, with the only constraint that they be not strongly interacting (that is, no jets or hadrons are allowed in $F$). The $X$ denotes (as usual) unresolved extra radiation. Our master formula~\cite{Catani:2013tia} for the \emph{fully differential} cross-section $\dif \sigma$ at low $q_T$ for the $h_1 h_2 \to F + X $ process is then:
\begin{equation}
\label{master0}
\begin{split}
& \frac{\dif\sigma_{res}^F(p_1,p_2,M^2,\mathbf{q}_T,y,\mathbf{\Omega})}{\dif M^2 \dif^2 \mathbf{q}_T \dif y \dif \mathbf{\Omega}} =  \int \frac{\dif^2 b}{(2\pi)^2} \int_{x_1}^1 \frac{\dif z_1}{z_1} \int_{x_2}^1 \frac{\dif z_2}{z_2} W^F(\mathbf{b},z_1,z_2, \ldots) \equiv \frac{M^2}{s} \times\\
& \ \left[\dif\hat\sigma_{c\bar{c}}^{F,\,\text{LO}}\right] \int \frac{\dif^2 b}{(2\pi)^2} e^{\ui \mathbf{b}\cdot\mathbf{q}_T} S_c(M^2,b_0^2/b^2) \int_{x_1}^1 \frac{\dif z_1}{z_1} \int_{x_2}^1 \frac{\dif z_2}{z_2}
\left[ H^F C_1 C_2 \right]_{c\bar c; a_1 a_2} f_{a_1/h_1}(x_1/z_1, b_0^2/b^2)
f_{a_2/h_2}(x_2/z_2, b_0^2/b^2) \; .
\end{split}
\end{equation}
Eq.~\eqref{master0} is given as a function of the hadron momenta $p_1$ and $p_2$, the total invariant mass ($M^2$, also denoted $q^2$ or ${qq}^2$ elsewhere), transverse momentum (a 2-dimensional vector $\mathbf{q}_T$, with $q_T = |\mathbf{q}_T|$, and similarly $b = \mathbf{b}$) and rapidity $y$ of the final state $F$, as well as on any additional variables (collectively denoted by $\mathbf{\Omega}$) internal to $F$ that may be needed to fully define $F$ phase space. The indices $a_1$, $a_2$ and $c$ are flavour indices ($\bar c$ denotes the anti-parton of $c$) and a sum over them is implicit. The $f_{a_i/h_i}$ are usual PDFs. $b_0 = 2 e^{-\gamma_E}$ (with $\gamma_E = 0.5722\ldots$), and the partonic momentum fractions $x_1$ and $x_2$ are fixed and given by
\begin{equation}
\label{x1x2}
x_1 = \frac{M}{\sqrt{s}} e^{+y} \ , \qquad x_2 = \frac{M}{\sqrt{s}} e^{-y} \; .
\end{equation}
As noted, eq.~\eqref{master0} is an approximation that holds for low values of $q_T$: more explicitly, this means that the expression is valid for $q_T/M \ll 1$ up to corrections formally of $\mathcal{O}(q_T^2/M^2)$ (in other words, the hard scale of the process is taken to be of order $M$). To keep in mind this restriction, a suffix \emph{res} (for ``resummed'') is appended to the differential cross-section symbol. To recover a description valid for arbitrary values of $q_T$, some prescription for a matching with a high-$q_T$ description for the same process (which in turn can be obtained via a usual fixed order expansion) must be provided. There are several different approaches to defining such a matching, with an additive approach typically being used in the $b$-space formalism~\cite{Bozzi:2005wk}. Note however, that no matching is required, if only the lower end of the $q_T$ spectrum is to be analysed. No matching scheme is implemented in the current version of {\tt reSolve}, rather {\tt reSolve} is aimed at computing the resummed (i.e. low $q_T$) part of the differential $q_T$ spectrum, which is often the most challenging piece.

\begin{figure}[!ht]
\begin{center}
\includegraphics[width=0.8\textwidth]{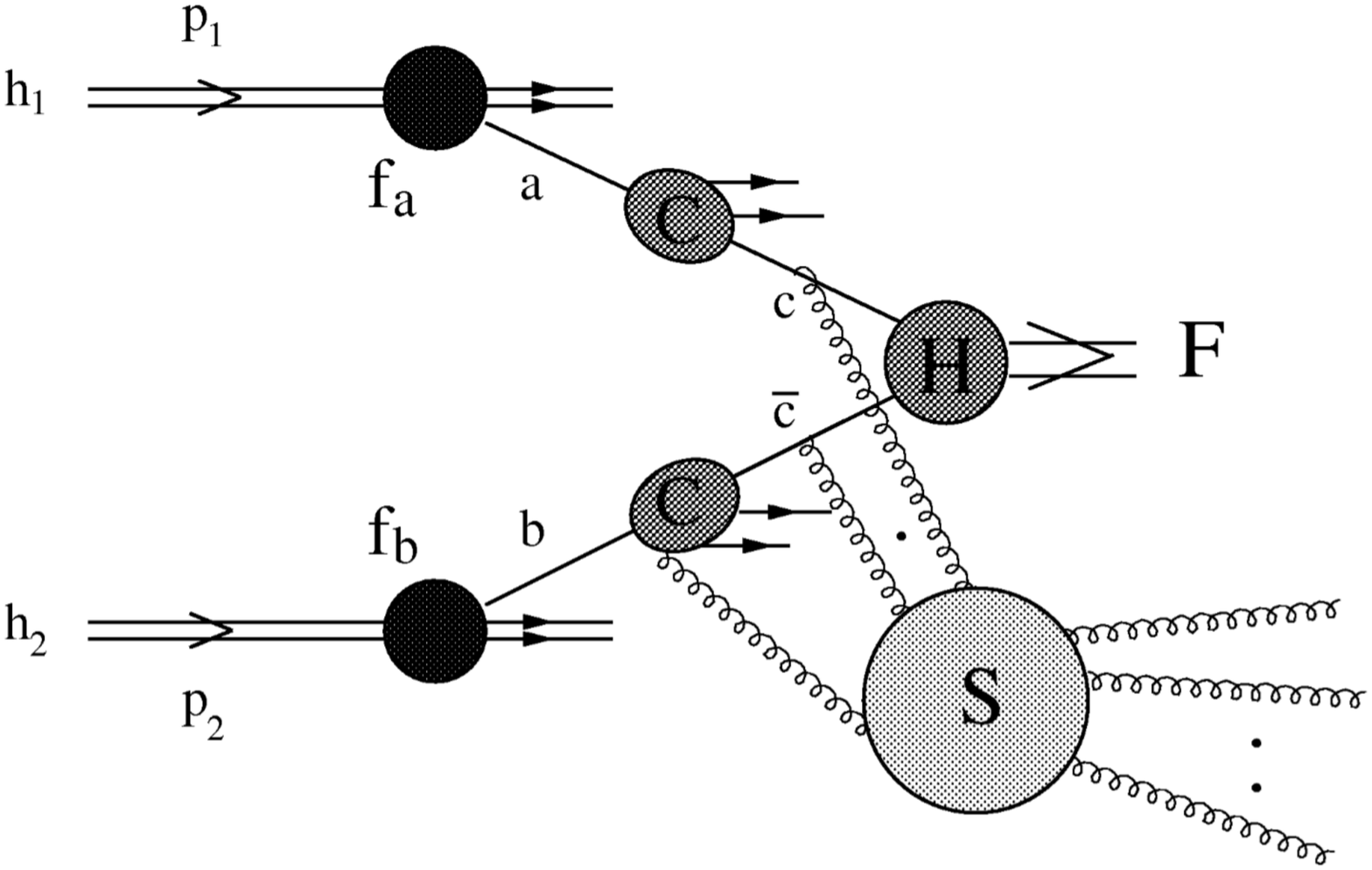}
\caption{\small{\textit{Pictorial version of eq.~\eqref{master0}: a parton of momentum fraction $x_1$ is extracted from hadron $h_1$ in the upper leg, then it splits ($z_1$) via a collinear partonic sub process $a\to c$ at the upper $C$ blob. A specular process happens at the lower leg: in the end the momenta that enter the hard process $H$ are $x_1 z_1 p_1$ and $x_2 z_2 p_2$, where $p_{1,2}$ are the momenta of the initial hadrons. Soft partons can be emitted anywhere (except inside $H$ itself) and contribute to the Sudakov form factor $S_c$. }}}
\label{fig:master0}
\end{center}
\end{figure}

The physical meaning of eq.~\eqref{master0} is nicely illustrated by fig.~\ref{fig:master0}, which also gives an intuitive understanding of how the various pieces of the formula arise.

Several comments are needed in order to fully appreciate the structure of eq.~\eqref{master0}, as a number of its ingredients are written in a rather implicit form.
$S_c$ is the Sudakov form factor, which is \emph{universal}, that is, process-independent (the two terms will be used interchangeably throughout this section) and given by~\cite{Collins:1984kg}
\begin{equation}
\label{sudakov}
S_a (\mu_2^2,\mu_1^2) = \exp\left\{{-\int_{\mu_1^2}^{\mu_2^2} \frac{\dif q^2}{q^2} \left[ A_a(\as(q^2)) \log\frac{\mu_2^2}{q^2} + B_a(\as(q^2))  \right]} \right\} \;,
\end{equation}
where $\mu_1$, $\mu_2$ are any two scales. The functions $A_a$ and $B_a$ depend only on $\as$ and admit the perturbative expansion
\begin{equation}
\label{ABexp}
A_a(\as) = \sum_{n=1}^\infty \left(\frac{\as}{\pi}\right)^n A_a^{(n)} \;, \qquad B_a(\as) = \sum_{n=1}^\infty \left(\frac{\as}{\pi}\right)^n B_a^{(n)} \; ,
\end{equation}
and the coefficients up to $A_a^{(3)}$ and $B_a^{(2)}$ are explicitly known. The expressions for these and other resummation coefficients will be collected, along with the references they are originally collected from, in \ref{app:coeffs}. The Sudakov form factor~\eqref{sudakov} resums to all orders, in eq.~\eqref{master0}, logarithmically-enhanced contributions of the form $\log(M^2 b^2)$ (the region $q_T/M \ll 1$ corresponds, in impact parameter space, to $Mb \gg 1$). To see this explicitly, the $(\dif q^2)/q^2 = \dif \log q^2$ integral in eq.~\eqref{sudakov} must be carried out; the most straightforward way to do this is to use the $\as$ evolution equation to express $\as(q^2)$ in terms of $\as(M^2)$ via the $\as$ evolution equation
\begin{equation}
\label{asevol}
\begin{split}
\frac{\dif \log(\as(q^2))}{\dif \log q^2} & = \beta(\as(q^2)) \equiv -\sum_{n=0}^\infty \beta_n \as(q^2)^{n+1} \\
\To \as(q^2) & = \frac{\as(M^2)}{l} - \left( \frac{\as(M^2)}{l} \right)^2 \frac{\beta_1}{\beta_0} \log l + \ldots \; ,
\end{split}
\end{equation}
where $l = 1+ \beta_0 \as(M^2) \log(q^2/M^2)$. This way, the scale $q^2$ only appears through the logarithmic term $\as(M^2) \log(q^2/M^2)$ in the integrand and, carrying out the integration, one then obtains an expression of the form
\begin{equation}
-\int_{b_0^2/b^2}^{M^2} \frac{\dif q^2}{q^2} \left[ A_a(\as(q^2)) \log\frac{\mu_2^2}{q^2} + B_a(\as(q^2)) \right] =
\left(\frac{\as(M^2)}{\pi}\right)^{-1} \bar{g}^{(1)} + \left(\frac{\as(M^2)}{\pi}\right)^{0} \bar{g}^{(2)} + \left(\frac{\as(M^2)}{\pi}\right) \bar{g}^{(3)} + \ldots \; ,
\end{equation}
where the coefficients $\bar{g}^{(n)}$ contain contributions to all orders in the combination $\as\log(b^2/b_0^2 M^2)$, which is formally of $\mathcal{O}(1)$, but that are suppressed by growing powers of the small parameter $\as(M^2)/\pi$. The first order in this expansion is usually called Leading Log (LL), the following one Next-to-Leading Log (NLL), and so on. The explicit expressions of the first three terms are
\begin{align}
\label{LL}
\bar{g}^{(1)} & = \frac{A^{(1)}}{\beta_0} \frac{\lambda + \log(1-\lambda)}{\lambda} \; , \\
\label{NLL}
\begin{split}
\bar{g}^{(2)} & = \frac{B^{(1)}}{\beta_0}\log(1-\lambda) - \frac{A^{(2)}}{\beta_0^2}
\left(\frac{\lambda}{1-\lambda} + \log(1-\lambda) \right) \\
& + \frac{A^{(1)} \beta_1}{\beta_0^3} \left( \frac{1}{2}\log^2(1-\lambda) + \frac{\log(1-\lambda)}{1-\lambda}
+ \frac{\lambda}{1-\lambda} \right) \; ,
\end{split} \\
\label{NNLL}
\begin{split}
\bar{g}^{(3)} & = -\frac{A^{(3)}}{2\beta_0^2} \frac{\lambda^2}{1-\lambda} -\frac{B^{(2)}}{\beta_0}
\frac{\lambda}{1-\lambda} + \frac{A^{(2)} \beta_1}{\beta_0^3} \left( \frac{\lambda(3\lambda -2)}{2(1-\lambda)^2}
- \frac{(1-2\lambda)\log(1-\lambda)}{(1-\lambda)^2} \right) \\
& + \frac{B^{(1)} \beta_1}{\beta_0^2} \left( \frac{\lambda}{1-\lambda} + \frac{\log(1-\lambda)}{1-\lambda} \right)
+ A^{(1)} \left( \frac{\beta_1^2}{2\beta_0^4} \frac{1-2\lambda}{(1-\lambda)^2} \log^2(1-\lambda) \right. \\
& \left. +\log(1-\lambda) \left( \frac{\beta_0\beta_2 - \beta_1^2}{\beta_0^4} +
\frac{\beta_1^2}{\beta_0^4(1-\lambda)} \right) \right)\; ,
\end{split}
\end{align}
where the $\log(M^2 b^2)$ terms are contained in the variable $\lambda$:
\begin{equation}
\lambda = \frac{1}{\pi} \beta_0 \alpha_s(M^2) \log(M^2 b^2/b_0^2) \; .
\end{equation}
These are also the terms currently implemented in the program, which thus can perform resummations up to NNLL.

It is important to note that the Sudakov form factor~\eqref{sudakov} is not the only source of $b$-dependence -- and thus of large logarithms -- in eq.~\eqref{master0}. One more contributions comes from the PDFs, which are also functions of $b$ through their scale dependence. This dependence can be extracted, and the PDFs evaluated at the usual (hard) factorization scale $\mu_F \sim M$ via the standard PDF evolution equation, whose integral solution reads
\begin{equation}
f_{a/h}(x, b_0^2/b^2) = \int_x^1 \dif z  U_{ab} (x/z; b_0^2/b^2, \mu_F^2)f_{a_1/h_1}(z, \mu_F^2) \; ,
\end{equation}
and the kernels $U_{ab} (x/z; b_0^2/b^2, \mu_F^2)$ also contribute to the all-order resummation of logarithmic $\as\log(b^2/b_0^2 \mu_F^2) \sim \as\log(b^2/b_0^2 M^2)$. The $U_{ab}$ evolution kernels are also implemented in the program in a dedicated routine.

As already stated, the Sudakov factor is universal, as of course are the PDFs. The process dependence is in fact completely contained in the last two remaining factors which build up eq.~\eqref{master0}, the first of which, $\left[\dif\hat\sigma_{c\bar{c}}^{F,\,\text{LO}}\right]$, is essentially trivial, being simply the Born-level partonic cross-section for the process $c\bar c \to F$. More explicitly
\begin{equation}
\left[\dif\hat\sigma_{c\bar{c}}^{F,\,\text{LO}}\right] = \frac{\dif\hat\sigma_{c\bar{c}}^{F,\,\text{LO}}}{M^2 \dif\mathbf{\Omega}}(x_1 p_1, x_2 p_2, \mathbf\Omega, \as(M^2)) \; .
\end{equation}
This term obviously contains no $b$ dependence.
Notice that since all particles in $F$ are colourless, the process at parton-level can only be initiated by opposite-coloured partons, that is, either $q_i \bar{q}_j$ or $gg$, which is the reason for the label $c\bar c$ rather than a more general $c c'$.

The remaining process-dependent factor in eq.~\eqref{master0} is $\left[ H^F C_1 C_2 \right]_{c\bar c; a_1 a_2}$, and its explicit expression is different depending on whether the Born-level partonic process is initiated by $q\bar q$ or $gg$.
In the former case, its explicit form is
\begin{equation}
\label{HCCqq}
\left[ H^F C_1 C_2 \right]_{q \bar{q}; a_1 a_2} = H_q^F(x_1 p_1, x_2 p_2, \mathbf\Omega, \as(M^2), \mu_R) C_{q a_1}(z_1, \as(b_0^2/b^2)) C_{\bar q a_2}(z_2, \as(b_0^2/b^2)) \; ,
\end{equation}
where the $H^F_q$ and $C_{q a}$ admit a perturbative expansion similar to the ones in~\eqref{ABexp}:
\begin{equation}
\begin{split}
H_q^F(x_1 p_1, x_2 p_2, \mathbf\Omega, \as(M^2), \mu_R) & = 1 + \sum_{n=1}^\infty \left(\frac{\as}{\pi}\right)^n H_q^{F(n)}(x_1 p_1, x_2 p_2, \mathbf\Omega,\mu_R) \\
C_{q a}(z,\as) & = \delta_{q a} \delta(1-z) + \sum_{n=1}^\infty \left(\frac{\as}{\pi}\right)^n C_{q a}^{(n)}(z)
\end{split}
\end{equation}
(note that a dependence of the coefficients $H_a$ and $C_{qa}$ on the specific flavours of the quark $q_i$ and anti quark $\bar{q}_j$ never arises from QCD corrections thanks to flavour symmetry, since quark mass effects are neglected in eq.~\eqref{master0}. In $H_a$ however such a dependence can in principle still arise due to, for instance, charge effects).
The coefficients $C_{q a}^{(n)}$ are all known up to order 2 and are process-independent. The explicit expressions for the $C_{q a}^{(n)}$ are given in \ref{app:coeffs}. The $C_{q a}(z, \as)$ also contribute one last piece of $b$-dependence, since they are evaluated at $\as(b_0^2/b^2)$. To make this dependence more explicit, we can use the general relation
\begin{equation}
h(\as(\mu_1^2)) = h(\as(\mu_2^2)) \exp\left[ \int_{\mu_1^2}^{\mu_2^2} \frac{\dif q^2}{q^2} \beta(\as(q^2)) \frac{\dif \log(h(\as(q^2)))}{\dif \log(\as(q^2))} \right]\; ,
\end{equation}
valid for any function of $\as$. With this, we can rewrite
\begin{equation}
\label{Cexp}
C_{q a}(z, \as(b_0^2/b^2)) = C_{q a}(z, \as(M^2)) \exp\left[ \int_{b_0^2/b^2}^{M^2} \frac{\dif q^2}{q^2} \beta(\as(q^2)) \frac{\dif \log(C_{q a}(z, \as(q^2)))}{\dif \log(\as(q^2))} \right]
\end{equation}
(this is not a matrix equation: it holds element by element for $C_{q a}$), so that each $C_{q a}$ coefficient is now expressed as a function of $\as$ at the \emph{perturbative} scale $M^2$ and an integral of the same form as~\eqref{sudakov}, which gives the final contribution to the large $\log(M^2 b^2)$ logarithm resummation. Meanwhile, the process dependence comes only from the ``hard factor'' $H_q^{F(n)}$, which is however $b$-independent and does not contain any large logarithms.

In the case of a $gg$-initiated process, the term $\left[ H^F C_1 C_2 \right]_{gg; a_1 a_2}$ has a richer structure, due to correlations~\cite{Catani:2010pd} produced by the evolution of the colliding partons $a_1$, $a_2$ into gluons. The general considerations are however the same. Explicitly we have:
\begin{equation}
\label{HCCgg}
\begin{split}
\left[ H^F C_1 C_2 \right]_{gg; a_1 a_2} &= \sum_{h_1 h_1 \lambda_1 \lambda_2} H_g^{F (h_1 \lambda_1)(h_2 \lambda_2)}(x_1 p_1, x_2 p_2, \mathbf\Omega, \as(M^2), \mu_R) \\
& C_{g a_1}^{h_1\lambda_1}(z_1, p_1, \mathbf{b}, \as(b_0^2/b^2)) \, C_{g a_2}^{h_2\lambda_2}(z_2, p_2, \mathbf{b}, \as(b_0^2/b^2)) \; ,
\end{split}
\end{equation}
where $h_{1,2}$ and $\lambda_{1,2}$ are helicity indices. The terms $C_{g a_i}^{\lambda_i h_i}$ can be decomposed into \emph{helicity conserving} and \emph{helicity flipping} components
\begin{equation}
C_{ga_i}^{\lambda_i h_i}(z_i,p_i,\mathbf{b},\as) = C_{ga_i}(z_i,\as) \delta^{\lambda_i, h_i} + G_{ga_i}(z_i,\as) D^{(\lambda_i)}(p_i, \mathbf{b}) \delta^{\lambda_i, -h_i} \; , \quad i=1,2 \; ;
\end{equation}
here $D^{(\lambda_i)}(p_i, \mathbf{b})$, which embodies the full $\mathbf{b}$ and $p_i$ dependence of the $C_{g a_i}^{\lambda_i h_i}$ coefficient, is in fact a simple phase factor
\begin{equation}
D^{(\lambda_i)}(p_i, \mathbf{b}) = -e^{\pm 2\ui \lambda_i (\varphi(\mathbf{b}) - \varphi_i)} \; ,
\end{equation}
where $\varphi(\mathbf{b})$, $\varphi_1$, $\varphi_2$ are the azimuthal angles of $\mathbf{b}$, $p_1$ and $p_2$ respectively, and the overall sign of the exponent is ``$+$'' if $i=1$ and ``$-$'' if $i=2$. Again, the $H_g^{F (h_1 \lambda_1)(h_2 \lambda_2)}$, $C_{ga}$ and $G_{ga_i}$ admit perturbative expansions
\begin{equation}
\begin{split}
& H_g^{F (h_1 \lambda_1)(h_2 \lambda_2)}(x_1 p_1, x_2 p_2, \mathbf\Omega, \as(M^2), \mu_R) = 1 + \sum_{n=1}^\infty \left(\frac{\as}{\pi}\right)^n H_g^{F (h_1 \lambda_1)(h_2 \lambda_2)(n)}(x_1 p_1, x_2 p_2, \mathbf\Omega,\mu_R) \\
& C_{q a}(z,\as) = \delta_{q a} \delta(1-z) + \sum_{n=1}^\infty \left(\frac{\as}{\pi}\right)^n C_{g a}^{(n)}(z) \; , \quad G_{g a}(z,\as) = \sum_{n=1}^\infty \left(\frac{\as}{\pi}\right)^n G_{g a}^{(n)}(z) \; ,
\end{split}
\end{equation}
where it is significant to notice that the helicity flipping coefficients $G_{g a}$ have no $\mathcal{O} (\as)^0$ term. Both the $C_{ga}^{(n)}$ and the $G_{ga}^{(n)}$ are, as their analogues $C_{q a}$, universal and known up to order 2 (with their explicit expressions given in ~\ref{app:coeffs}), and can be expressed as functions of $\as(M^2)$ times exponential terms which contribute to large $\log$ resummation. The process dependence is, in this case, embodied in the hard factor $H_g^{F (h_1 \lambda_1)(h_2 \lambda_2)}$ which is, as in the $q\bar q$ case, $b$-independent.

It is useful to further clarify the explicit form of the hard factors  $H_q^F$ and $H_g^{F (h_1 \lambda_1)(h_2 \lambda_2)}$: in fact, apart from the complication arising from helicity correlation in the latter, their structure is very similar. In both cases, these terms include the contributions from all \emph{purely virtual} corrections to the $c\bar c \to F$ partonic process, up to a specified order. The explicit definition of the $H_q^F$ and $H_g^{F (h_1 \lambda_1)(h_2 \lambda_2)}$ factors can be given in terms of a \emph{UV-renormalised} and \emph{IR subtracted} helicity scattering amplitude, $\tilde{\mathcal{M}}_{c\bar c \to F}^{h_1 h_2}$. The UV renormalisation is carried out by standard means in a specified renormalisation scheme, typically $\overline{\text{MS}}$ (this is the case for {\tt reSolve}). The \emph{UV-renormalised} scattering amplitude $\mathcal{M}_{c\bar c \to F}^{h_1 h_2}$ is then transformed to $\tilde{\mathcal{M}}_{c\bar c \to F}^{h_1 h_2}$ via a subtraction operation, which can be written to all orders as
\begin{equation}
\label{subtraction}
\tilde{\mathcal{M}}_{c\bar c \to F}^{h_1 h_2}(x_1 p_1, x_2 p_2,\Omega, \mu_R) = \left(1 - \tilde{I}_c(\epsilon, M^2, \mu_R)\right) \mathcal{M}_{c\bar c \to F}^{h_1 h_2}(x_1 p_1, x_2 p_2,\Omega, \mu_R, \epsilon) \; ,
\end{equation}
where $\tilde{I}_c(\epsilon, M^2, \mu_R)$ is an universal subtraction operator and $\mu_R$ the renormalisation scale. Here we supposed $\mathcal{M}_{c\bar c \to F}^{h_1 h_2}$ to be evaluated in dimensional regularization, and $\epsilon = d-4$ is the corresponding dimensional regularization parameter. $\mathcal{M}_{c\bar c \to F}^{h_1 h_2}$ will be in general $\epsilon$ dependent because, even though it is renormalised, it will still contain divergences of IR origins. For the sake of brevity, we will not report the specific form (and perturbative expansion) here, referring to the reader to the original paper~\cite{Catani:2013tia}. The important point is that the subtraction in eq.~\eqref{subtraction}, exactly cancels all $\epsilon$ poles in $\mathcal{M}_{c\bar c \to F}^{h_1 h_2}$, leaving the left-hand side of eq.~\eqref{subtraction} finite.

In terms of $\tilde{\mathcal{M}}_{c\bar c \to F}^{h_1 h_2}$, the hard factors $H_q^F$ and $H_g^{F (h_1 \lambda_1)(h_2 \lambda_2)}$ can now be written as
\begin{equation}
\label{Htruedef}
H_q^F = \frac{|\tilde{\mathcal{M}}_{q\bar q \to F}|^2}{|\mathcal{M}_{q\bar q \to F (0)}|^2} \; , \qquad
H_g^{F (h_1 \lambda_1)(h_2 \lambda_2)} = \frac{\left[ \tilde{\mathcal{M}}_{q\bar q \to F}^{h_1 h_2} \right]^* \tilde{\mathcal{M}}_{q\bar q \to F}^{\lambda_1 \lambda_2}}{|\mathcal{M}_{q\bar q \to F (0)}|^2} \; ,
\end{equation}
where the squared matrix elements have an implicit summation over helicity indices, that is
\begin{equation}
|\tilde{\mathcal{M}}_{q\bar q \to F}|^2 = \sum_{h_1 h_2} \left[ \tilde{\mathcal{M}}_{q\bar q \to F}^{h_1 h_2} \right]^* \tilde{\mathcal{M}}_{q\bar q \to F}^{h_1 h_2}
\end{equation}
and $\mathcal{M}_{q\bar q \to F (0)}^{h_1}$ is the Born scattering amplitude (the tilde is omitted since the Born term is finite by itself).

Eq.~\eqref{Htruedef} makes the relation between the gluon-initiated process hard factor $H_g^{F (h_1 \lambda_1)(h_2 \lambda_2)}$ and the quark-initiated one $H_q^F$ explicit and easy to understand: if we could turn off the spin flipping coefficients $G_{ga}$, substituting eq.~\eqref{Htruedef} in eqs.~\eqref{HCCqq} and~\eqref{HCCgg} respectively would yield exactly specular expressions. By contrast, in the actual full expression for $\left[ H^F C_1 C_2 \right]_{gg; a_1 a_2}$, extra contributions involving the $G_{ga}$ appear. It is interesting to note~\cite{Catani:2010pd}, though we will not show it here, that whenever an azimuthally-averaged observable (such as the $q_T$ spectrum) is considered, any cross-terms involving both $C_{ga}$ and $G_{ga}$ vanish to all orders. Even in those cases, however, an extra contribution involving two $G_{ga}$ terms always survives, though it is suppressed (since $G_{ga}$ has no order 0 contribution) by at least two extra powers of $\as(\mu_R^2)$ with respect to the leading term in $\left[ H^F C_1 C_2 \right]_{gg; a_1 a_2}$.

\subsection{The resummation scale}

The factorization of logarithmic terms in eqs.~\eqref{sudakov} and~\eqref{Cexp} involves some degree of arbitrariness (see for instance~\cite{Bozzi:2005wk}). The argument of the formally large logarithms can always be rescaled as
\begin{equation}
\log(M^2 b^2) = \log(\mu^2 b^2) + \log(M^2 / \mu^2) \; ,
\end{equation}
where $\mu$ is any scale such that $\mu \sim M$. In order to parametrize this arbitrariness, an extra scale $\mu_S$ can be introduced~\cite{Bozzi:2005wk}. The large logarithms are then defined to be
\begin{equation}
L \equiv \log\frac{\mu_S^2 b^2}{b_0^2} \; ,
\end{equation}
and eqs.~\eqref{sudakov} and~\eqref{Cexp} are modified accordingly, in a straightforward way. The function of the $\mu_S$ is to estimate the intrinsic uncertainty in the resummation formula~\eqref{master0}; in this sense, the scale is analogous to the usual renormalisation and factorization scales $\mu_R$ and $\mu_F$. By purely empirical considerations, the central value of the resummation scale is set to be $\mu_R/2$ by default in {\tt reSolve}, even though of course the user can set it to any value he or she chooses from the input file.

\subsection{The $\mathbf{b}$ and $z_1$, $z_2$ integrals}
\label{sec:FourierAndMellin}

The evaluation of the differential cross-section $\dfrac{\dif\sigma_{res}^F}{\dif M^2 \dif^2 \mathbf{q}_T \dif y \dif \mathbf{\Omega}}$ involves four integrations, over $\mathbf{b}$, $z_1$ and $z_2$. The dependence on the azimuthal angle $\varphi_b$ is actually simple enough that the corresponding integration can always be done analytically, reducing the number of integrals which have to be dealt with numerically to three. In particular, for $q\bar q$-initiated processes, the $\dif \varphi_b$ integration just implies the replacement:
\begin{equation}
\int \frac{\dif^2 \mathbf{b}}{(2\pi)^2} e^{-\ui \mathbf{q}_T \cdot \mathbf{b}}\to \int_0^\infty \frac{b}{2\pi} J_0(q_T b) \; ,
\end{equation}
where $J_0$ is the $0^{th}$-order Bessel function, since nothing in the main integrand $W^F$ depends on $\varphi_b$ and
\begin{equation}
J_0(x) = \int_0^{2\pi} \frac{\dif\varphi}{2\pi} e^{\pm\ui x \cos \varphi} \; .
\end{equation}
In the $gg$-initiated case, due to the presence of azimuthal correlations, after the $\dif \varphi_b$ integration the integral is split into two terms still involving Bessel functions, the first proportional to $J_0(q_T b)$ and the second to $J_2(q_T b)$; we refer the reader to~\cite{Catani:2010pd} for details.

The surviving integration over $b = |\mathbf{b}|$ is the most delicate one from both a numerical and theoretical point of view. Numerically, it involves a semi-infinite integrand over a rapidly oscillating (asymptotically) function. To perform this integral in {\tt reSolve}, we use an external package based on ref.~\cite{OOURA1991353} which is specifically designed in order to deal with this kind of integral.

From a theoretical point of view, the issue is that the main function $W^F(\mathbf{b}, \ldots)$ actually becomes ill-defined both at very small and very large values of $b$. Very small values $\mu_S b \ll 1$ are actually spurious from the point of view of resummation, since they correspond in transverse momentum space to large $q_T$ values $q_T / \mu_S \gg 1$ which are outside the range of validity of eq.~\eqref{master0}. To neatly cut-off these contributions, and thus to make the integrand well-defined in the small $b$ region as well as to reduce the impact of unjustified resummed logarithms in the large $q_T$ region, we implement, following ref.~\cite{Bozzi:2005wk}, the replacement:
\begin{equation}
\label{tildeL}
L \to \tilde{L} \equiv \log\left(\frac{\mu_S^2 b^2}{b_0^2} + 1\right) \; ,
\end{equation}
everywhere inside the $b$-dependent coefficients. We see that $\tilde{L} \rightarrow L$ in the resummation region $\mu_S b \gg 1$, so the replacement is legitimate, while $\tilde{L} \to 0$ in the $q_T/\mu_S$ region where resummation is not justified. The replacement is also useful in implementing a matching procedure at intermediate and large values of $q_T$, see for instance ref.~\cite{Bozzi:2005wk} for details.

When approaching the opposite region, $\mu_S b \gg 1$, the functions $\bar{g}^{(n)}$ in eqs.~\eqref{LL}-\eqref{NNLL} become singular, namely at $ \bar\lambda \to 1 \ \left(\bar\lambda \equiv (1/\pi)\beta_0 \as(\mu_S^2)\log(\mu_S^2 b^2/b_0^2) \right)$. In terms of the impact parameter, this gives:
\begin{equation}
b^2 \to b^2_L = \frac{b^2}{\mu_S^2} \exp\left( \frac{\pi}{\beta_0 \as(\mu_S^2)} \right) \; ,
\end{equation}
which in turns corresponds to $b_L \sim 1/\Lambda_{QCD}$, where $\Lambda_{QCD}$ is the scale of the Landau pole in QCD. This singularity signals the onset of non-perturbative phenomena at very small transverse momentum scales $q_T \sim \Lambda_{QCD}$, where even formula~\eqref{master0} breaks down. This kind of singularity is a common feature of all-order resummation formulae of soft gluon contributions (see also~\cite{Bozzi:2005wk}), and has to be regularised. In the current {\tt reSolve} implementation, we follow the prescription suggested in ref.~\cite{Collins:1981va}, which freezes the integration over $b$ below a fixed upper limit via the substitution:
\begin{equation}
\begin{split}
& b \to b_* = \frac{b}{\sqrt{1+b^2/b_{lim}^2}}, \\
\text{with} \quad b_{lim} & = \dfrac{b_{0}^{'}}{q}\exp\left(\dfrac{1}{2\alpha_{s}\beta_0}\right) \ \text{ and} \quad b_{0}^{'}=2\exp\left[\gamma \dfrac{q}{\mu_{S}}\right] \; .
\end{split}
\end{equation}
Admittedly, if non-perturbative contributions are sizeable, they should be properly included. In {\tt reSolve}, we include them according to the simple parametrization suggested in~\cite{Bozzi:2005wk}, multiplying the $b$-space integrand function by a non-perturbative factor, $S_{NP}$, which includes a Gaussian smearing:
\begin{equation} \label{NonPTfactors}
S_{NP} = \exp \left(- g_{NP}^c b^2 \right) \; ,
\end{equation} 
where $c = q, g$, that is, different factors can be used for $q\bar q$- and $gg$-initiated processes. Variations of the $g_{NP}^c$ (these variables are called {\tt ggnp} and {\tt gqnp} in the code) can then be used to estimate the impact of non-perturbative corrections.

Finally, let us comment briefly about the $z_1$ and $z_2$ integrals. Following refs.~\cite{Bozzi:2005wk,Bozzi:2007pn}, we don't do these directly, but rather switch to Mellin space, defining
\begin{equation}
W_{N_1,N_2}^F(\mathbf{b}, \ldots) = \int_0^1 \dif x_1 x_1^{N_1-1} \int_0^1 \dif x_2 x_2^{N_2-1} W^F(\mathbf{b} \; ,\ldots)
\end{equation}
where remember $x_1$ and $x_2$ were defined in eq.~\eqref{x1x2}. The double Mellin transformation turns the $\dif z_1 \dif z_2$ convolution integrals in eq.~\eqref{master0} into simple products~\cite{Bozzi:2007pn}, making them trivial to perform. At the end, the $W_{N_1, N_2}^F$ function is transformed back to the physical $x_1$, $x_2$ space using a simple and rapidly converging Gaussian quadrature. The only downside of this approach is that the Mellin transforms of the PDFs evaluated at arbitrary complex values of the Mellin variables $N_1$ and $N_2$ are needed to fully define $W_{N_1,N_2}^F$. Since the PDF collaborations typically provide the $f_{a/h}$ functions as numerical grids in momentum fraction (as opposed to Mellin) space, the PDF Mellin moments have to be calculated inside {\tt reSolve} itself, which requires a fit of the PDFs themselves to an analytic form whose Mellin transform is known or calculable. The strategy used in this case by {\tt reSolve} was described in more detail in section~\ref{Multipddfits}.

\subsection{The resummation scheme}
\label{sec:resscheme}

The various coefficients $A_a^{(n)}$, $B_a^{(n)}$, $C_{ab}^{n}$ and $H_c^{(n)}$ in eq.~\eqref{master0} are not actually uniquely defined. Indeed, the formula is invariant under the transformations~\cite{Catani:2013tia}
\begin{equation}
\label{resscheme}
\begin{split}
H_c^F(\as, \ldots) & \to H_c^F(\as, \ldots) (h_c(\as))^-1 \; , \\
B_c^F(\as) & \to B_c(\as) - \beta(\as) \frac{\dif \log h_c(\as)}{ \dif \log \as} \; , \\
C_{ab}(z, \as) & \to C_{ab}(z, \as) (h_c(\as))^{1/2} \; , \\
G_{gb}(z, \as) & \to G_{gb}(z, \as) (h_g(\as))^{1/2} \; ,
\end{split}
\end{equation}
where $h_c(\as) = 1 + \mathcal{O}(\as)$ are arbitrary perturbative functions (with $h_q(\as) = h_{\bar q}(\as)$). A specific choice of the coefficients is dubbed a \emph{resummation scheme}. The most often used scheme is the so-called \emph{hard scheme}, which is defined by the fact that, order by order in perturbation theory, the coefficients $C_{ab}^{(n)}(z)$ for $n \geqslant 1$ do not contain any $\delta(1-z)$ term. An alternative possibility is using the freedom in eq.~\eqref{resscheme} to set $H_q^{F(n)}=0$ and $H_g^{F'(n)}=0$ for $n \geqslant 1$ in two specific processes, a $q\bar q$-initiated $F$ and a $gg$-initiated one $F'$. This can be dubbed the \emph{F-$F'$ scheme}. For reasons related to its historical development, the current version of {\tt reSolve} is defined in the \emph{DY-H scheme}, meaning that we set $H_q^{(n)}$ to $0$ for the Drell-Yan process ($DY$) and $H_g^{(n)}$ to $0$ for Higgs production ($H$). We plan to switch to the arguably more physical \emph{hard-scheme} in one of the upcoming releases in the near future.

\subsection{Phase-space definition of the final state $F$}
The definition of the full phase space of the final state $F$, obviously needed for a full calculation, implies a subtlety. In fact, the four variables $M$, $\mathbf{q}_T$ and $y$ uniquely define the \emph{total} 4-momentum of $F$. The $\Omega$ variables then define the individual momenta of the particles in $F$ (for instance, in the specific case of the diphoton production process, the azimuthal and polar angles $\theta_{CM}$ and $\varphi_{CM}$ of one of the photons in the diphoton center-of-mass are used to fully define both photons momenta). At this point, however, in order to obtain the $\dif\hat\sigma_{c\bar{c}}^{F,\,\text{LO}}$ and $H_c^F$ coefficients, also the momenta of the initial state partons $c$, $\bar c$ are needed. As formally the initial state partons have vanishing transverse momenta, this creates a mismatch between initial and final state total momentum. This is related to the fact that the all-order emissions which are resummed in~\eqref{master0} are factorized in the limit $q_T \to 0$, so that the ``hard'' matrix elements they multiply (which give rise to $\dif\hat\sigma_{c\bar{c}}^{F,\,\text{LO}}$ and $H_c^F$) nominally have $q_T = 0$. This is another ambiguity which is unavoidable in any transverse momentum resummation calculation, and in fact only has an impact when $q_T$ is sizeable, that is, when the resummed approximation to the differential cross-section is losing validity anyway. From the practical point of view of the code, various approaches can be used. Arguably the simplest is to forgo momentum conservation altogether and simply set $q_T = 0$ when defining initial state momenta. In {\tt reSolve}, we follow the slightly more involved (and arguably more physical) strategy of refs.~\cite{Balazs:2007hr} and~\cite{Cieri:2015rqa}, where the initial state partons are assigned a fictitious transverse momentum proportional to $q_T$ in order to preserve momentum conservation.

\newpage

\section{Validation and Results} \label{validation}

In order to validate this completely new code, great care and attention was used during its development to compare the outputs of as many of the various routines included as possible against similar routines and calculations in private programs. These private programs include the {\tt 2gres} code (version 2.2), which has been used in the production of results for previous papers in refs.~\cite{Cieri:2015rqa} and~\cite{Cieri:2017kpq}, and the {\tt DYRes} code (version 1.0) \cite{Bozzi:2010xn,Catani:2015vma}. In addition to these checks, extensive physical checks were performed - ensuring the correct events were cut, ensuring no negative cross-sections were produced for any event (which would be a sign of running into regimes where our transverse momentum resummation formulation was invalid), analysing the sudakovs and hard factors produced, and many other checks. A selection of these checks are presented here, for both the diphoton case and the Drell-Yan case. We conclude this section by commenting on the speed of the program, which is often a limiting factor for such Monte Carlo reliant programs.

\subsection{Diphoton Results} \label{DiphotonResults}

For the validation of the program for the diphoton process ($process=1$ in the input file), first of all we ensure that the program produces the correct output for the Born-level process by comparison against known results, this is described in Section~\ref{BornCheckDiph}. Following this, we compare differential cross-sections in both invariant mass and in transverse momentum, for two main test files, with the private {\tt 2gres} (version 2.2) program. These are detailed in Section~\ref{resurevampcomp}. Finally, we demonstrate the validation of {\tt reSolve} for a setup for which we compare against the {\tt 2gres} program and experimental data, this is described in Section~\ref{expvalid}.

\subsubsection{Born Check} \label{BornCheckDiph}

The first validation is the comparison against known results in previous papers, in particular in Table~1 in \cite{Catani:2011qz} for the Born cross-section. This reference details the production cross section for diphotons plus jets given a typical set of kinematical cuts applied in ATLAS and CMS analyses \cite{ATLASref} \cite{CMSref}. These inputs are $\sqrt{s} = 14$TeV, $2\mu_S = \mu_R = qq$, $\mu_F = 113$GeV, $qq_{min} = 20$GeV, $qq_{max} = 250$GeV, $q_T$ is unimportant as this is at leading-order (LO), $-2.5 < \eta < 2.5$, no gg box as we are at leading order, {\tt etacut}$= 2.5$, no ``crack'' in the detector, {\tt pT1cut}$=40$GeV, {\tt pT2cut}$=25$GeV and {\tt Rcut}$=0.4$. The input file used is provided with the code as {\tt input/Diphoton\_Born\_LHC.dat}. We can only use this as a test of the LO result as beyond LO not only the resummed piece of the cross-section is required, but also the finite piece. Nonetheless for LO we obtain a total cross-section of $5.708 \pm 0.008$ pb whilst the value given in the paper was $5.712 \pm 0.002$ pb, therefore we are consistent. 

\subsubsection{Validation of {\tt reSolve} against {\tt 2gres} code} \label{resurevampcomp}
Differential cross-sections in both invariant mass and transverse momentum were produced for two main test files and compared with the private code {\tt 2gres}, this has been previously used for similar transverse momentum resummation calculations for diphotons. It should be noted that a small error was found in the old {\tt 2gres} code around the Jacobian and subsequently corrected before undertaking this comparison. The test files chosen are the {\tt Diphoton\_NNLO\_test\_1.dat} and {\tt Diphoton\_NNLO\_test\_2.dat} provided with the program and with the paper. The inputs of these two tests are summarised in Table~\ref{testinputstable}. These test files reflect common cuts, invariant mass ranges and transverse momentum ranges used for diphoton measurements, at $14$TeV and $8$TeV for {\tt test1} and {\tt test2} respectively. Therefore this comparison is of results produced by the new {\tt reSolve} program in the expected regions of application.
\begin{center}
\begin{table} [!htbp]
\centering
\def\arraystretch{1.5}
\begin{tabular}{|c|c|c|} \hline
Test file & test\_1 & test\_2 \\ \hhline{|=|=|=|}
Process & 1 & 1 \\ \hline
Order & 2 & 2 \\ \hline
CM\_energy ($TeV$) & 14 & 8 \\ \hline
$\mu_S$, $\mu_R$, $\mu_F$ ($GeV$) & $qq/2$, $qq$, 113 & $qq/2$, $qq$, 85 \\ \hhline{|=|=|=|}
QQ\_Min, QQ\_Max ($GeV$) & 80, 160 & 50, 110 \\ \hline
QT\_Min, QT\_Max ($GeV$) & 0, 120 & 0, 100 \\ \hline
$\eta$\_Min, $\eta$\_Max & -2.5, 2.5 & -2.37, 2.37\\ \hhline{|=|=|=|}
gg box ({\tt boxflag}) & No (0) & No (0) \\ \hhline{|=|=|=|}
etaCut & 2.5 & 2.37 \\ \hline
crack1, crack2 & 1.37, 1.37 & 1.37, 1.37 \\ \hline
pT1cut, pT2cut ($GeV$) & 40, 25 & 40, 30\\ \hline
Rcut & 0.4 & 0.4 \\ \hhline{|=|=|=|}
\end{tabular}
\caption{The two test files used for validation for the diphoton process in {\tt reSolve} both against the private code {\tt 2gres} and internally, ensuring different numbers of PDF fits, different integrators and different numbers of iterations all produce consistent results. The files are the {\tt Diphoton\_NNLO\_test\_1.dat} and {\tt Diphoton\_NNLO\_test\_2.dat} provided with the program.} 
\label{testinputstable}
\end{table}
\end{center}

\begin{figure}
  \centering
  \subfloat[Invariant Mass]{\includegraphics[height = 9.5cm]{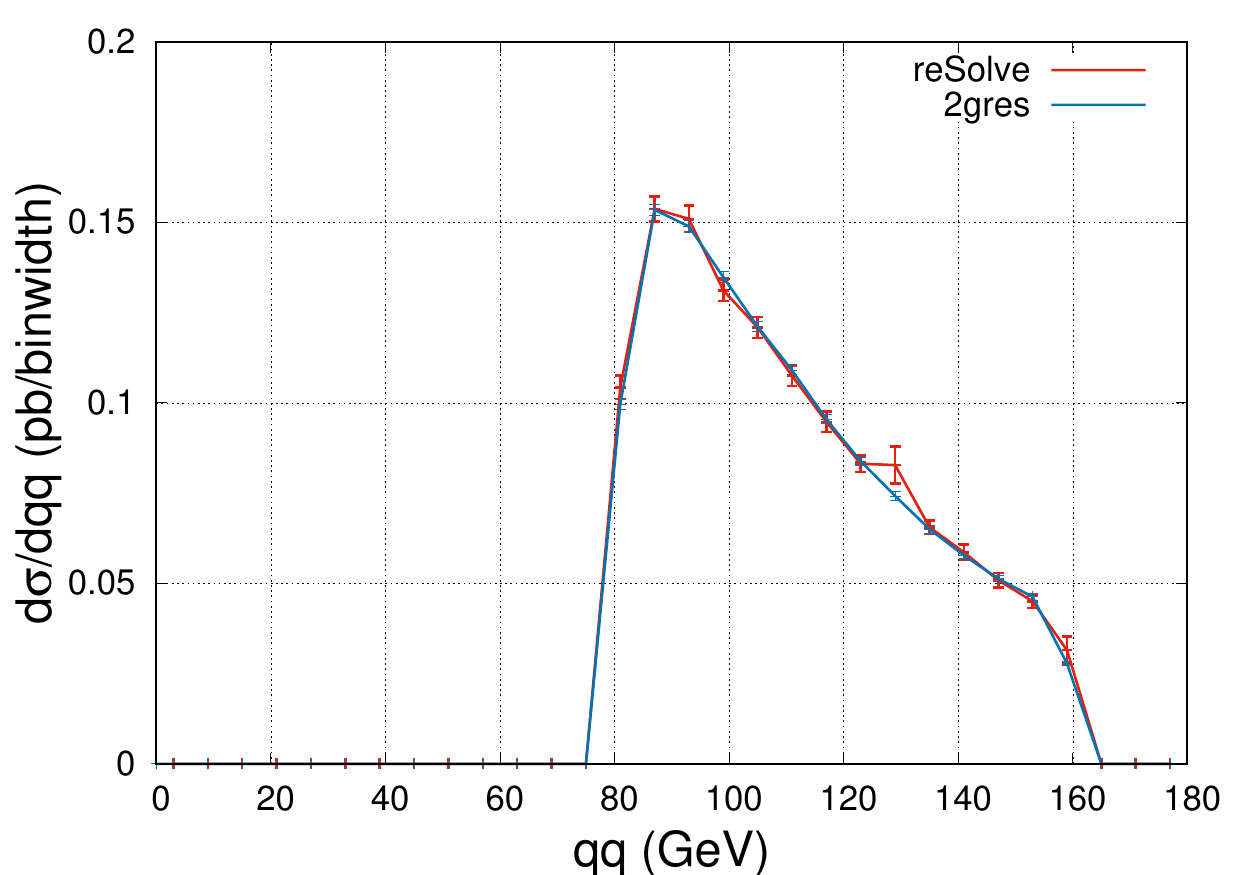}\label{qqresurevamptest1_500000}}
  \hfill
  \subfloat[Transverse Momentum]{\includegraphics[height = 9.5cm]{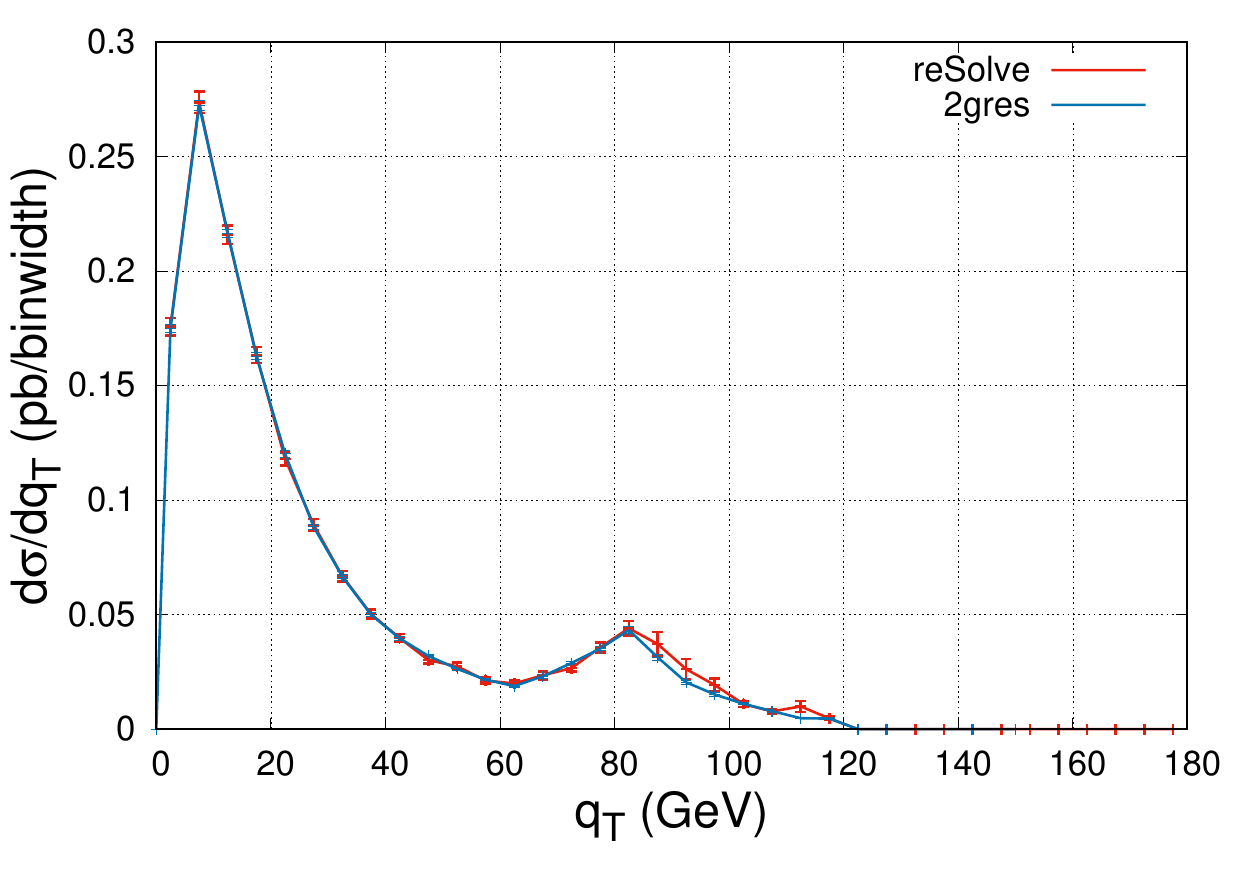}\label{qTresurevamptest1_500000}}
  \caption{Comparison plots of the diphoton differential cross-sections with invariant mass and transverse momentum for the {\tt test1} inputs as listed in Table~\ref{testinputstable} for the {\tt reSolve} program against the previous private program {\tt 2gres}. These just include the resummed part of the differential cross-section, not the finite piece.} \label{test1_5000000resurevampcomp}
\end{figure}

Figure~\ref{test1_5000000resurevampcomp} shows the comparison between {\tt reSolve} and the previous private program {\tt 2gres} for the {\tt test1} inputs with 500,000 events, with figure~\ref{qqresurevamptest1_500000} showing the invariant mass spectrum and figure~\ref{qTresurevamptest1_500000} showing the transverse momentum spectrum. Excellent agreement is seen in both cases with the two programs agreeing within the errors shown. First of all consider the invariant mass spectrum, the invariant mass region for the {\tt test1} inputs is $QQ\_Min = 80$GeV to $QQ\_Max = 160$GeV, this is exactly the region over which we have non-zero cross-section, demonstrating the cuts are being implemented correctly. Meanwhile, the shape of the distribution is as expected, rising sharply above $80$GeV once within the invariant mass region and peaking at the lower end of the invariant mass range. This occurs because as the invariant mass increases the phase space for the production of a higher diphoton invariant mass decreases. Meanwhile in figure~\ref{qTresurevamptest1_500000} we obtain the characteristic transverse momentum spectrum shape, with the spectrum approaching zero at $q_T = 0$ GeV, peaking sharply just above $0$GeV in the region where transverse momentum resummation is most important, and again reducing as the $q_T$ increases. Furthermore we see a slight rise from $65GeV$ peaking around $80$GeV, this is a kinematical effect caused by the $q_T$ cuts applied on each photon being $40$GeV and $25$GeV. The $q_T$ spectrum is then correctly cut off at the $q_T = 120$GeV upper bound. Again the agreement between the new public {\tt reSolve} program and the previous private {\tt 2gres} code is excellent within the errors. Moreover, for this {\tt test1} input file the total cross-section is also in agreement; {\tt reSolve} obtains $7.68 \pm 0.03pb$, whilst {\tt 2gres} obtains $7.67 \pm 0.03pb$.

\begin{figure}
  \centering
  \subfloat[Invariant Mass]{\includegraphics[height = 9.5cm]{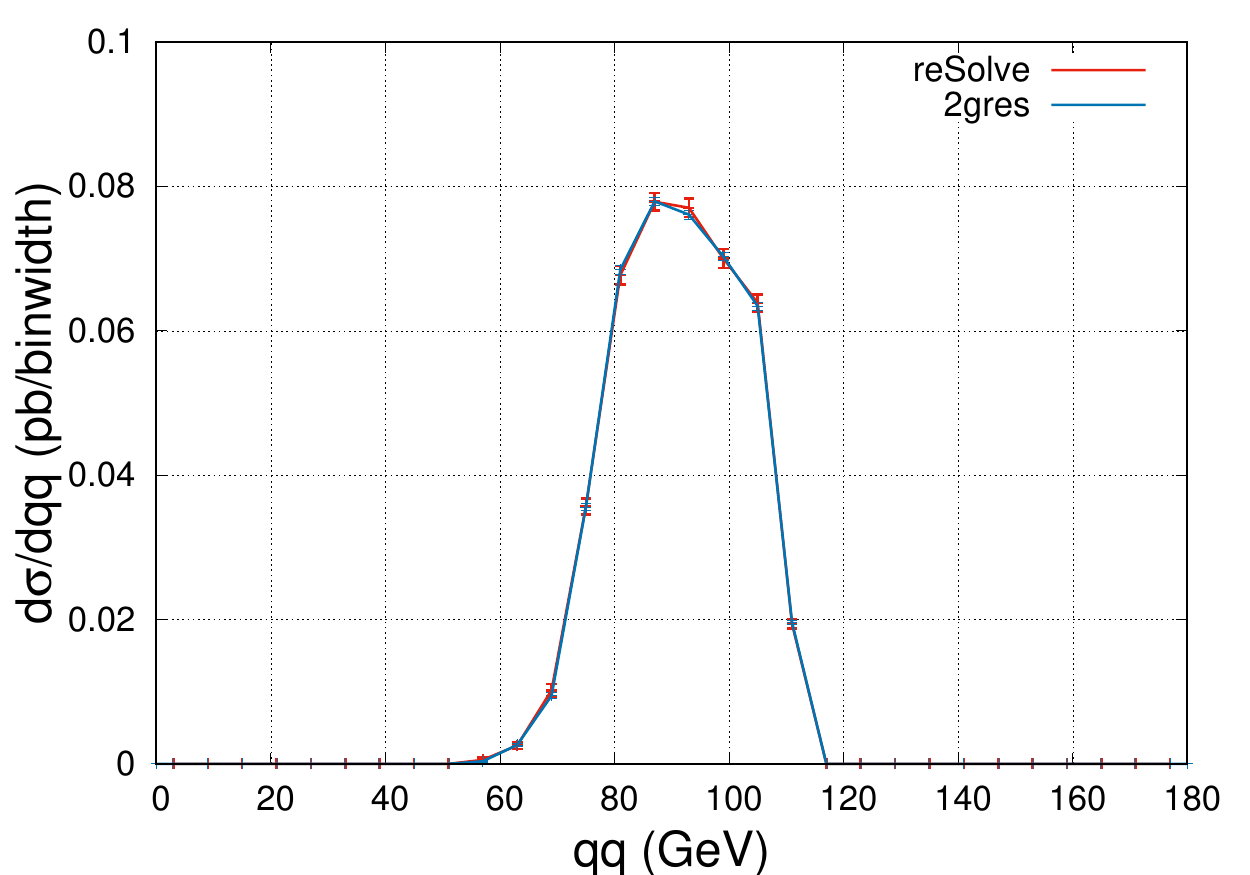}\label{qqresurevamptest2_500000}}
  \hfill
  \subfloat[Transverse Momentum]{\includegraphics[height = 9.5cm]{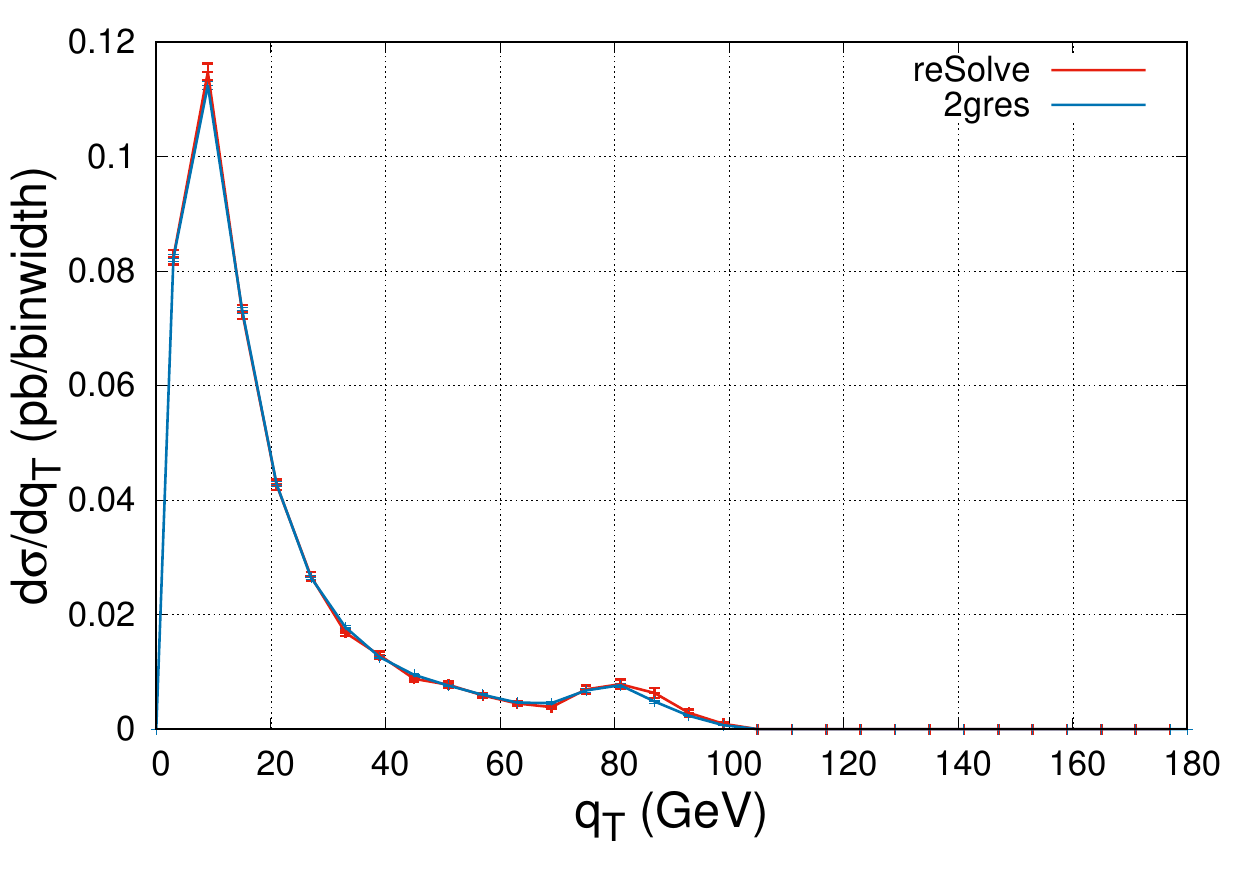}\label{qTresurevamptest2_500000}}
  \caption{Comparison plots of the diphoton differential cross-sections with invariant mass and transverse momentum for the {\tt test2} inputs as listed in Table~\ref{testinputstable} for the {\tt reSolve} program against the previous private program {\tt 2gres}. These just include the resummed part of the differential cross-section, not the finite piece.} \label{test2_5000000resurevampcomp}
\end{figure}

Figure~\ref{test2_5000000resurevampcomp} illustrates a similar comparison for the {\tt test2} inputs, again with 500,000 events. Figure~\ref{qqresurevamptest2_500000} shows the invariant mass spectrum comparison, whilst figure~\ref{qTresurevamptest2_500000} is the transverse momentum spectrum comparison. For {\tt test2} the invariant mass range is $50$GeV to $110$GeV as given in Table~\ref{testinputstable}, this region is clearly visible in figure~\ref{qqresurevamptest2_500000}. The transverse momentum spectrum in figure~\ref{qTresurevamptest2_500000} again shows the correct behaviour, going towards 0 at $q_T = 0$GeV, peaking sharply just above 0 and then falling away and cutting off at the edge of the $q_T$ region at $110$GeV. Again the peak around $80$GeV results from an increase in the phase space at this point, due to the cuts applied, as illustrated in the invariant mass spectrum in figure~\ref{qqresurevamptest2_500000}. Once more the total cross-section is also in good agreement between the two programs, with {\tt reSolve} obtaining $2.54 \pm 0.01pb$ and {\tt 2gres} $2.56 \pm 0.01pb$.

\begin{figure}
  \centering
  \subfloat[50 bins]{\includegraphics[height = 6.2cm]{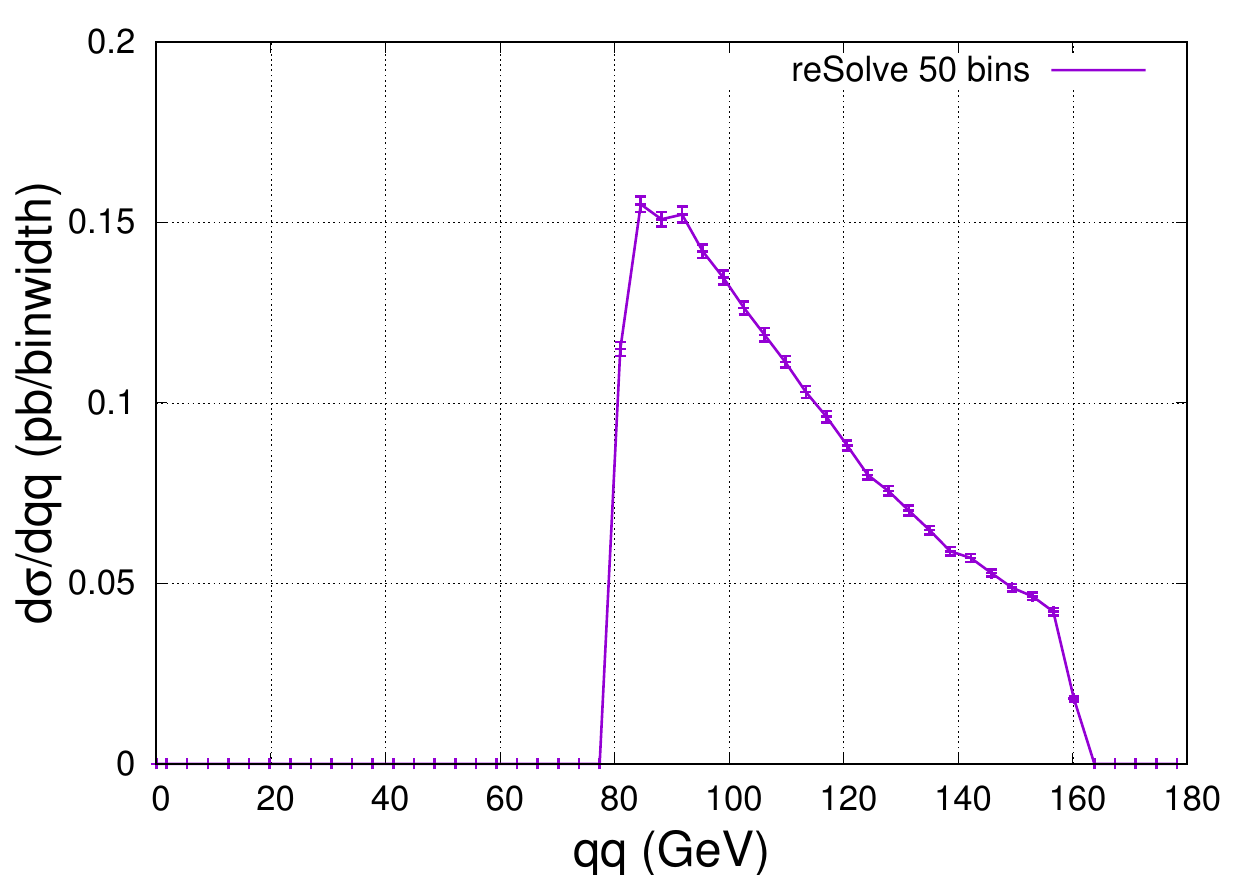}\label{qqresutest1_onemillion_50bins}}
  \hfill
  \subfloat[100 bins]{\includegraphics[height = 6.2cm]{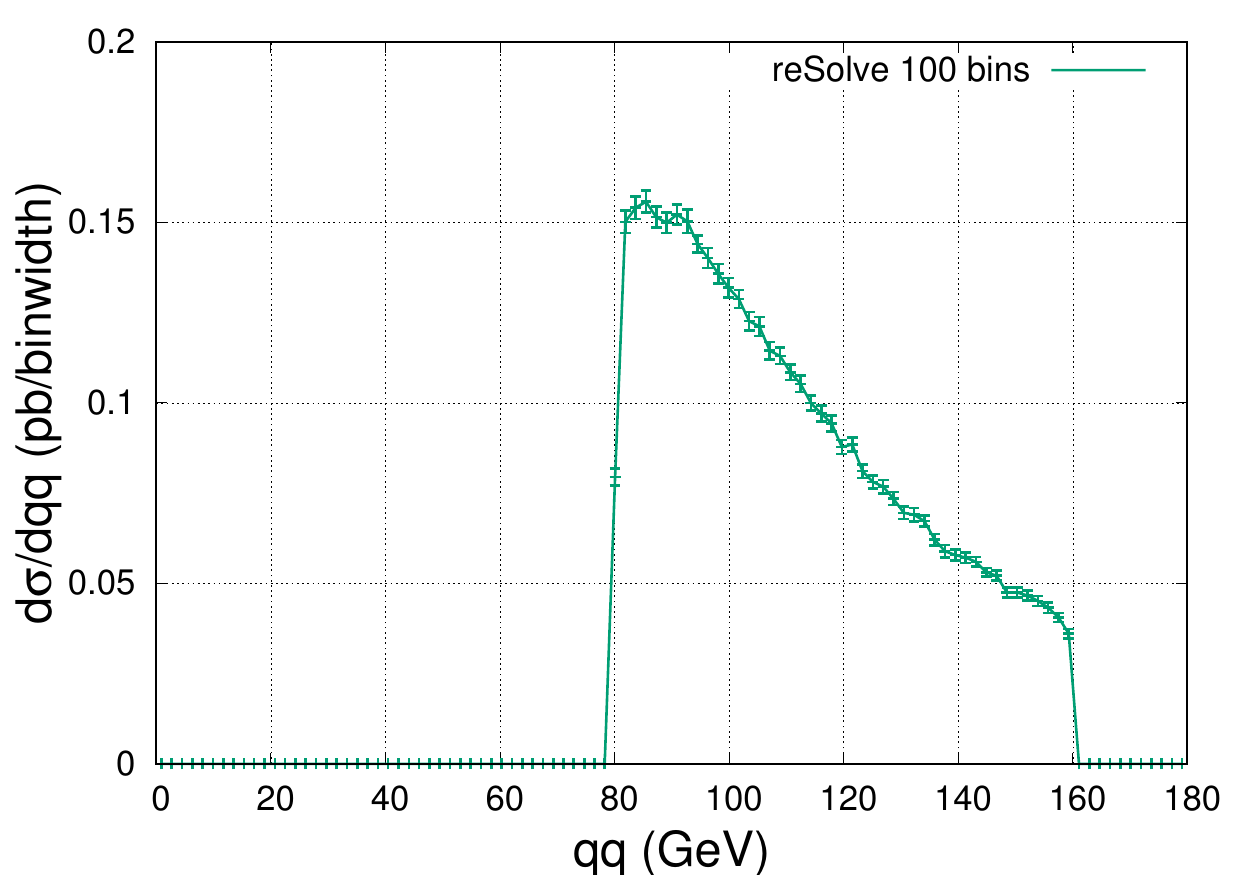}\label{qqresutest1_onemillion_100bins}}
   \hfill
  \subfloat[200 bins]{\includegraphics[height = 6.2cm]{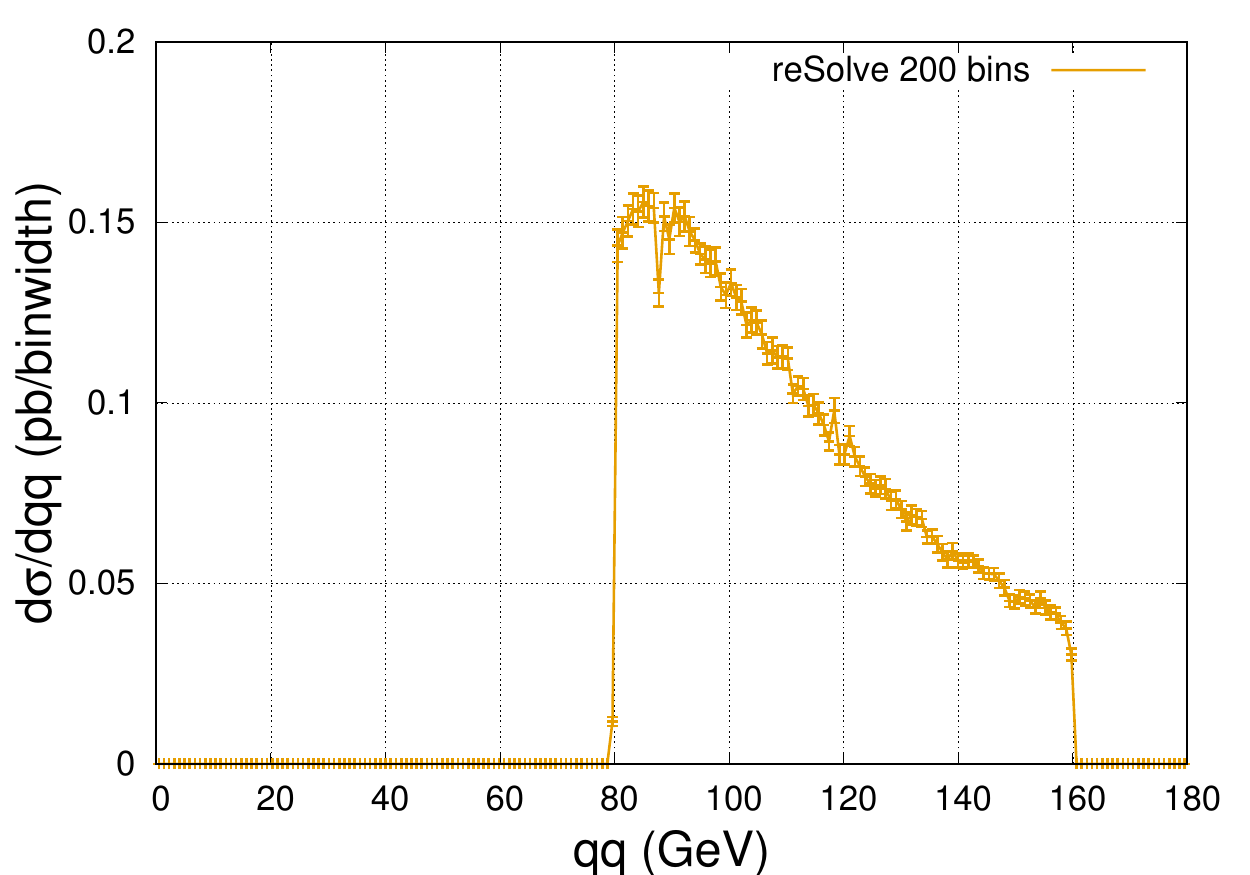}\label{qqresutest1_onemillion_200bins}}
  \caption{Comparison of the behaviour of the diphoton differential cross-section with invariant mass histograms for the test1 input file with different binnings; with 50, 100 and 200 bins respectively spread across the invariant mass range $0$GeV to $180$GeV.} \label{test1_onemillion_binningscomp}
\end{figure}

Another check that can be performed, is to compare the spectra seen with different width bins for a given input file, such a comparison is presented briefly here and validates the events and the histogram routine, see figure~\ref{test1_onemillion_binningscomp}. Figures \ref{qqresutest1_onemillion_50bins}, \ref{qqresutest1_onemillion_100bins}, \ref{qqresutest1_onemillion_200bins} illustrate the invariant mass spectrum obtained for the {\tt test1} input file for 50, 100 and 200 bins respectively spread evenly over the region $0 \rightarrow 180$GeV. Given the events appear in a mass range $80$GeV to $160$GeV this means there are approximately 22, 44 and 89 bins over this region for the respective binnings. All three plots of course demonstrate the same overall shape given they are different presentations of the same generated events. In general the behaviour of the histogrammer here is exactly as expected, with more consistent size errors seen in figure~\ref{qqresutest1_onemillion_50bins} which has the largest bins. Figures~\ref{qqresutest1_onemillion_100bins} and \ref{qqresutest1_onemillion_200bins} show yet narrower bins and as a result we start to see statistical variations in the shape of the distribution, by figure~\ref{qqresutest1_onemillion_200bins} in particular the errors on the bins are becoming larger as a result of fewer events per bin, and as a result we see occasional bins - such as that around $88GeV$ - which show large discrepancies from the surrounding bins. To determine the optimal bin size, one must balance the competing needs for small enough bins not to smear out any interesting behaviour in the spectra, and for large enough bins that there are sufficient events per bin to make the statistical fluctuations and errors per bin palatable, these themselves can be reduced by longer run times. This behaviour is all exactly as one would expect, providing some confirmation of the events generated and, in particular, of the histogram package included within the {\tt reSolve} program.

\subsubsection{ATLAS Validation} \label{expvalid}

An additional validation that was performed was to use {\tt reSolve} to reproduce the events and corresponding total cross-section, invariant mass spectrum and transverse momentum spectrum for the inputs listed in \cite{Cieri:2017kpq}. These inputs are provided in the {\tt reSolve} program package in {\tt ``input/Diphoton\_Atlas\_A.dat''} and are: diphoton process at NNLO+NNLL, MSTW NNLO PDFs ({\tt PDF\_flag = 82}), centre of mass energy 8TeV, pp collisions; square factorisation, resummation and renormalisation scales set proportional to $qq^2$ - i.e. $\mu_F^2 = \mu_R^2 = qq^2$ and $\mu_S^2 = \frac{qq^2}{4}$; 2,000,000 iterations with $nstart = nincrease = 10000$, $0 < QQ < 500$ GeV, $ 0 < QT < 150$ GeV, $-2.37< \eta < 2.37$, gg box included, {\tt etacut}$=2.37$, {\tt crack1}$= 1.37$, {\tt crack2}$= 1.56$, {\tt pT1cut}$= 40$GeV, {\tt pT2cut}$= 30$GeV and {\tt Rcut}$= 0.4$. As a result of the large invariant mass range considered, 5 PDF fits are used across the allowed range to improve accuracy. For these inputs the {\tt 2gres} program has been previously validated against experimental data.

\begin{figure} [!htbp]
  \centering
\includegraphics[height = 8.5cm]{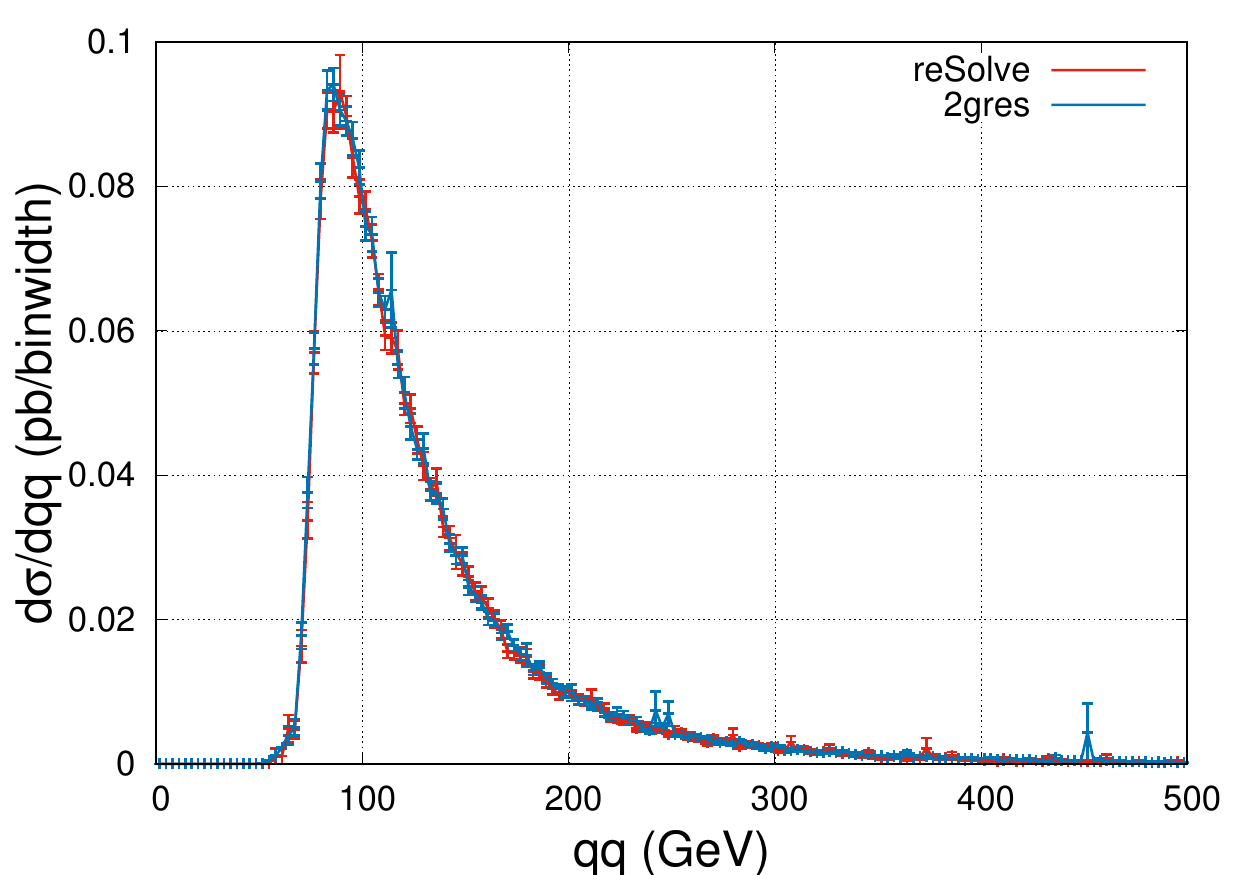}
\caption{Invariant mass spectrum for diphoton production for the {\tt Diphoton\_Atlas\_A.dat} input file provided with the program, and whose inputs are also listed in the text, as produced by the {\tt reSolve} program and the previous private program {\tt 2gres}, which was used in the work in \cite{Cieri:2015rqa} and \cite{Cieri:2017kpq}. Only the resummed part of the differential cross-section is shown.} \label{qqFrancescocompplot}
\end{figure}

\begin{figure} [!htbp]
  \centering
\includegraphics[height = 8.5cm]{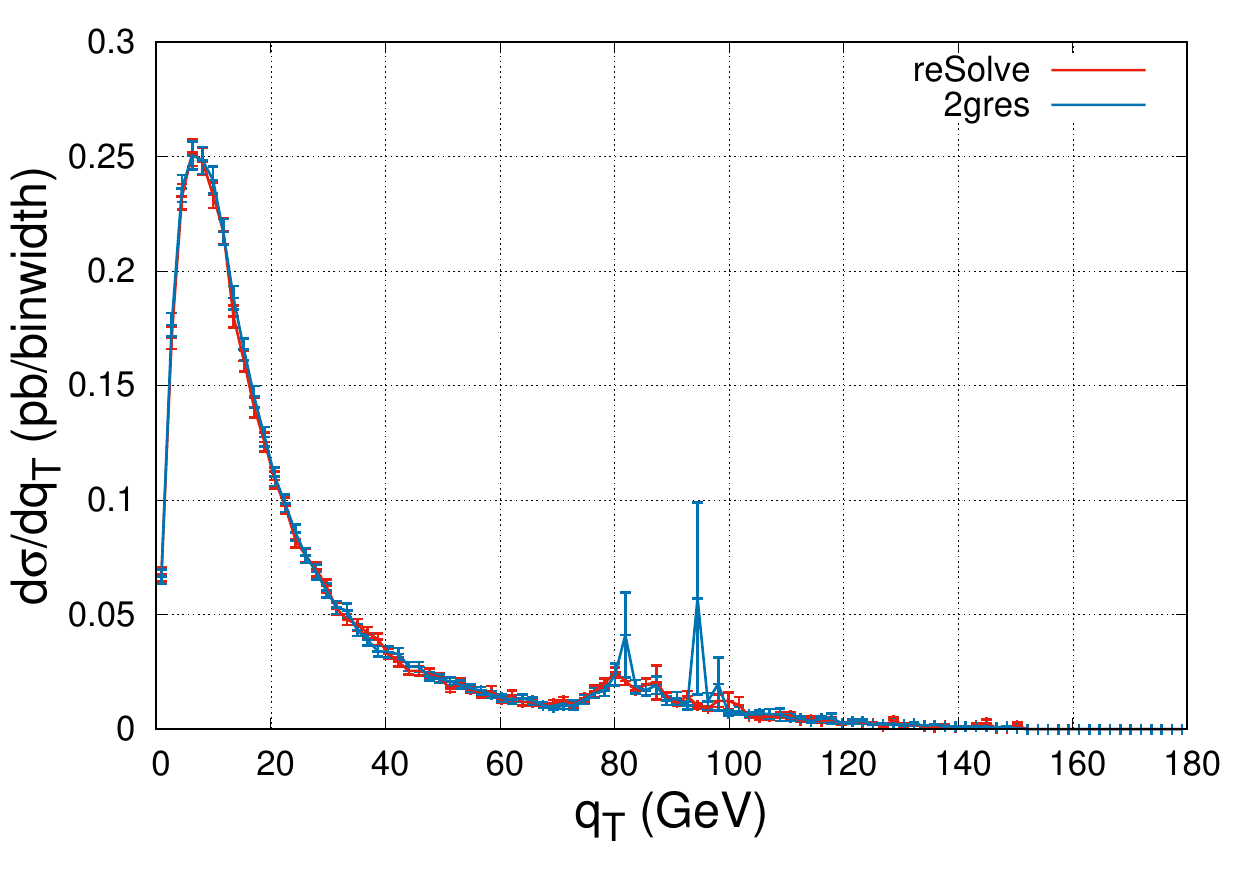}
\caption{Transverse momentum spectrum for diphoton production for the {\tt Diphoton\_Atlas\_A.dat} input file provided with the program, and whose inputs are also listed in the text, as produced by the {\tt reSolve} program and the previous private program {\tt 2gres}, which was used in the work in \cite{Cieri:2015rqa} and \cite{Cieri:2017kpq}. Only the resummed part of the differential cross-section is shown.} \label{qTFrancescocompplot}
\end{figure}

The total cross-section produced by the {\tt reSolve} program for these inputs was $6.188 \pm 0.013$, compared with $6.18 \pm 0.02$ from {\tt 2gres}, this therefore indicates very good agreement. The invariant mass and transverse momentum spectra also are consistent and are given in figure~\ref{qqFrancescocompplot} and figure~\ref{qTFrancescocompplot} respectively. The $q_T$ plot shows agreement at both the low $q_T$ end and higher $q_T$ end, with the position and height of the peak in the spectrum  agreeing within the errors, meanwhile as one increases $q_T$ the differential cross-section reduces as expected. The spectrum peaks again slightly around 80GeV as a result of increasing phase space available beyond this $q_T$, as demonstrated in the invariant mass spectrum raising rapidly above 80GeV, this is a threshold effect caused by the cuts - with cuts on {\tt pT1} and {\tt pT2} of 40GeV and 30GeV invariant masses of less than 80GeV are difficult to attain.

\begin{figure} [!htbp]
  \centering
\includegraphics[height = 12cm]{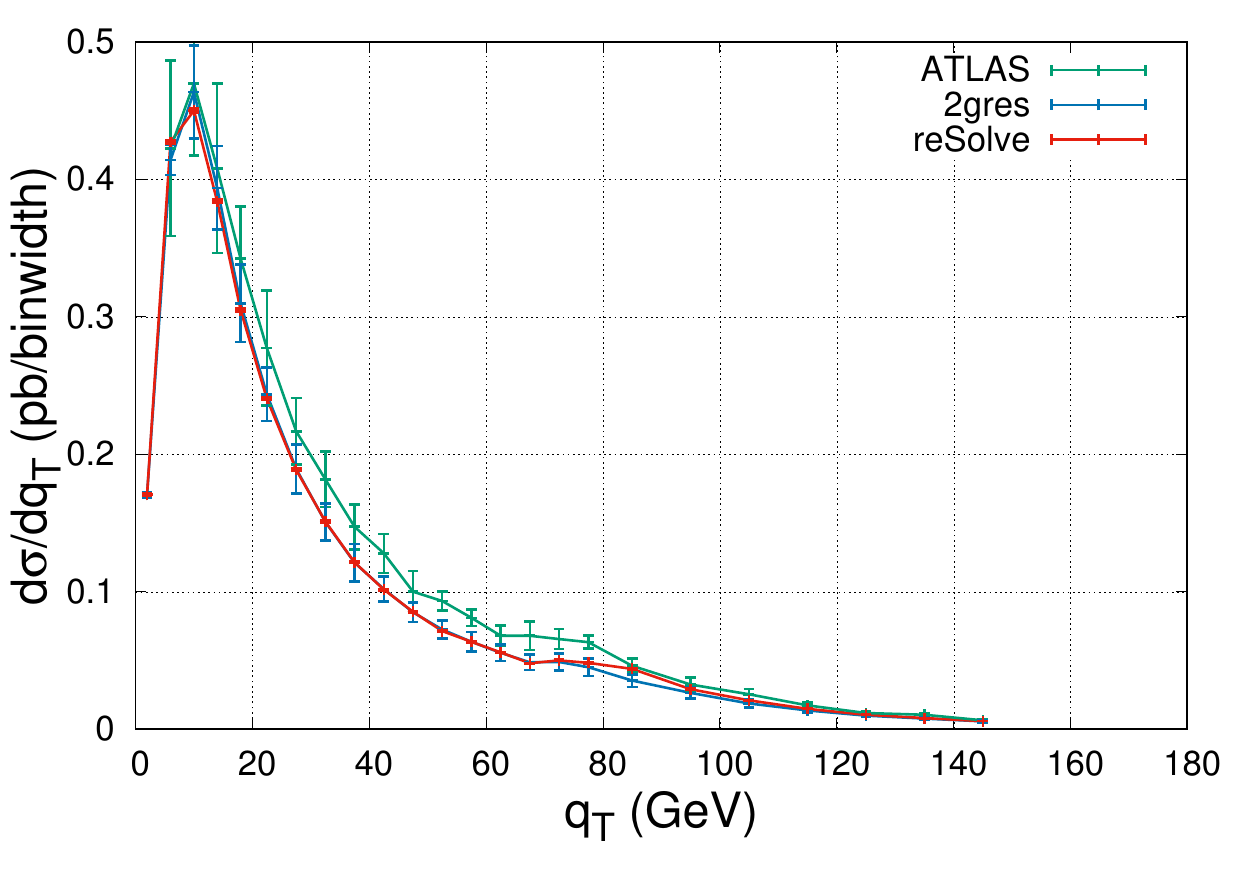}
\caption{Transverse momentum spectrum, including resummed and finite pieces, for diphoton production for the {\tt Diphoton\_Atlas\_A.dat} input file provided with the program, and whose inputs are also listed in the text, as produced by the {\tt reSolve} program (with finite pieces from {\tt 2gNNLO} \cite{Catani:2011qz}) and the previous private program {\tt 2gres}, which was used in the work in \cite{Cieri:2015rqa} and \cite{Cieri:2017kpq}, also shown are the ATLAS experimental results and corresponding errors. The error bars for {\tt 2gres} show the scale variation error, which is the dominant error. This is not calculable in {\tt reSolve} as we do not include the finite part of the cross-section in this first version of the program, therefore the {\tt reSolve} error bars are the Monte Carlo errors from the resummed part only.} \label{resolve2gresAtlasqT}
\end{figure}

As well as validation against just the resummed part of the transverse momentum differential cross-section, we added the finite piece - as previously calculated by {\tt 2gNNLO} \cite{Catani:2011qz} - in order to validate against the total transverse momentum differential cross-section. We then compared our {\tt reSolve} results with the matched finite piece added (with Monte Carlo error only and only for the resummed part of the differential cross-section) against those of {\tt 2gres} (with errors shown indicating scale variation) and ATLAS experimental results \cite{Aaboud:2017vol}. The comparison plot is shown in figure~\ref{resolve2gresAtlasqT}. The figure demonstrates the excellent agreement between the {\tt reSolve} program and the data previously calculated with {\tt 2gres}, the main difference coming in the 80GeV - 100GeV region where we observe a bump in the resummed cross-section (see for example figure~\ref{qTFrancescocompplot}), the difference here is because the previous {\tt 2gres} program had a small bug in the Jacobian which suppressed the effect of the bump. We use the old {\tt 2gres} data to demonstrate its effect - indeed the new {\tt reSolve} program now shows better agreement with the experimental data in this region than the {\tt 2gres} code did previously. The bump, as described previously, is a threshold effect arising from the cuts applied. There are small differences between the {\tt reSolve} and {\tt 2gres} predictions and the experimental results at intermediate transverse momenta, we expect these are within errors once all errors - including those from the pdffit, scale variation, Monte Carlo, matching and other sources are taken into account; in any case the current version of {\tt reSolve} determines only the resummed piece of the differential cross-section and so it is the excellent agreement at low tranverse momenta which is the focus of our validation.


\hfill \break

\pagebreak

\newpage

\subsection{Drell-Yan Results} \label{Drell-Yan Results}

In order to validate the Drell-Yan process, we perform similar checks as for the diphoton case. First we focus on the Born cross-section, comparing against known results in Section~\ref{BornCheckDY}. We then present a comparison against the program {\tt DYRes} (version 1.0) \cite{Catani:2015vma} in various differential variables, including the transverse momentum spectrum and rapidity($\eta$) spectrum; in Section \ref{DYDifferentialPlots} we provide further differential plots, comparing the rapidity, transverse mass ($m_T$), and $p_T^{min}$ and $p_T^{max}$ distributions against those in \cite{Catani:2009sm}. Here further plots of the invariant mass spectrum for the Drell-Yan pair and additional differential distributions are also provided in order to confirm the qualitative behaviour of the results is as expected.

\subsubsection{Born Validation} \label{BornCheckDY}

In order to add a new process (in this case all the Drell-Yan processes) to the {\tt reSolve} program, as described in Section \ref{AddProcess}, we needed only to alter the non-resummation part of the code, providing a new Born-level cross-section for the added process. For this reason, the key test to perform to validate the added process is to confirm it produces the Born cross-section correctly. In order to confirm this, {\tt reSolve} was run for three different setups, setup 1 is for $Z/\gamma^*$ at the Tevatron, setup 2 is for $Z$ On-Shell at the 14TeV LHC, and setup 3 is for $W^{\pm}$ at the Tevatron; the full invariant mass ranges, rapidity ranges, scale and cut setups are listed in Table \ref{DYtestinputstable}. These benchmarks were also used for NLO+NLL and NNLO+NNLL comparisons and plots in later sections (where the {\tt resum\_flag} must be changed to $1$ and the {\tt order} changed accordingly), therefore the transverse momentum ranges are given, however note that there is no transverse momentum ($q_T$) at Born-level so the $q_T$ range set is unimportant for Born comparisons. These benchmarks were chosen as they reflect the full range of Drell-Yan processes added and there are results quoted in \cite{Catani:2009sm} to compare against, in addition results were also obtained from the private DYRes (version 1.0) program \cite{Catani:2015vma} and from the MCFM (version 8.1) (``Monte Carlo for FeMtobarn processes'') program \cite{Campbell:1999ah,Campbell:2011bn,Campbell:2015qma,Boughezal:2016wmq} for comparison.

\begin{center}
\begin{table} [!htbp]
\centering
\begin{tabular}{|c|c|c|c|} \hline
Test file & setup 1 & setup 2 & setup 3 \\ \hhline{|=|=|=|=|}
Process & 2 & 2 & 2 \\ \hline
resum\_flag & 0 & 0 & 0 \\ \hline
DYProcess & 5 & 4 & 3 \\ \hline
DYnarrowwidthapprox & 0 & 1 & 0 \\ \hline
Order & 0 & 0 & 0 \\ \hline
pdf\_flag & 80 & 80 & 80 \\ \hline
CM\_energy ($GeV$) & 1960 & 14000 & 1960 \\ \hline
ih1 & 1 & 1 & 1\\ \hline
ih2 & -1 & 1 & -1\\ \hline
$\mu_S$, $\mu_R$, $\mu_F$ ($GeV$) & All $m_Z = 91.187$ & All $m_Z = 91.187$ & All $m_W = 80.398$ \\ \hhline{|=|=|=|=|}
QQ\_Min, QQ\_Max ($GeV$) & 70, 110 & 70, 110 & 0, 200 \\ \hline
QT\_Min, QT\_Max ($GeV$) & 0, 200 & 0, 200 & 0, 200 \\ \hline
$\eta$\_Min, $\eta$\_Max & -3, 3 & -10, 10 & -3, 3\\ \hhline{|=|=|=|=|}
crack1, crack2 & 1.37, 1.37 & 1.37, 1.37 & 1.37, 1.37 \\ \hline
pT1cut, pT2cut ($GeV$) & 20, 20 & 0, 0 & N.A., N.A.\\ \hline
eta1Cut = eta2cut & 2 & 10 & N.A. \\ \hline
pTecut ($GeV$) & N.A. & N.A. & 20\\ \hline
pTmisscut ($GeV$) & N.A & N.A. & 25\\ \hline
etaecut & N.A & N.A. & 2\\ \hline
tmasscut ($GeV$) & N.A & N.A. & 0\\ \hhline{|=|=|=|=|}
\end{tabular}
\caption{The three test files used for validation of the Born cross-section for the Drell-Yan added processes in {\tt reSolve} against the results in \cite{Catani:2009sm} and results from the program MCFM \cite{Campbell:1999ah,Campbell:2011bn,Campbell:2015qma,Boughezal:2016wmq}, as well as against those of DYRes \cite{Catani:2015vma}. The pdf sets used are the MSTW2008 LO PDFs, therefore pdf\_flag = 80. The files are the {\tt yZ\_Born\_Tevatron.dat}, {\tt Z\_OnShell\_Born\_LHC.dat} and {\tt Wpm\_Born\_Tevatron.dat} provided with the program. Similar inputs were used for the NLO and NNLO tests, these files are also provided with the {\tt reSolve} program.} 
\label{DYtestinputstable}
\end{table}
\end{center}

With these inputs as listed in Table \ref{DYtestinputstable}, the {\tt reSolve} program obtains the following Born cross-sections: for the $Z/\gamma^*$ Tevatron setup 1, {\tt reSolve} calculates $\sigma_{LO} = 103.37\pm 0.06pb$, compare this with the results in \cite{Catani:2009sm} where $\sigma_{LO} = 103.37 \pm 0.04pb$, whilst MCFM \cite{Campbell:1999ah,Campbell:2011bn,Campbell:2015qma,Boughezal:2016wmq} obtains $\sigma_{LO} = 103.34 \pm 0.04pb$; for the on-shell $Z$ LHC $14TeV$ setup 2, {\tt reSolve} calculates $\sigma_{LO} = 1758.9\pm1.1pb$, for comparison with $\sigma_{LO} = 1761\pm1pb$ in \cite{Catani:2009sm} and $\sigma_{LO} (pp \rightarrow Z \rightarrow l^+ l^-) = \sigma_{LO} (pp \rightarrow Z) \times BR(Z \rightarrow l^+ l^- ) = 52.3197\pm0.0049 nb \times 0.03366 = 1761.1 \pm 0.1pb$; finally for the $W^{\pm}$ Tevatron setup 3, {\tt reSolve} calculates $\sigma_{LO} = 1160.4\pm0.7pb$ for comparison with $\sigma_{LO} = 1161\pm1pb$ in \cite{Catani:2009sm}, whilst MCFM obtains $\sigma_{LO}(p pb \rightarrow W^{\pm}) = 1187.87\pm0.43pb$. Therefore there is good agreement between the {\tt reSolve} program for these Born cross-sections with known calculations for all three setups.

Given this incorporates the majority of the process dependence of the formalism used for the {\tt reSolve} program, this indicates the new Drell-Yan processes are working correctly. Nonetheless, we demonstrate many further results and validations over the next few sections.

\subsubsection{Total Cross-Sections at LO, NLO and NNLO} \label{DYTotalsigmas}

Before any of the differential cross-sections are analysed, first we check that the total cross-sections are sensible for each order; leading order (LO) Born cross-section without resummation, next-to-leading order (NLO) cross-section with next-to-leading logarithm (NLL) resummation, and next-to-next-to leading order (NNLO) with next-to-next-to-logarithm (NNLL) resummation. The input files used are those for the 3 benchmark setups of Table~\ref{DYtestinputstable} at each of the 3 orders; they are provided with the {\tt reSolve} program and are {\tt yZ\_Born\_Tevatron.dat}, {\tt yZ\_NLO\_Tevatron.dat}, {\tt yZ\_NNLO\_Tevatron.dat}, {\tt Z\_OnShell\_Born\_LHC.dat}, {\tt Z\_OnShell\_NLO\_LHC.dat}, {\tt Z\_OnShell\_NNLO\_LHC.dat}, {\tt Wpm\_Born\_Tevatron.dat}, {\tt Wpm\_NLO\_Tevatron.dat} and {\tt Wpm\_NNLO\_Tevatron.dat} . The results obtained from {\tt reSolve} are compared with known results calculated in \cite{Catani:2009sm}, the difference being that, as-of-yet, {\tt reSolve} does not include the finite part of the cross-section, just the resummed part, therefore we expect our beyond $LO$ results to be lower than in \cite{Catani:2009sm} but showing the same trend with NNLO $>$ NLO $>$ LO. The results are summarised in Table~\ref{DYTotalsigmascomptable}, those for LO were also given in the previous section. The agreement shown at leading-order is good for all three benchmark setups. Meanwhile the NLO and NNLO results behave as expected, with each successive order increasing the total cross-section as more contributions are added, whilst still being smaller than the known results which include the additional finite contributions. Moreover, the increases in the total cross-section are significant in going from LO to NLO(+NLL) but much smaller upon going to NNLO(+NNLL), exactly as seen in the known results.

\begin{center}
\begin{table} [!htbp]
\centering
\begin{tabular}{|c|c|c|c|} \hline	
 & & {\tt reSolve} /pb & Known Result /pb\\ \hline
 \multirow{3}{*}{$Z/\gamma^*$ setup 1} & LO & $103.37\pm 0.06$ & $103.37 \pm 0.04$ \\ \cline{2-4}
 & NLO(+NLL) & $130.37\pm0.10$ & $140.43\pm 0.07$\\ \cline{2-4}
 & NNLO(+NNLL) & $130.40\pm0.10$ & $143.86\pm 0.12$ \\ \hline
  \multirow{3}{*}{$Z$ On-shell setup 2} & LO & $1758.9\pm 1.1$  & $1761\pm 1$  \\ \cline{2-4}
 & NLO(+NLL) & $2009.1\pm0.5$ & $2030\pm 1$\\ \cline{2-4}
 & NNLO(+NNLL) & $2056.2\pm3.0$ & $2089\pm 3$ \\ \hline
  \multirow{3}{*}{$W^{\pm}$ setup 3} & LO & $1160.4\pm 0.7$ & $1161\pm 1$\\ \cline{2-4}
 & NLO(+NLL) & $1438.6\pm 1.2$ & $1550\pm 1$ \\ \cline{2-4}
 & NNLO(+NNLL) & $1465.4 \pm 1.3$ & $1586\pm 1$ \\ \hline
\end{tabular}
\caption{Summary of the total cross-sections as calculated by {\tt reSolve} compared with known results \cite{Catani:2009sm} for the three setups described in Table~\ref{DYtestinputstable}. The agreement is good between reSolve and the known results at LO and the behaviour beyond LO is expected, {\tt reSolve} results beyond leading order are smaller as {\tt reSolve} only includes the resummed part of the total cross-section, not the finite part which is important at larger transverse momentum $q_T$. The errors indicated in the table are Monte Carlo errors only and so are largely determined by the length of the run used and not indicative of the precision of the program. The {\tt reSolve} input files used are included with the program.} 
\label{DYTotalsigmascomptable}
\end{table}
\end{center}

\subsubsection{Validation of {\tt reSolve} against {\tt DYRes} code} \label{DYresurevampcomp}

In order to validate the new Drell-Yan process added to reSolve, we can compare with the program {\tt DYRes}, which is from the same series of programs as the {\tt 2gres} program against which we compared the diphoton, and allows also for the calculation of the resummed contribution only.

First we consider the $Z/\gamma^*$ Tevatron case of setup 1 (given previously in table~\ref{DYtestinputstable}). {\tt DYRes} produces data for the transverse momentum spectrum and rapidity distribution, amongst others, so we may compare these directly with these distributions as produced by {\tt reSolve}, the comparison plots are shown below in figures~\ref{qTcompDYRESNNLOZ} and \ref{etacompDYRESNNLOZ} respectively, both figures show the NNLO(+NNLL) spectra and demonstrate very good agreement with the {\tt DYRes} program, thereby validating {\tt reSolve} for this process. The total cross-section also agrees well, {\tt reSolve} obtains $130.1 \pm 0.4pb$ as detailed in Table \ref{DYTotalsigmascomptable}, whilst {\tt DYRes} (for the resummed piece only) obtains $130.3 \pm 1.2pb$.

\begin{figure} [!htbp]
  \centering
\includegraphics[height = 9cm]{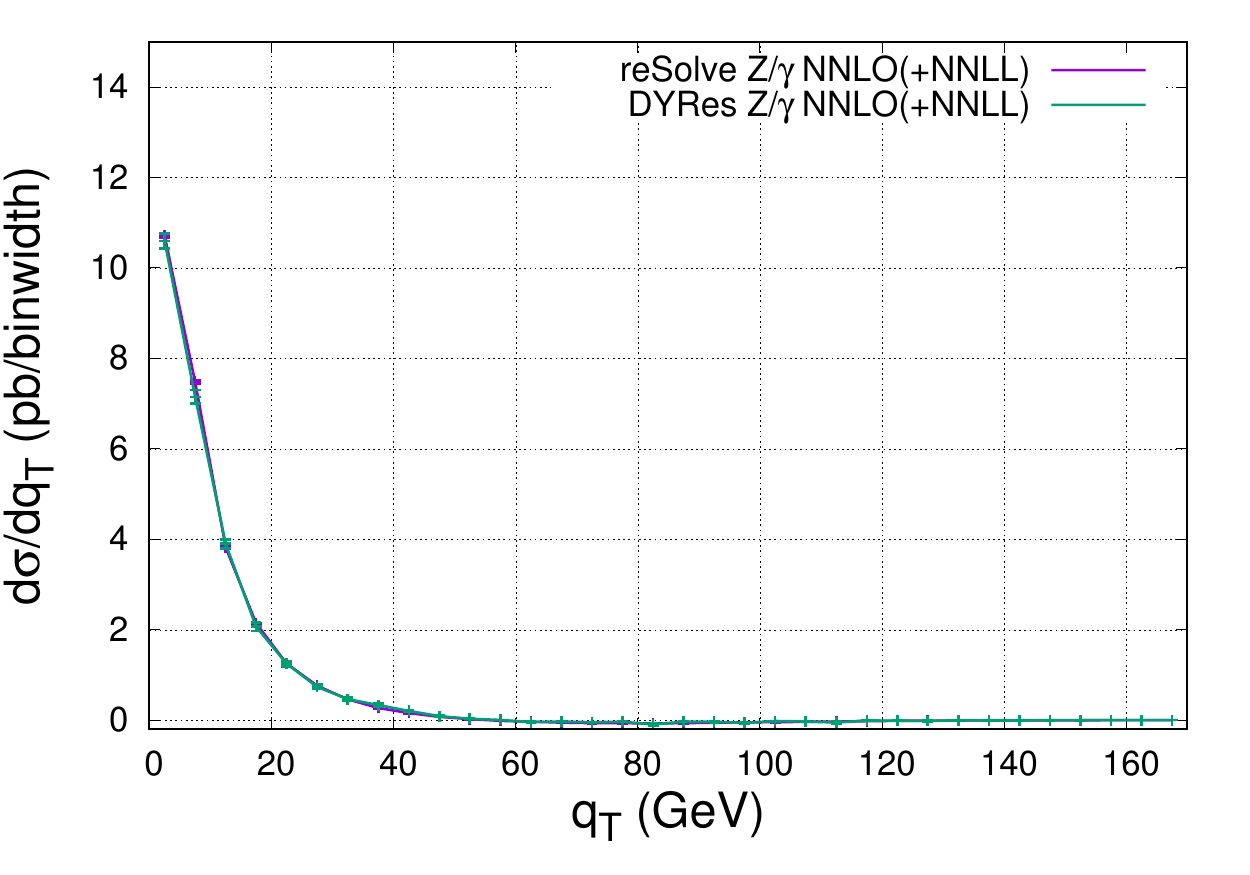}
\caption{Transverse momentum spectrum, including only the resummed piece, for Drell-Yan production via neutral current $Z$ or $\gamma^*$   at NNLO (with NNLL resummation) for the setup 1 benchmark, as given in table \ref{DYtestinputstable}. The agreement between the two programs is excellent, validating  {\tt reSolve}. The error bars are the Monte Carlo errors from the resummed part only and are largely a reflection on the length of the runs performed.} \label{qTcompDYRESNNLOZ}
\end{figure} 
\begin{figure} [!htbp]
  \centering
\includegraphics[height = 9cm]{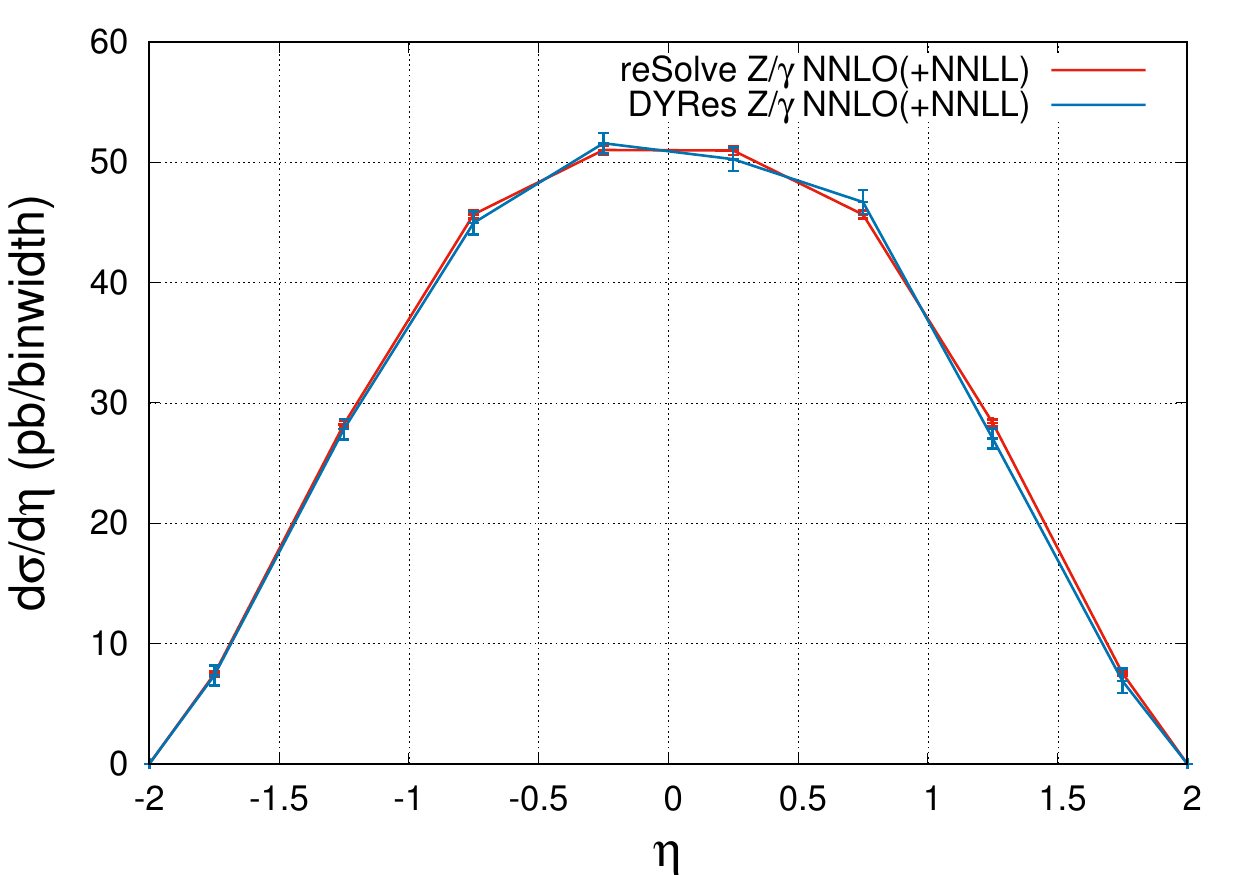}
\caption{Rapidity spectrum, including only the resummed piece, for Drell-Yan production via neutral current $Z$ or $\gamma^*$ at NNLO (with NNLL resummation) again for the setup 1 benchmark, as given in table \ref{DYtestinputstable}. The agreement between the two programs is good, further validating  {\tt reSolve}. The error bars are the Monte Carlo errors from the resummed part only and are largely a reflection on the length of the runs performed.} \label{etacompDYRESNNLOZ}
\end{figure} 

\begin{figure} [!htbp]
  \centering
\includegraphics[height = 9cm]{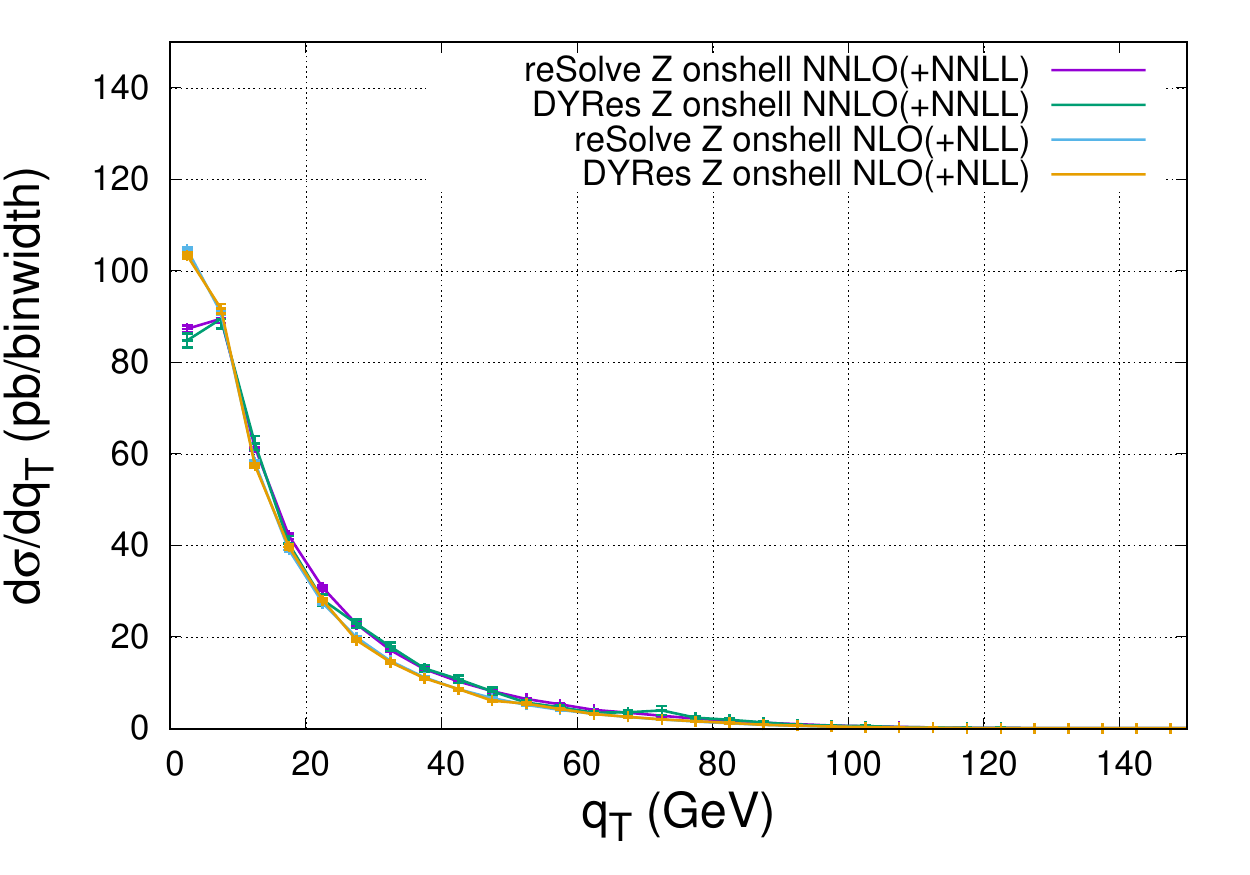}
\caption{Transverse momentum spectrum, including only the resummed piece, for Drell-Yan production via neutral current $Z$ onshell for the setup 2 benchmark, as given in table \ref{DYtestinputstable}. The agreement between the two programs is good at both NLO(+NLL) and NNLO(+NNLL), further validating  {\tt reSolve}. The error bars are the Monte Carlo errors from the resummed part only and are largely a reflection on the length of the runs performed.} \label{setup2qTreSolvevsDYRes}
\end{figure}

Similar comparisons can be performed for the on-shell Z LHC $14TeV$ case of setup 2 (given previously in table~\ref{DYtestinputstable}). Figure~\ref{setup2qTreSolvevsDYRes} demonstrates the agreement in the NLO(+NLL) and NNLO(+NNLL) transverse momentum spectra between {\tt reSolve} and {\tt DYRes}. The total cross-sections are also in agreement between {\tt reSolve} and {\tt DYRes} with the programs obtaining  $2056.2 \pm 3.0pb$ and $2050.5 \pm 2.1pb$ respectively for the resummed pieces only. The transverse momentum spectra in both figures~\ref{qTcompDYRESNNLOZ} and \ref{setup2qTreSolvevsDYRes} do not show the expected behaviour of the differential cross-section tending to 0 at zero transverse momentum solely because of the binning (set to match DYRes), figure~\ref{WpmqTfulll} demonstrates this expected behaviour in reSolve for the $W^{\pm}$ case of setup 3, where finer binning has been used for the low $q_T$ end of the spectrum to demonstrate this behaviour.

Finally, similar comparisons may be undertaken for the $W^{\pm}$ channel, but are not included here. The overall resummed piece only NNLO results for {\tt reSolve} and {\tt DYRes} are $1465.4 \pm 1.3pb$ and $1487 \pm 10pb$, this agrees at the $2\%$ level.

In general, in producing the theoretical predictions beyond leading-order there are a multitude of choices and methodology-linked effects which affect the precise output values of the two programs. In order to produce these comparisons of the {\tt reSolve} and {\tt DYRes} results we have sought to minimise these differences and thereby demonstrate the level of agreement of the codes. In general, results may show larger differences down to the exact choices of the running method for $\alpha_s$, the nature of how higher transverse momenta are dealt with ({\tt reSolve} essentially uses a step function by allowing the user to specify a $q_T$ range, whereas {\tt DYRes} gradually reduces the effects of higher $q_T$s via an arbitrarily-defined ``switch''), the precise generation of phase-space points, how the $\eta$ range is limited at its extremities, the precise nature of the pdf-fitting function (here we refer to not the pdfs - which of course are another source of variation in theoretical predictions - but to the form of the fitting function itself chosen) and many others, which are chosen differently in the two programs. These choice differences were eliminated as much as possible in the comparisons presented here, nonetheless these effects tend to result in differences of order $5\%$ and so this should be considered the accuracy of the predictions for a generic input\footnote{In particular, the effects of the $\alpha_s$ running method and the $q_T$ switch themselves are the largest differences seen and may cause differences themselves of up to $5\%$, the choices in {\tt DYRes} raise the predictions by around this amount relative to the default choices in {\tt reSolve}.}.

\subsubsection{Further Differential Spectra} \label{DYDifferentialPlots}

Finally, we provide several spectra obtained for each of the three setups to demonstrate the agreement with the corresponding figures provided in \cite{Catani:2009sm} and to illustrate that the qualitative behaviour of the reSolve program is as expected.

First again consider the case of $Z/\gamma^*$ in our benchmark setup 1, the $p_T^{min}$ and $p_T^{max}$ distributions for this setup were given in figure~2 of \cite{Catani:2009sm}. The corresponding spectrum, as produced by the new {\tt reSolve} program, is provided below in figure~\ref{ZgammapTmaxmin}, we provide only the NLO(+NLL) and NNLO(+NNLL) spectra as the transverse momentum of the system at LO is zero and so is meaningless within the formalism. The qualitative agreement between the {\tt reSolve} spectra and the previous spectra is good, meanwhile the behaviour of the spectra is exactly as expected with the NNLO spectrum having a larger peak. Both the $p_T^{min}$ and $p_T^{max}$ distributions cut-off at $20 GeV$ due to the applied pTcut, the $p_T^{min}$ spectrum peaks just below $\frac{m_Z}{2}$ and the $p_T^{max}$ spectrum at just above $\frac{m_Z}{2}$. The $p_T^{min}$ spectrum also cuts-off at $55GeV$ as the $q_T$ range had an upper limit of $110 GeV$ whilst the $p_T^{max}$ spectrum continues above $55GeV$, all this behaviour is exactly as expected. Small differences in the figures occur as our calculations include the resummed piece of the cross-section only.

For the on-shell $Z$ case of benchmark setup 2, figure~1 of \cite{Catani:2009sm} provides the rapidity distribution, the corresponding distribution calculated by {\tt reSolve} is given below in our figure~\ref{Zonshelleta}, the agreement between the two is excellent, with the effects of the resummation beyond LO significantly increasing the cross-section between LO and NLO(+NLL), with NNLO(+NNLL) only offering a small additional correction.

Finally, for the case of $W^{\pm}$ in our benchmark setup 3 the differential distribution provided in \cite{Catani:2009sm} is the transverse mass distribution in their figure~3, compare this with the same transverse mass distribution produced by {\tt reSolve} in figure~\ref{WpmmT}. Again, there is good agreement between the results, the leading order $m_T$ distribution turns on at $50GeV$ because the $W^{\pm}$ is produced at zero net transverse momentum, so without any additional radiation we require $p_T^{lepton} = p_T^{miss}$ therefore the $p_{T}^{miss}$ cut of $25GeV$ sets the lower limit of the LO $m_T$ distribution to $50 GeV$. Of course, this limit is not a hard limit beyond LO as additional radiation can carry away transverse momentum. For LO, NLO(+NLL) and NNLO(+NNLL) the $m_T$ distribution peaks just below the W mass at around $80GeV$ as expected.

\begin{figure} [!htbp]
  \centering
\includegraphics[height = 9cm]{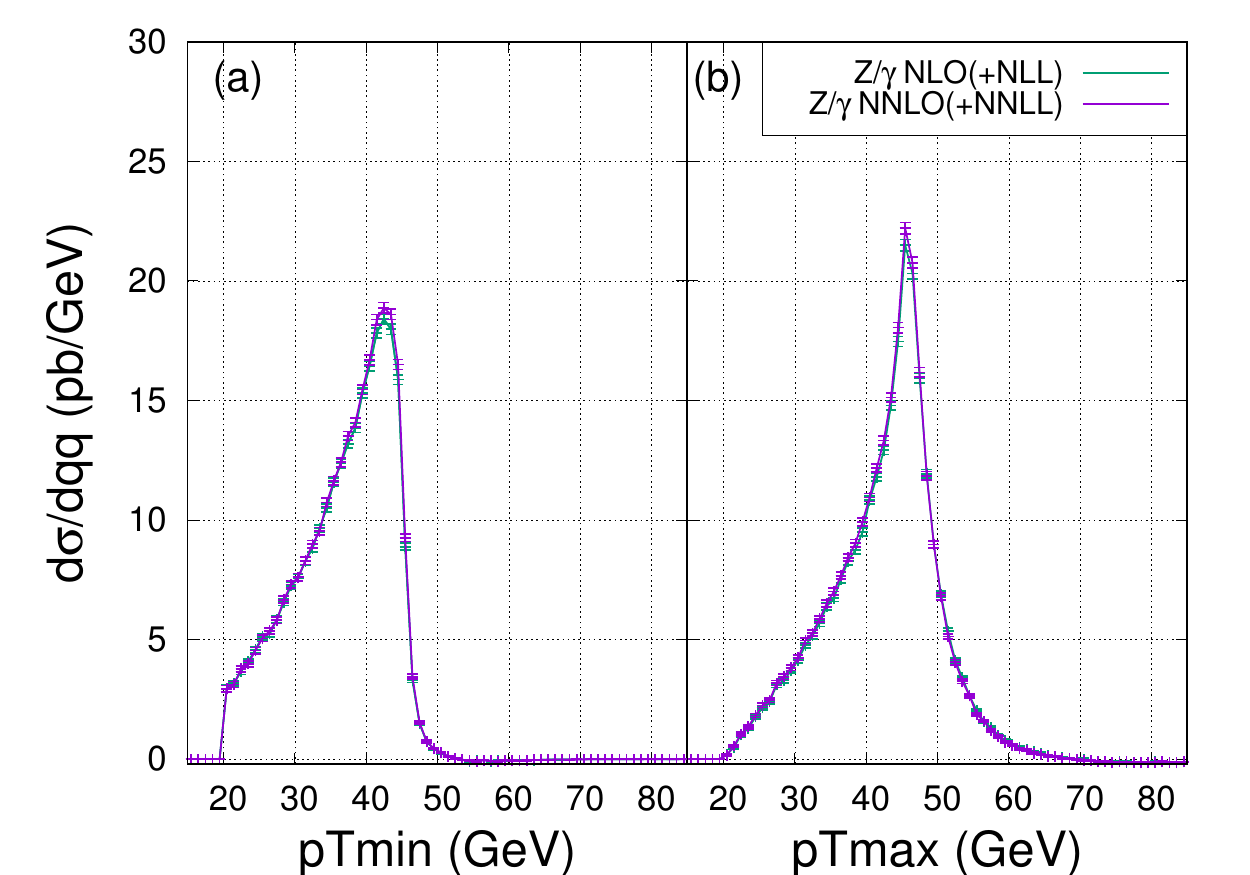}
\caption{Minimum and maximum transverse momenta spectrum produced by {\tt reSolve} for the two outgoing leptons produced, including only the resummed piece, for Drell-Yan production via neutral current $Z$ or $\gamma^*$ for the setup 1 benchmark, as given in table \ref{DYtestinputstable}. This figure should be compared with figure~2 of \cite{Catani:2009sm}, the agreement between the two results is good, with slight differences arising due to the implementation of the formalism, the results are consistent within uncertainties. The error bars are the Monte Carlo errors from the resummed part only and are largely a reflection on the length of the runs performed.} \label{ZgammapTmaxmin}
\end{figure}

\begin{figure} [!htbp]
  \centering
\includegraphics[height = 8.5cm]{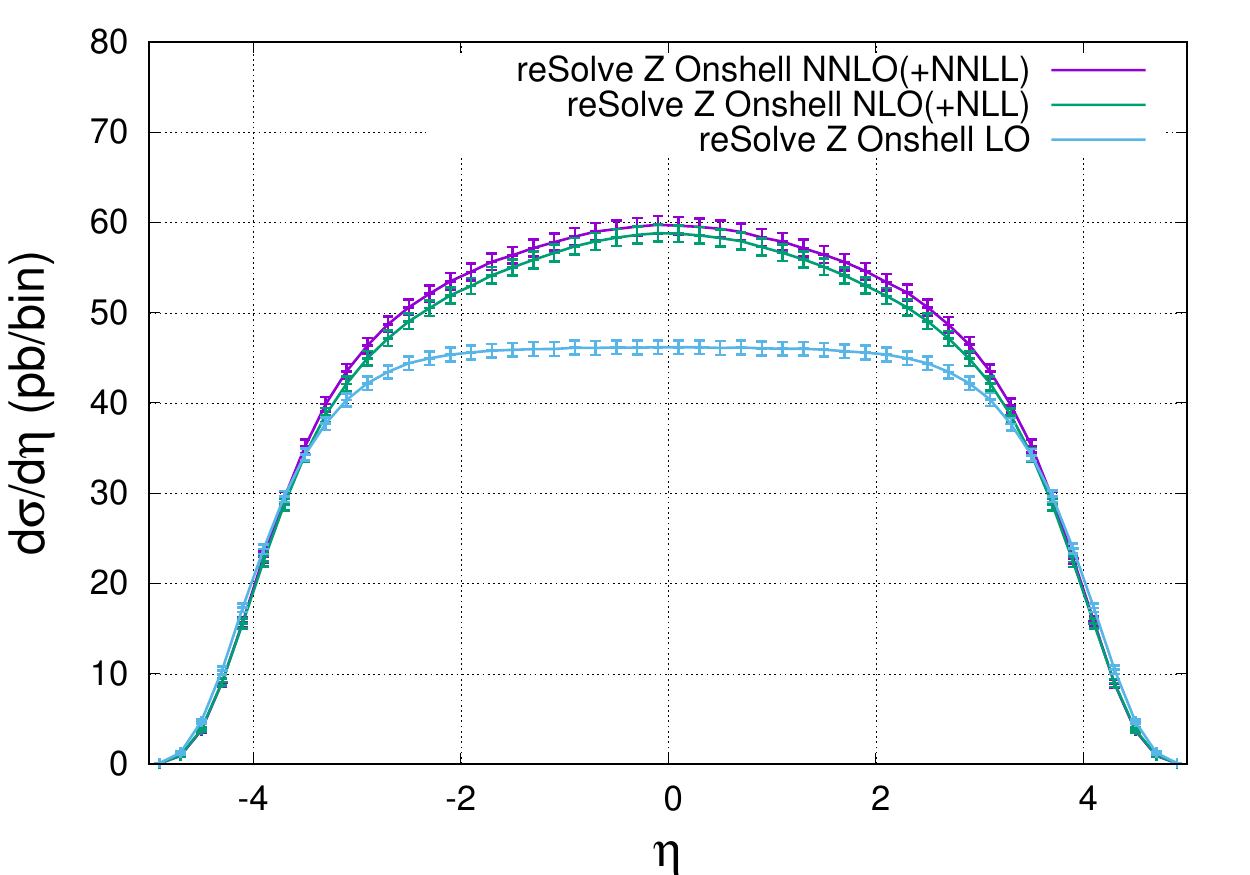}
\caption{Rapidity distribution for the two outgoing leptons produced by the on-shell Z boson, including only the resummed piece, for Drell-Yan production via neutral current $Z$ on-shell for the setup 2 benchmark, as given in table \ref{DYtestinputstable}. This figure should be compared with figure~1 of \cite{Catani:2009sm}, the agreement between the two results is excellent. The LO includes no resummation, whilst for beyond LO resummation is included, so NLO includes NLL resummation and NNLO includes NNLL resummation of logarithms. The error bars are the Monte Carlo errors from the resummed part only and are largely a reflection on the length of the runs performed.} \label{WpmmT}
\end{figure}

\begin{figure} [!htbp]
  \centering
\includegraphics[height = 8.5cm]{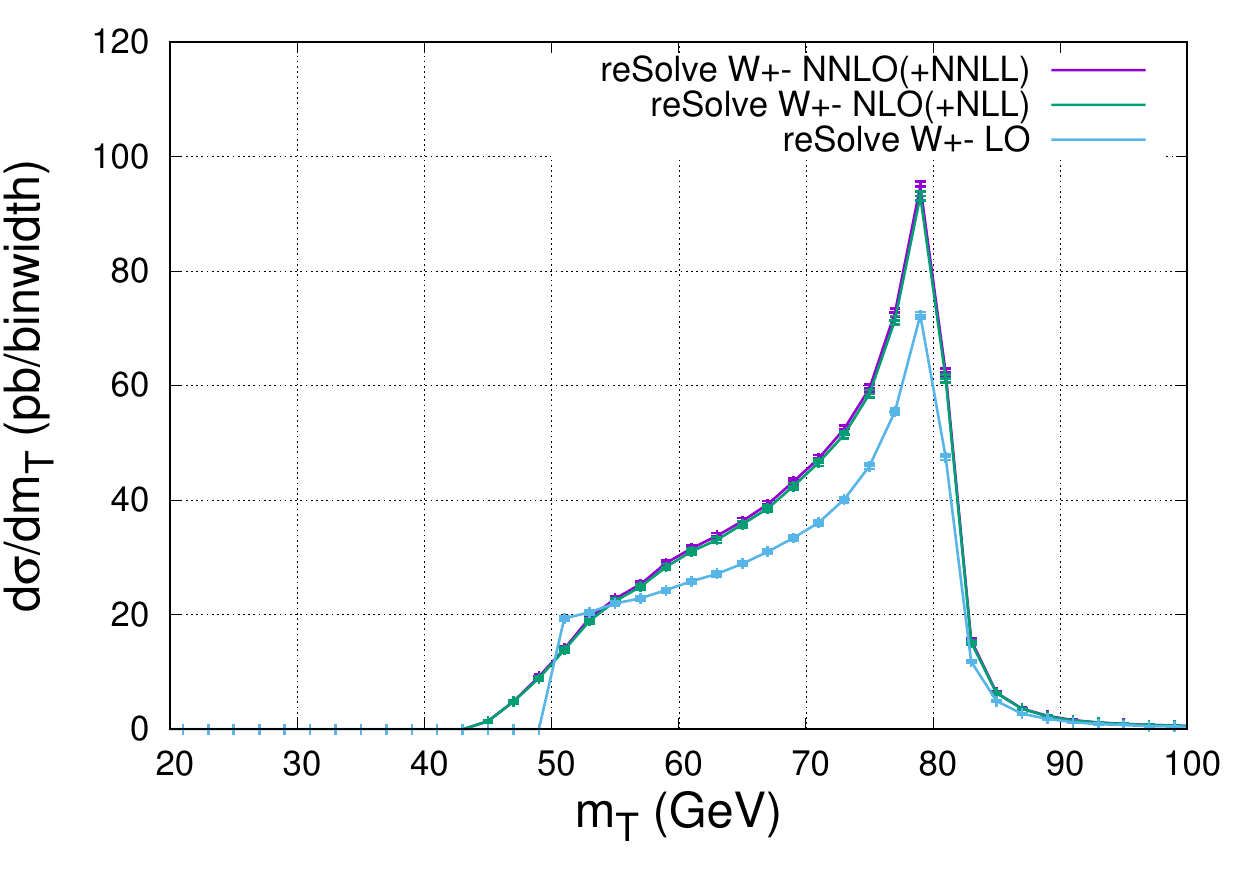}
\caption{Transverse mass distribution for the $W^{\pm}$ case of the setup 3 benchmark, including only the resummed piece, as given in table \ref{DYtestinputstable}. This figure should be compared with figure~3 of \cite{Catani:2009sm}, the agreement between the two results is good. The LO includes no resummation, whilst for beyond LO resummation is included, so NLO includes NLL resummation and NNLO includes NNLL resummation of logarithms. The error bars are the Monte Carlo errors from the resummed part only and are largely a reflection on the length of the runs performed.} \label{Zonshelleta}
\end{figure} 

\hfill \break

\pagebreak

\begin{figure} [!htbp]
  \centering
\includegraphics[height = 7.5cm]{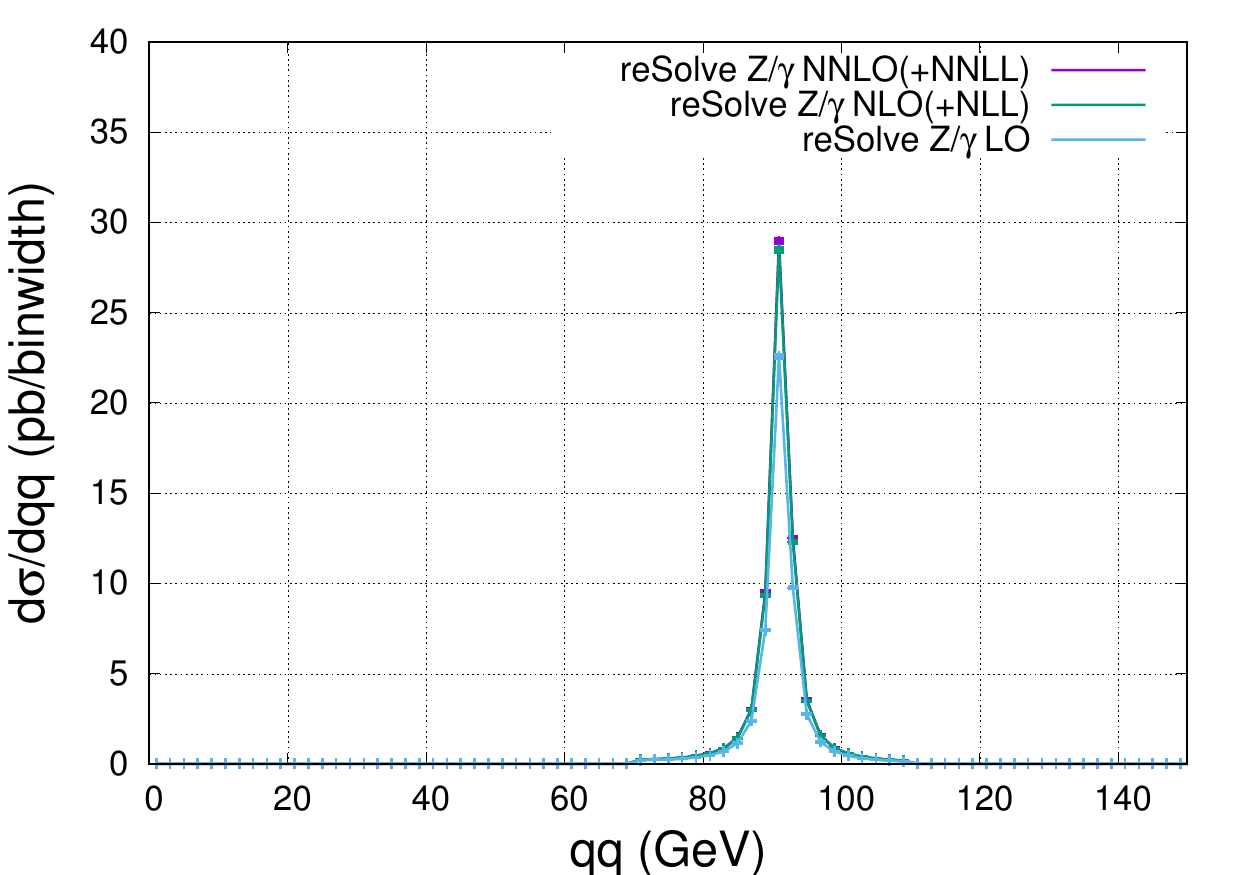}
\caption{Invariant mass distribution for the neutral current $Z/\gamma^*$ case of the setup 1 benchmark, including only the resummed piece. This figure illustrates the expected peaking of the invariant mass distribution around $m_Z$ even in the case of Drell-Yan production which isn't specifically onshell, due to the enhancment around $q^2 = m_Z^2$ from the Z propagator. As previously, the NLO and NNLO include resummation to their corresponding orders (NLL and NNLL respectively). The error bars are the Monte Carlo errors from the resummed part only and are largely a reflection on the length of the runs performed.} \label{Zgammaqq}
\end{figure}

\begin{figure} [!htbp]
  \centering
\includegraphics[height = 7.5cm]{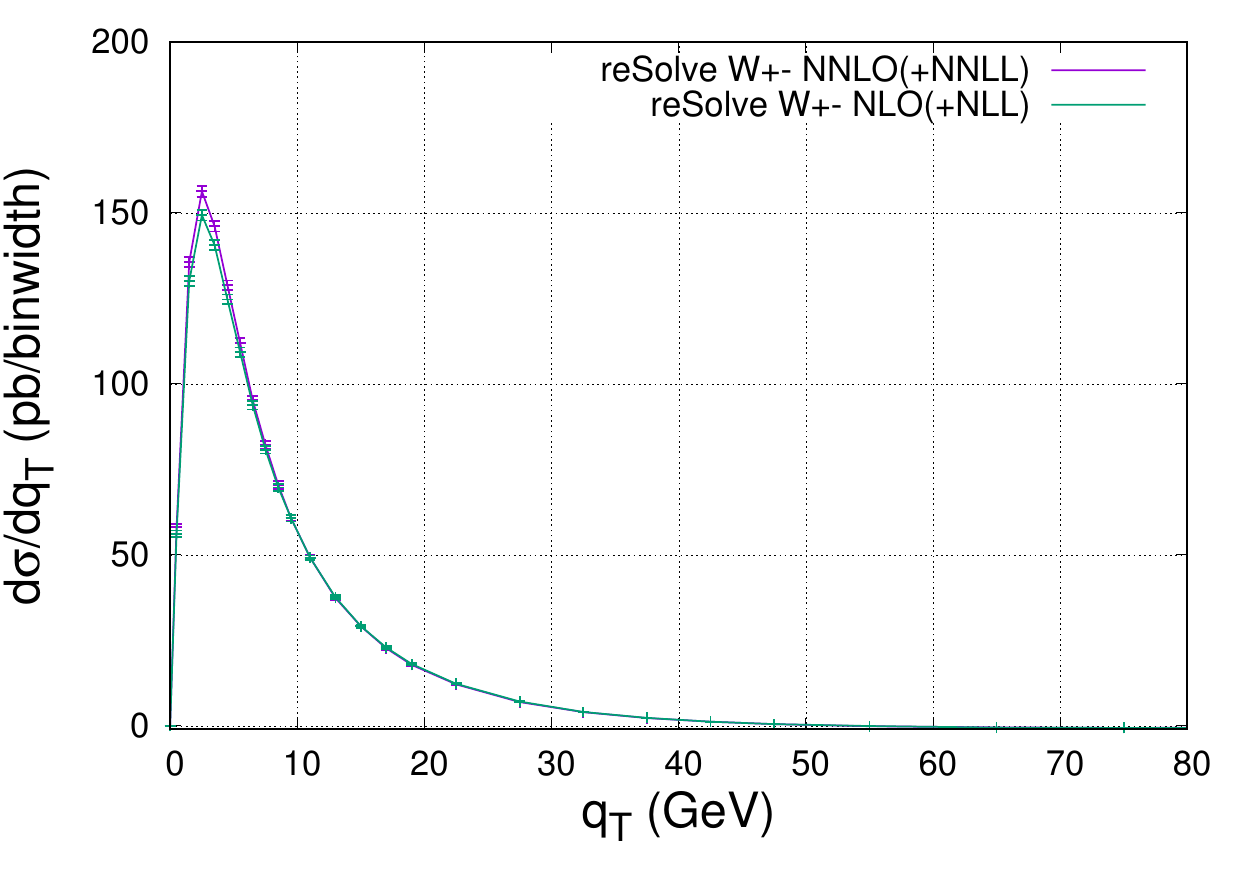}
\caption{Transverse momentum distribution for the charged current $W^{\pm}$ case of the setup 3 benchmark, including only the resummed piece. This figure illustrates the expected peaking of the tranverse mass distribution near 0 and the fact that, despite this, the differential cross-section is still 0 at zero transverse momentum. As previously, the NLO and NNLO include resummation to their corresponding orders (NLL and NNLL respectively). The error bars are the Monte Carlo errors from the resummed part only and are largely a reflection on the length of the runs performed.} \label{WpmqTfulll}
\end{figure}

\begin{figure} [!htbp]
  \centering
\includegraphics[height = 8cm]{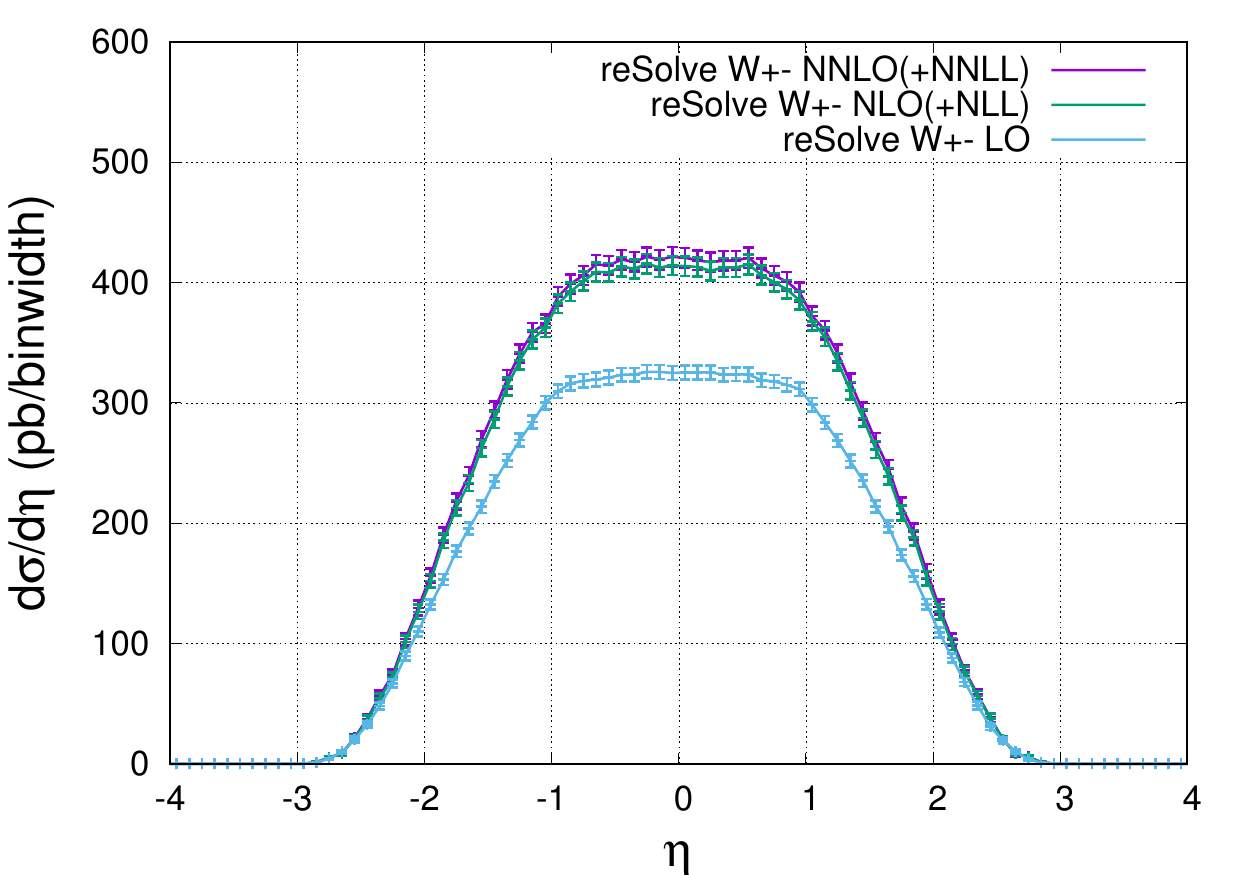}
\caption{Rapidity distribution for the charged current $W^{\pm}$ case of the setup 3 benchmark, including only the resummed piece. The distribution is correctly zero outside the range $-3<\eta<3$ as a result of the $\eta$ range set in the input file. The addition of resummation and the additional virtual corrections in going from leading order to next-to-leading order (+next-to-leading-logarithm resummation) significantly increases the amplitude of the rapidity distribution and the total cross-section, contrastingly however the increase between NLO(+NLL) and NNLO(+NNLL) is insignigicant. The error bars are the Monte Carlo errors from the resummed part only and are largely a reflection on the length of the runs performed.} \label{Wpmeta}

\end{figure}

\begin{figure} [!htbp]
  \centering
\includegraphics[height = 8cm]{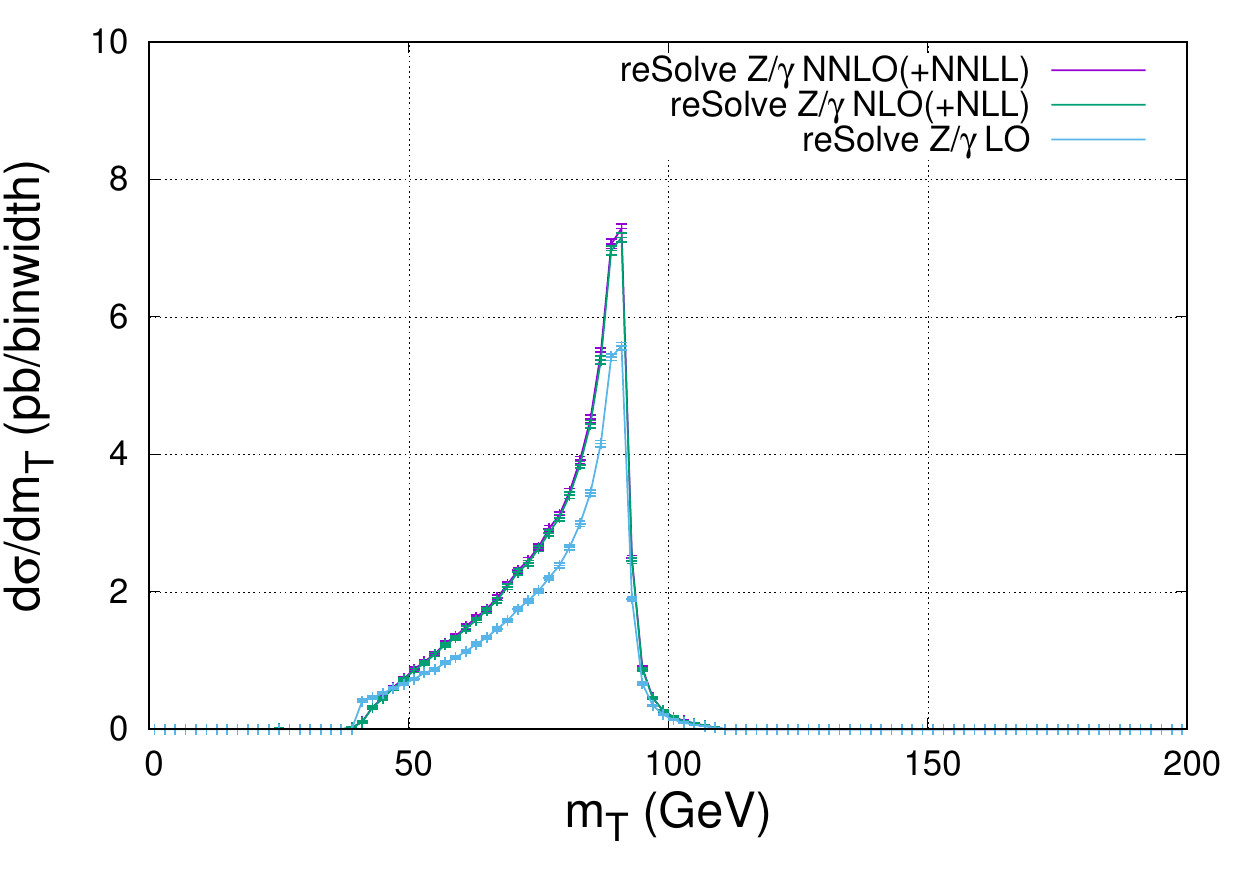}
\caption{Transverse mass distribution for the neutral current $Z/\gamma^*$ case of the setup 1 benchmark, including only the resummed piece. The distribution has a kinematic boundary at leading order at $40GeV$ as each of the outgoing leptons has a $p_T > 20 GeV$ due to the cuts and at LO there are no other particles to carry away any transverse momentum, therefore conservation of transverse momentum imposes this kinematic constraint upon the $m_T$ distribution. As in many other cases shown, the addition of resummation and the additional virtual corrections in going from leading order to next-to-leading order (+next-to-leading-logarithm resummation) significantly increases the amplitude of the distribution and the total cross-section, however the increase between NLO(+NLL) and NNLO(+NNLL) is not signigicant. The error bars are the Monte Carlo errors from the resummed part only and are largely a reflection on the length of the runs performed.} \label{ZgammamT}

\end{figure}

In addition to these comparisons against the figures in the literature, many plots of the various differential distributions produced by {\tt reSolve} for all three benchmark setups of table~\ref{DYtestinputstable}, and others, were extensively checked for any obvious problems during our validation of the program, figures~\ref{Zgammaqq}-\ref{ZgammamT} are a small subset of these, again all illustrate the exact expected behaviour.

\pagebreak

\subsection{Multiple PDF fits validation} \label{multipdffits_validation}

\begin{figure}[ht!]
\centerline{\includegraphics[height = 9cm, width = 14cm]{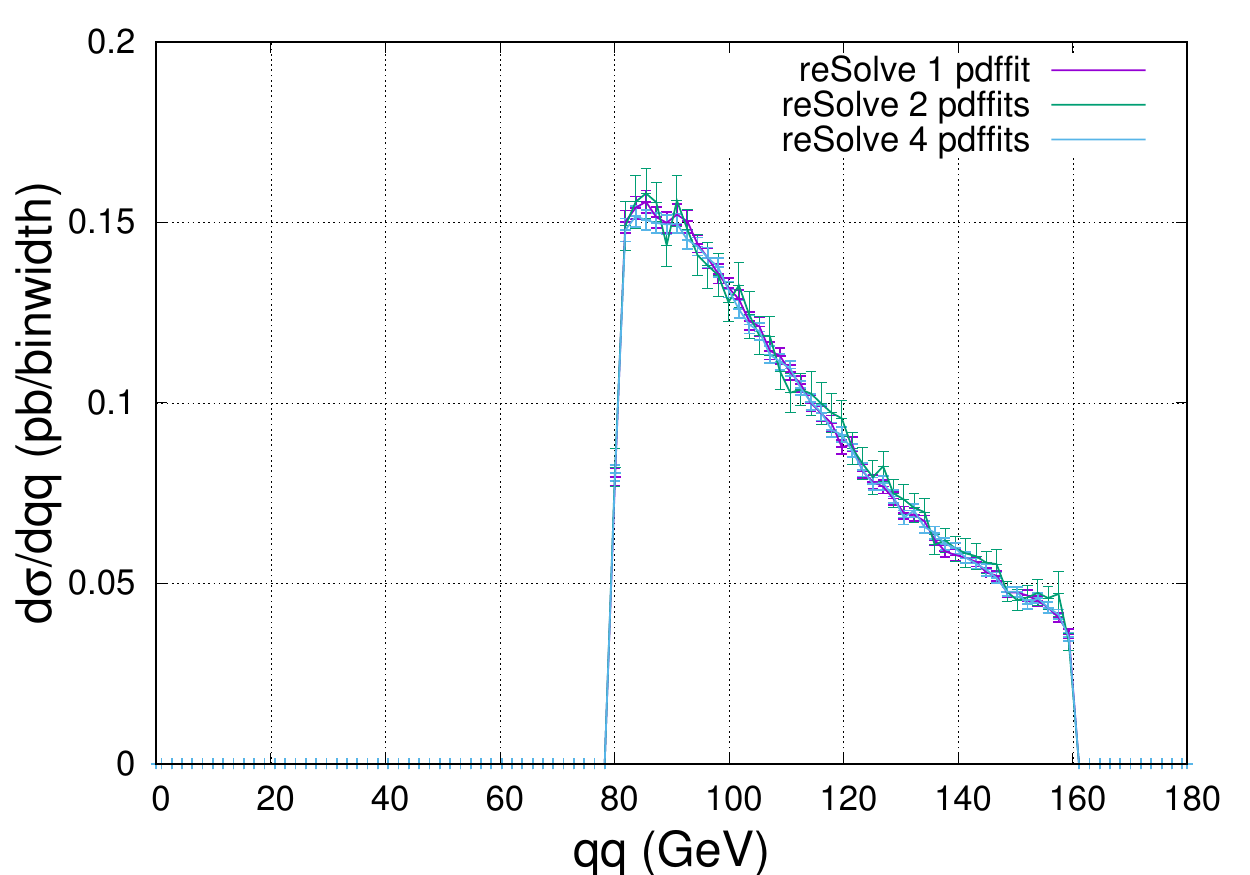}} 
\caption{The invariant mass spectrum for diphoton production for the {\tt test1} input parameters provided in table~\ref{testinputstable} for 1, 2 and 4 pdffits made.  The input files are again provided with the {\tt reSolve} program and are {\tt Diphoton\_NNLO\_test\_1.dat}, {\tt Diphoton\_NNLO\_test1\_twopdffits.dat} and  {\tt Diphoton\_NNLO\_test1\_fourpdffits.dat}. The results show excellent agreement, validating the program for use with multiple PDF fits.} \label{multipdffits_qq} 
\end{figure}

\begin{figure}[ht!]
\centerline{\includegraphics[height = 9cm, width = 14cm]{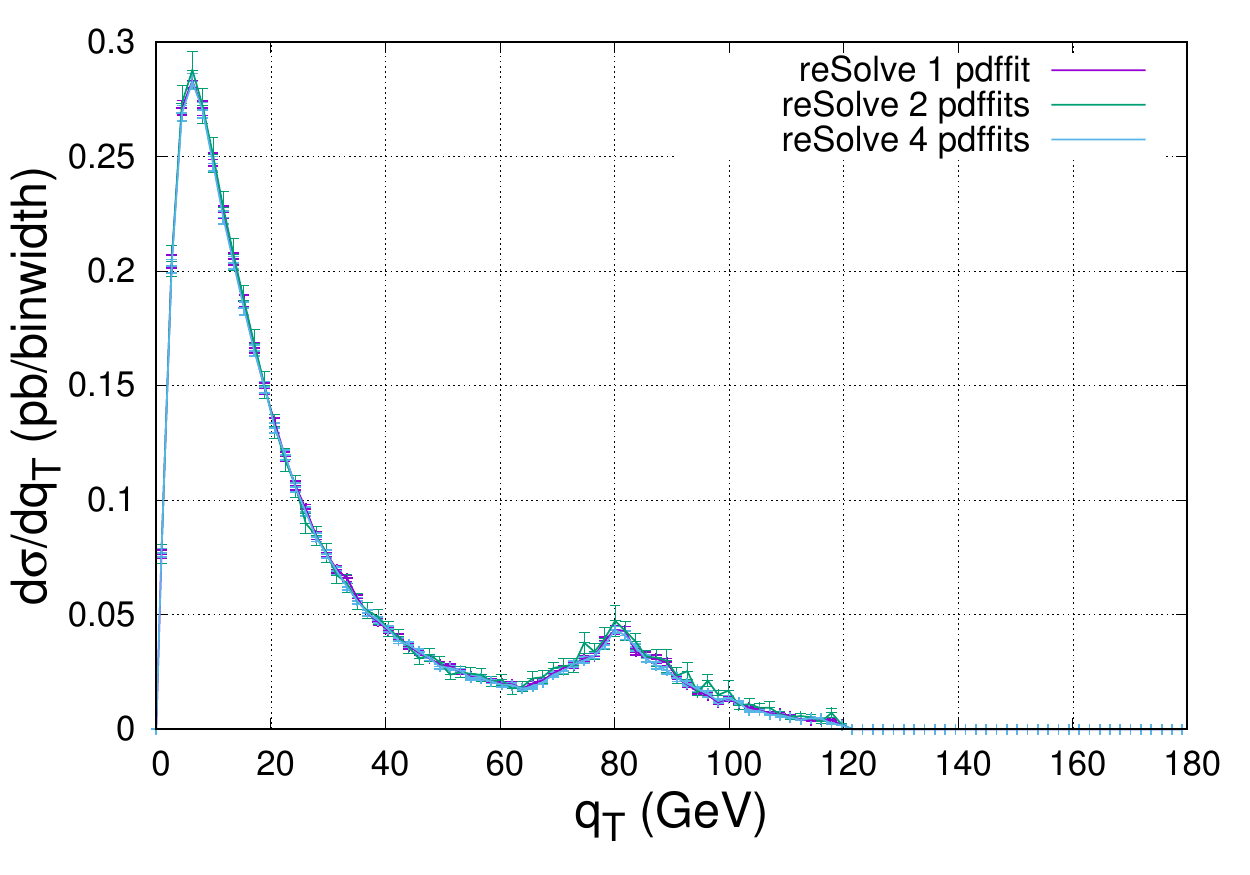}} 
\caption{The transverse momentum spectrum for diphoton production for the {\tt test1} input parameters provided in table~\ref{testinputstable} for 1, 2 and 4 pdffits made.  The input files are again provided with the {\tt reSolve} program and are {\tt Diphoton\_NNLO\_test\_1.dat}, {\tt Diphoton\_NNLO\_test1\_twopdffits.dat} and  {\tt Diphoton\_NNLO\_test1\_fourpdffits.dat}. The results show excellent agreement, validating the program for use with multiple PDF fits.} \label{multipdffits_qT} 
\end{figure}

In order to validate the use of multiple PDF fits, we used the {\tt test1} input parameters for the diphoton process, detailed in table~\ref{testinputstable}, and carried out the same resummation calculation of the cross-section with 1 PDF fit at $113.14GeV$; 2 PDF fits at $97.98$GeV and $138.56$GeV (corresponding to {\tt en\_sec\_multiplier} $= 1.5$); and 4 PDF fits at $87.64$GeV, $105.16$GeV, $126.20$GeV and $148.72$GeV (corresponding to {\tt en\_sec\_multiplier} $= 1.2$). Given the invariant mass range for the diphoton here was narrow, from $80$GeV to $160$GeV, we expect no particular benefit from using multiple PDF fits, therefore we use this as a check that all 3 runs produce a consistent differential cross-section. Figure~\ref{multipdffits_qq} and figure~\ref{multipdffits_qT} provide the invariant mass spectrum and transverse momentum spectrum comparison respectively between the 1, 2 and 4 PDF fit results. The input files are again provided with the {\tt reSolve} program and are {\tt Diphoton\_NNLO\_test\_1.dat}, {\tt Diphoton\_NNLO\_test1\_twopdffits.dat} and  {\tt Diphoton\_NNLO\_test1\_fourpdffits.dat}. The results show very good agreement, within the errors, thereby validating the program for use with multiple PDF fits. The total cross-sections evaluated for 1, 2 and 4 PDF fits are given in table~\ref{timetablepdffits} in section~\ref{speed} as part of the discussion of the speed of the program and are also in agreement. Meanwhile, the invariant mass and transverse momentum spectrum produced for the diphoton ATLAS validation in section~\ref{expvalid} in figures~\ref{qqFrancescocompplot}-\ref{resolve2gresAtlasqT} for the {\tt reSolve} data used 5 PDF fits across the large invariant mass ($qq$) range, thereby also providing validation of the multiple PDF fit running of {\tt reSolve} against experimental data.

\subsection{Performance} \label{speed}

Given the number of Monte Carlo iterations it is necessary to perform for the phase space integral to produce the desired accuracy, and the fact that for each phase space point an inverse Fourier transform and double inverse Mellin transform is required - each of which require around 20, 50 and 50 points respectively, speed can be very important for transverse momentum resummation programs in this formalism. As a demonstration, should one require 1,000,000 phase space points, one can expect parts of the resummation code (parts of {\tt inverse\_mellinresummed.cc} and {\tt hardfns.cc}) to be called $1,000,000 \times 20 \times 50 \times 50 = 5\times 10^{10}$ times. Therefore particular care has been taken, even within this first main version of the code, to ensure it runs quickly. Whilst the run time varies significantly depending upon the input file, as certain points may require more $b$ points to be evaluated in the inverse Fourier transform, large $\eta$ values have more points on each Mellin contour, and different processes require different numbers of non-zero contributions to the cross-section to be summed; we seek here to give a guide as to the speed of this first main version of the {\tt reSolve} program. In particular, in table~\ref{timetabletests} we compare it against the private code {\tt 2gres} used also for the validations in Section~\ref{resurevampcomp}, again for the diphoton {\tt test1} and {\tt test2} inputs at NNLO(+NNLL) listed in table~\ref{testinputstable} with 550,000 phase space points. As demonstrated, the {\tt reSolve} program is consistently almost twice as quick as the previous {\tt 2gres} program, completing the same run in $53\%$ of the time. This speed up is important, allowing a greater number of evaluations to be performed and thereby attaining a greater accuracy with the same computer resources.

\begin{center}
\begin{table} [!htbp]
\centering
\begin{tabular}{|c|c|c|c|c|} \hline
Program & \multicolumn{2}{c|}{Diphoton\_NNLO\_test\_1} & \multicolumn{2}{c|}{Diphoton\_NNLO\_test\_2} \\ \hline
&  $\sigma$(pb) & time(min:s) & $\sigma$(pb) & time(min:s) \\ \hline
{\tt reSolve} & $7.68 \pm 0.03$ & 1678m:22s & $2.536 \pm 0.009$ & 1370m:11s \\ \hline
{\tt 2gres} & $7.68 \pm 0.03$ & 3178m:35s & $2.556 \pm 0.008$ & 2763m:46s \\ \hline
\end{tabular}
\caption{Comparison of the time taken to evaluate 550,000 phase space points in the new public code {\tt reSolve} and the old private code {\tt 2gres}. The times listed are total core times, summing those across all cores. The total cross-sections are also given, demonstrating good consistency between the programs. Note {\tt reSolve} here used one PDF fit, as that is all that was available in the previous {\tt 2gres} program, and the integrator {\tt Cuba} was used by both programs to allow a fair comparison. The test files are the {\tt Diphoton\_NNLO\_test\_1} and {\tt Diphoton\_NNLO\_test\_2} files used previously and listed in table~\ref{testinputstable}.} 
\label{timetabletests}
\end{table}
\end{center}

Running with multiple PDF fits will also slow down the running of the {\tt reSolve} program; multiple PDF fits therefore should only be used for a wide invariant mass range, where the adoption of multiple PDF fits at different scales through the invariant mass range may offer increased precision. In order to demonstrate this slowdown, table~\ref{timetablepdffits} provides the run-times of the {\tt reSolve} program using 1, 2 and 4 PDF fit files respectively, once more for the {\tt Diphoton\_NNLO\_test\_1} inputs listed previously. Again the specific input files are available pre-setup for 1, 2 and 4 PDF fits as {\tt Diphoton\_NNLO\_test\_1.dat}, {\tt Diphoton\_NNLO\_test1\_twopdffits.dat} and {\tt Diphoton\_NNLO\_test1\_fourpdffits.dat}. The invariant mass and transverse momentum spectra for these runs were given previously in section~\ref{multipdffits_validation} in figures~\ref{multipdffits_qq} and \ref{multipdffits_qT}. Note the time listed does not include that to produce the fits, the PDF fit files were provided here. The comparison demonstrates that, as one might expect, the runtime is significantly longer for multiple PDF fits, indeed it is more than twice as long, however between 2 and 4 PDF fits the runtime does not increase, demonstrating that the main difference comes when one starts to use multiple PDF fits. Note however that even with multiple PDF fits, the {\tt reSolve} program is similar in speed to the old private {\tt 2gres} program, which only used one PDF fit.

Many of these issues of speed and time taken to evaluate the phase space points required are further ameliorated by the ability of the {\tt reSolve} program to allow parallelisation across many cores in many machines. A detailed description of how to perform the parallelisation and how the program works when parallelised using {\tt k\_vegas} was given in Section~\ref{parallelisation}. A comparison of the physical time elapsed when running more than 500,000 phase space points on one core, on 4 cores with {\tt Cuba}, and on many cores on many machines with {\tt k\_vegas} is provided in table~\ref{timetableparallelised}. The run-times clearly demonstrate how much physical time may be saved using the parallelisation option in {\tt reSolve}. The results obtained are for the {\tt Diphoton\_NNLO\_test\_1} inputs used previously and were also consistent with those provided in table~\ref{timetabletests}, with the 12 cores parallelisation across 3 machines obtaining  $7.67 \pm 0.08$ pb. Note the error is however larger here as the same number of total iterations were performed in smaller batches. To obtain an error of similar magnitude will require larger batches for each core, and therefore a longer runtime than indicated here, however the combination of batches from different cores and machines still means the total runtime required for a given error will also of course be significantly smaller than when unparallelised. Also it should be noted that, as the {\tt reSolve} program when parallelised using {\tt k\_vegas} waits for all cores to be complete at a given iteration, the speed of the program will be governed by the slowest core - this is why the core time in the parallelised setup is longer than for the unparallelised or {\tt Cuba} parallelised over one machine implementations. This is necessary so that after each iteration the overall grid of points and weights for the Monte Carlo integration is updated and used by all cores for the next iteration. We therefore recommend parallelising across cores and machines of similar speeds.

\begin{center}
\begin{table}
\centering
\begin{tabular}{|c|c|c|} \hline
Number of PDF fits & \multicolumn{2}{c|}{Diphoton\_NNLO\_test\_1} \\ \hline
&  $\sigma$(pb) & time(min:s) \\ \hline
1 & $7.68 \pm 0.03$ & 1678m:22s \\ \hline
2 & $7.67 \pm 0.03$ & 3636m:35s \\ \hline
4 & $7.63 \pm 0.03$ & 3617m:52s \\ \hline
\end{tabular}
\caption{Comparison of the time taken to evaluate 550,000 phase space points in the {\tt reSolve} program for different numbers of PDF fits. Again the times listed are total core times, summing those across all cores.The total cross-sections are also given, demonstrating consistency between the fits. As expected multiple PDF fits take much longer, however 4 PDF fits took no longer than 2 PDF fits. The test files used were the {\tt Diphoton\_NNLO\_test\_1.dat}, {\tt Diphoton\_NNLO\_test1\_twopdffits.dat} and  {\tt Diphoton\_NNLO\_test1\_fourpdffits.dat} provided with the {\tt reSolve} program. The general inputs for this setup were listed in table~\ref{testinputstable} in the case of 1 PDF fit.} 
\label{timetablepdffits}
\end{table}
\end{center}
\begin{center}
\begin{table}
\centering
\begin{tabular}{|c|c|c|c|c|c|} \hline
\multicolumn{6}{|c|}{Time elapsed(min:s)} \\ \hline
\multicolumn{2}{|c|}{One Core} & \multicolumn{2}{c|}{4 cores (one machine) using {\tt Cuba}} & \multicolumn{2}{c|}{12 cores across 3 machines using {\tt k\_vegas}} \\ \hline
  Core time   & Physical time &   Core time   & Physical time &   Core time   & Physical time \\ \hline
 $\sim$1678m:22s & $\sim$1678m:22s & 1678m:22s & 474m:34s & 2756m:12s & 229m:41s
 \\ \hline
\end{tabular}
\caption{The time taken to evaluate over 500,000 phase space points in the {\tt reSolve} program with different degrees of parallelisation. ``One Core'' indicates either {\tt k\_vegas} or {\tt Cuba} used with one core only, the second column shows {\tt Cuba} used parallelising across the 4 cores of the machine used as standard, finally the time taken parallelising across 12 cores across 3 machines using the {\tt k\_vegas} parallelisation routine {\tt multi\_machine\_parallel\_local} provided with {\tt reSolve} is given. Note the time given here for one core is approximate as this test was not run, it is an indication based on extrapolation from 4 cores on the same machine. This comparison was performed using the {\tt Diphoton\_NNLO\_test\_1} inputs given in table~\ref{testinputstable}.} 
\label{timetableparallelised}
\end{table}
\end{center}

\section{Future Developments} \label{future}

This version of the {\tt reSolve} program is the first main version of many, therefore we intend to optimise and extend this version further, undertaking an ongoing development program. There are many areas for improvement in the program, a few of those we consider the most important are listed here:
\begin{itemize}
\item Add the finite parts for the diphoton and Drell-Yan spectra. Given the program currently only includes the resummed (i.e. low $q_T$) part of the transverse momentum differential cross-section, the obvious extension is to add the finite piece which is dominant at high $q_T$. This will require matching of the low and high $q_T$ pieces at intermediate $q_T$ using the matching procedure outlined in \cite{Bozzi:2005wk}. This extension to the program will be performed in the near future and will enable the production of the transverse momentum spectrum across the whole $q_T$ range.
\item Extend to additional processes, Higgs production foremost amongst them, perhaps also Beyond Standard Model contributions in cases such as $Z'$.
\item Inclusion of the gg hard factors to higher orders, this will enable Higgs production to be incorporated. With this extension the program will include both signal and background (diphoton) for the Higgs, therefore signal-background interference could be examined \cite{Cieri:2017kpq}.
\item Currently the PDFs are fitted, and used at given scales, we would like to examine the possibility of interpolating and using them directly at the scale desired.
\item Currently the only PDF sets available to use are the MSTW PDFs \cite{Martin:2009iq}, in future versions we will broaden to allow any PDF set to be used. In order to do so we will allow Les Houches Accord PDF formats~\cite{Whalley:2005nh} to be read in.
\item There is substantial scope to further speed up the program significantly, the speed of the program is currently hampered by memory considerations which slows the program down by a further $20\%$, this could be reduced substantially.
\item The formalism could be relatively easily adapted to allow the implementation of a jet veto, which may be of interest in certain applications.
\item The program can be relatively easily extended to QED resummation, an area itself of increasing interest.
\end{itemize}

Beyond these shorter-term objectives, the universality of the formalism applied within {\tt reSolve}, along with the program's clearly-designed modularity, allow the potential to interface the code with existing more general packages in order to allow their extension to higher accuracy. This could incorporate interfacing with existing Matrix Element generators for automatic generation of resummed spectra for a much wider class of processes, for example NLO Matrix Element generators could be interfaced to allow semi-automatic production of differential spectra at next-to-next-to-leading order with next-to-next-to-leading logarithm resummation. In any case, whichever the precise longer-term direction taken, the properties of the {\tt reSolve} transverse momentum resummation program mean it can form a key part of current and future tools for precise theoretical predictions for collider processes.

\section{Conclusions} \label{conclusions}
We have presented a new transverse resummation program, {\tt reSolve}, specifically designed to allow generalisation to many processes. We described in detail the structure of the program and its modularisation, which allows different processes of interest to be added straightforwardly, and outlined how to do this. Currently the processes included are diphoton + jet and Drell-Yan + jet, both up to NNLO + NNLL, these processes have been extensively validated both internally and externally against known results and previous programs where available, as well as against experimental results. The code itself is publicly available with this paper and also on Github - the latter is where the most up-to-date version is guaranteed to be found. This paper provides details of the theoretical explanation of the b-space Mellin-space formalism applied, whilst also acting as the manual for users, we therefore hope it will prove useful. This version of {\tt reSolve} represents the first main incarnation of the development of this program and what we hope will become a ``go-to'' public code in this area for transverse momentum resummation for any appropriate process. {\tt reSolve} has several advantages over previous programs for transverse momentum resummation; being simple to use, transparently documented and modularised in a way to allow many different processes to be considered within the same program structure. Given the increasing interest in precision physics, both in order to increase understanding of the standard model and to search for hints of new physics, such tools are of vital importance to the ongoing program at the Large Hadron Collider and beyond. We therefore expect the {\tt reSolve} program to be key to many future phenomenological applications.

\section*{Acknowledgements}
This research has been partially supported by STFC consolidated grant 
ST/L000385/1 and by STFC consolidated grant ST/P000681/1, and also in part by the National Science Foundation under Grant No. NSF PHY-1748958 along with the Gordon and Betty Moore Foundation. We wish also to thank both the Cambridge SUSY working group and Leandro Cieri for helpful discussions. TC wishes to thank The Kavli Institute for Theoretical Physics, University of California Santa Barbara for their hospitality, supporting him for a KITP Graduate Fellowship, during which part of this work was completed; additionally he wishes to thank the Cambridge Philosophical Society and the Department of Applied Mathematics and Theoretical Physics at the University of Cambridge for their support during this work.

\pagebreak

\appendix

\section{Summary of {\tt reSolve} Usage} \label{app:usage}

A detailed description of how to use the {\tt reSolve} program is given in the main text of the paper in Section~\ref{usingprog}, however the details are summarised and condensed here for ease of reference.

\subsection{Where to obtain the {\tt reSolve} program}

The {\tt reSolve} program is available online with this paper, the latest version will also always be found on Github at {\tt https://github.com/fkhorad/reSolve}.

\subsection{Basic Usage}

\begin{itemize}
\item Upon downloading and extracting the program to your desired directory, {\tt reSolve} should work ``straight out of the box'' with the built-in {\tt k\_vegas} integrator, nonetheless if the user wishes to use {\tt Cuba} \cite{Hahn:2004fe} they must download this separately and alter the {\tt reSolve} makefile to detail where this is found on their PC.
\item First {\tt make} the program, then the {\tt reSolve} program can be used with a variety of built-in input files provided with the download - these are listed in Tables~\ref{Diphinputfilesincluded} and \ref{DYinputfilesincluded}. These may be setup to run with {\tt Cuba}, if so either download {\tt Cuba} or change ``integrator\_flag'' in the input file from $2$ to $1$.
\item To run an input file type in the terminal (in the {\tt reSolve} main working directory):

\begin{center}
{\tt ./reSolve.out input/\{name of input file\}}
\end{center}
For example:
\begin{center}
{\tt ./reSolve.out input/Diphoton\_Born\_LHC.dat}.
\end{center}

\item The {\tt reSolve} program will then run and generate events for this setup, with the overall total cross-section output into the working directory output folder (``{\tt workdir/}'') specified in the input file as {\tt reSolve\_main\_out.dat} (or {\tt reSolve\_main\_out\_END\_ITER.dat} for {\tt k\_vegas} parallelised cases). Meanwhile the events are also included in this specified output folder, along with any specified histogram data files that were required for the various differential cross-sections, in particular for the $q_T$ spectrum\footnote{Note histograms are only generated for those variables and binnings specified in the input file in the ``Histograms'' section, nonetheless additional histograms can be calculated for the events without rerunning the whole program by using it in {\tt hist\_only} mode - see Section~\ref{histogrammer}.}.
\item There are a plethora of input flags which can be used to alter the phase space region, the process or subprocess, the pdf-fitting, the various scales associated with the theoretical calculation, the cuts, the number of evaluations, the working directory, the seeds used for the random generation of the Monte Carlo, the histograms generated, and many other aspects; these are all described in Section~\ref{usingprog}.
\end{itemize}

\pagebreak

\subsection{Sample Input Files included with {\tt reSolve}} \label{inputsincluded}

\begin{center}
\begin{table} [!htbp]
\centering
\def\arraystretch{1.1}
\begin{tabular}{|p{6.8cm}|p{8.7cm}|} \hline
Diphoton Input Files Included & Description \\ \hline
{\tt Diphoton\_Born\_LHC.dat} & Leading-order diphoton production at the LHC at $14TeV$ \\ \hline
{\tt Diphoton\_Born\_LHC\_parallel.dat} & Leading-order diphoton production at the LHC at $14TeV$ setup for {\tt k\_vegas} parallelisation \\ \hline 
{\tt Diphoton\_NNLO\_test\_1.dat} & NNLO diphoton production at the LHC at $14TeV$ \\ \hline
{\tt Diphoton\_NNLO\_test\_1\_parallel.dat} & NNLO diphoton production at the LHC at $14TeV$ setup for {\tt k\_vegas} parallelisation on one machine\\ \hline
{\tt Diphoton\_NNLO\_test\_1\_parallel\_multi.dat} & NNLO diphoton production at the LHC at $14TeV$ setup for {\tt k\_vegas} parallelisation across many machines  \\ \hline
{\tt Diphoton\_NNLO\_test1\_twopdffits.dat} & NNLO diphoton production at the LHC at $14TeV$ using two PDF fits at different scales across the invariant mass range \\ \hline
{\tt Diphoton\_NNLO\_test1\_fourpdffits.dat} & NNLO diphoton production at the LHC at $14TeV$ using four PDF fits at different scales across the invariant mass range \\ \hline
{\tt Diphoton\_NNLO\_test\_2.dat} & NNLO diphoton production at the LHC at $8TeV$ \\ \hline
{\tt Diphoton\_Atlas\_A.dat} & NNLO diphoton production at the LHC at $8TeV$ setup for experimental comparison \\ \hline
\end{tabular}
\caption{The sample input files which are included with the {\tt reSolve} program download for the Diphoton process, and which have been used throughout this paper in validation and results generation, see Section~\ref{validation} for more information on the input files and for the corresponding results and histograms produced. Note the only difference between the files {\tt Diphoton\_NNLO\_test\_1\_parallel.dat} and {\tt Diphoton\_NNLO\_test\_1\_parallel\_multi.dat} is in the working directories to which they output, in order to allow the use of the {\tt single\_machine\_parallel} and {\tt multi\_machine\_parallel\_local} scripts for the Diphoton NNLO test 1 setup case without the need to delete events between runs.} 
\label{Diphinputfilesincluded}
\end{table}
\end{center}

\begin{center}
\begin{table} [!htbp]
\centering
\def\arraystretch{1.1}
\begin{tabular}{|p{6.3cm}|p{9.2cm}|} \hline
Drell-Yan Input Files Included & Description \\ \hline
{\tt Wpm\_Born\_Tevatron.dat} & Leading-order $W^{\pm}$ production at the Tevatron \\ \hline
{\tt Wpm\_NLO\_Tevatron.dat} & NLO $W^{\pm}$ production at the Tevatron \\ \hline
{\tt Wpm\_NNLO\_Tevatron.dat} & NNLO $W^{\pm}$ production at the Tevatron \\ \hline
{\tt yZ\_Born\_Tevatron.dat} & Leading-order $Z/\gamma^*$ production at the Tevatron \\ \hline
{\tt yZ\_Born\_Tevatron\_parallel.dat} & Leading-order $Z/\gamma^*$ production at the Tevatron setup for {\tt k\_vegas} parallelisation \\ \hline
{\tt yZ\_NLO\_Tevatron.dat} & NLO $Z/\gamma^*$ production at the Tevatron \\ \hline
{\tt yZ\_NNLO\_Tevatron.dat} & NNLO $Z/\gamma^*$ production at the Tevatron \\ \hline
{\tt yZ\_NNLO\_Tevatron\_parallel.dat} & NNLO $Z/\gamma^*$ production at the Tevatron setup for {\tt k\_vegas} parallelisation on one machine \\ \hline
{\tt yZ\_NNLO\_Tevatron\_parallel\_multi.dat} & NNLO $Z/\gamma^*$ production at the Tevatron setup for {\tt k\_vegas} parallelisation across many machines\\ \hline
{\tt Z\_OnShell\_Born\_LHC.dat} & Leading-order on-shell $Z$ production at the LHC \\ \hline
{\tt Z\_OnShell\_NLO\_LHC.dat} & NLO on-shell $Z$ production at the LHC \\ \hline
{\tt Z\_OnShell\_NNLO\_LHC.dat} & NNLO on-shell $Z$ production at the LHC \\ \hline
\end{tabular}
\caption{The sample input files which are included with the {\tt reSolve} program download for the Drell-Yan processes, and which have been used throughout this paper in validation and results generation, see Section~\ref{validation} for more information on the input files and for the corresponding results and histograms produced. Again the only difference between the files {\tt yZ\_NNLO\_Tevatron\_parallel.dat} and {\tt yZ\_NNLO\_Tevatron\_parallel\_multi.dat} is in the working directories to which they output so single and multiple machine parallelisations can be run separately without the need to delete events between running them.} 
\label{DYinputfilesincluded}
\end{table}
\end{center}

\section{Resummation coefficients} \label{rescoeffs}
\label{app:coeffs}

Here we report a collection of the various coefficients used in the master resummation formula~\eqref{master0}, along with the references they were extracted from. We start from process-independent coefficients $A_c^{(n)}$, $B_c^{n}$, $C_{ab}^{(n)}$ and $G_{ab}^{(n)}$, commenting on the $H_c^F$ for some specific processes in \ref{sec:Hcoeffs}.

A useful note regarding flavour dependence: all flavour indices in eq.~\eqref{master0} could in principle assume 13 different values (from $\bar t$ to $t$ plus $g$). However, since quark mass effects are not included in this formalism (that is, we work in a \emph{fixed flavour} scheme), the coefficients are related by flavour symmetry. In practice, this means that in coefficients $A_c^{(n)}$ and $B_c^{n}$, the subscripts can only take 3 different values: $c = q, \bar q$ or $g$. In the doubly-indexed objects $C_{ab}$ and $G_{ab}$, in general there is additionally the need to distinguish \emph{diagonal} and \emph{non-diagonal} flavour contributions, leading to the possible combinations $C_{qq'}^F$, $C_{\bar q \bar q'}$, $C_{q \bar q'}$ and $C_{\bar q q'}$ for the indices, with obvious meaning of the symbols. Furthermore, due to C symmetry, all coefficients are invariant under barring of all indices: this implies $A_q = A_{\bar q}$, $B_q^F = B_{\bar q}^F$ to all orders, and that there are only \emph{seven} independent combinations for the $C_{ab}$ and $G_{ab}$ coefficients: $C_{gg}^F$, $C_{qq}^F$, $C_{qg}^F$, $C_{gq}^F$, $C_{q\bar q}^F$, $C_{qq'}^F$ and $C_{q \bar q'}^F$, and similarly for $G_{ab}$.

First of all let us define the auxiliary constants:
\begin{align}
T_R&: \text{  tr}(t^a t^b) = T_R \, \delta_{ab} & \text{ $\to$ fundamental colour matrices normalisation} \\
C_A & \equiv C_g = N_c &\text{$\to$ SU($N_c$) adjoint representation Casimir}\\
C_F & \equiv C_q \equiv C_{\bar q} = T_R \frac{N_c^2 -1}{N_c} &\text{$\to$ SU($N_c$) fundamental representation Casimir}\\
N_f& &\text{$\to$ Number of \emph{active} (effectively massless) flavours}\\
\zeta_n &&\text{$\to$ Value of Riemann Zeta function on point $n$}
\end{align}
Notice that the expansion of these and the other perturbative coefficients in the existing literature sees different normalisation choices. For the sake of the generality, we use in this section the definition:
\begin{equation}
Z(\as) = \sum_n \left( \frac{\as}{k \pi} \right)^n Z^{(n)} \; ,
\end{equation}
where $k$ is an arbitrary numerical constant and $Z$ a generic perturbative function of $\as$. Choices actually used in the literature include at least $k=1$, $2$ and $4$. For completeness, we report the $\beta$ function (see eq.~\eqref{asevol}) coefficients (taken from ref.~\cite{Bozzi:2005wk} eq.~(28) and ref.~\cite{Becher:2010tm} eq.~(B5)):
\begin{align}
\beta_0 = & \frac{k}{12}(11C_A - 4 T_R N_f)\\
\beta_1 = & \frac{k^2}{24}(17 C_A^2 - 10 C_A T_R N_f - 6 C_F T_R N_f)\\
\begin{split}
\beta_2 = & \frac{k^3}{64}\left( \frac{2857}{54}C_A^3 -\frac{1415}{27}C_A^2 T_R N_f - \frac{205}{9} C_A C_F T_R N_f \right. \\
& \left.+ 2 C_F^2 T_R N_f + \frac{158}{27} C_A T_R^2 N_f^2 + \frac{44}{9} C_F T_R^2 N_f^2\right)
\end{split}
\end{align}

Now we list the actual resummation coefficients, $A_c$, $B_c$, $C_{ab}$ and $G_{ab}$. All coefficients are defined in \emph{hard scheme}(see Section~\ref{sec:resscheme}) unless otherwise specified. The $A_c^{(n)}$ coefficients (which are actually resummation scheme independent) are given by ($A_c^{(1,2)}$ from ref.~\cite{Bozzi:2005wk} eq.~(47) and $A_c^{(3)}$ from ref.~\cite{Becher:2010tm} eqs.~(51),~(74) and~(B3) -- see also the comment under eq.~(74)): 
\begin{align}
&A_c^{(1)} = k \, C_c\\
&A_c^{(2)} = \frac{k^2}{2} C_c\left[ \left( \frac{67}{18}-\frac{\pi^2}{6}  \right)C_A - \frac{10}{9} T_R N_f \right] \\
\begin{split}
A_g^{(3)} = & C_A\frac{k^3}{16}\left[ C_A^2 \left( \frac{245}{6} - \frac{134}{27}\pi^2 + \frac{11}{45}\pi^4 + \frac{22}{3}\zeta_3 \right) \right.\\
& \left. + C_A T_R N_f \left(-\frac{418}{27} + \frac{40}{27}\pi^2 - \frac{56}{3}\zeta_3 \right) + C_F T_R N_f \left( -\frac{55}{3} + 16\zeta_3 \right)
+ \frac{16}{27} T_R^2 N_f^2 \right]\\
\end{split}\\
&A_q^{(3)} = A_{\bar q}^{(3)} = \frac{C_F}{C_A} A_g^{(3)} + 2 \beta_0 \frac{k^2}{16} C_F \left[ C_A\left( \frac{808}{27} - 28\zeta_3 \right) -\frac{224}{27} T_R N_f \right]
\end{align}
The $B_c^{(n)}$ coefficients, from ref.~\cite{Bozzi:2005wk} eq.~(49) for $B^{(1)}$ and ref.~\cite{Catani:2013tia} eq.~(34)-(36) for $B^{(2)}$:
\begin{align}
& B_g^{(1)} = -\frac{k}{6}(11C_A - 4T_R N_f) \\
& B_q^{(1)} = B_{\bar q}^{(1)} = -k \, \frac{3}{2} \, C_F \\
& B_c^{(2)} = k \left( \frac{\gamma_c^{(1)}}{16} + \pi \beta_0 C_c \zeta_2 \right), \quad \text{with} \\
\begin{split}
&\gamma_q^{(1)} = \gamma_{\bar q}^{(1)} = k \left[ C_F^2(-3+24\zeta_2-48\zeta_3) + C_A C_F \left( -\frac{17}{3} - \frac{88}{3} \zeta_2 + 24\zeta_3 \right) \right. \\
& \left. \qquad + \, C_F T_R N_f \left( \frac{2}{3} + \frac{16}{3}\zeta_2 \right) \right]\\
& \gamma_g^{(1)} = k \left[ C_A^2\left( -\frac{64}{3} - 24\zeta_3 \right) + \frac{16}{3}C_A T_R N_f + 4 C_F T_R N_f \right] \; ;
\end{split}
\end{align}
the $\gamma_c^{(1)}$ coefficients are the $\delta(1-z)$ parts of the first order Altarell-Parisi splitting kernels.

Now we list the $C_{ab}^{(n)}$ and $G_{ab}^{(n)}$ coefficients. We start with the first order coefficients; the explicit expressions for these are taken from ref.~\cite{Catani:2013tia}, eqs.~(29)-(33):
\begin{align}
& C_{qq}^{(1)}(z) = \frac{k}{2} C_F (1-z)\\
& C_{gq}^{(1)}(z) = \frac{k}{2} C_F z \\
& C_{qg}^{(1)}(z) = \frac{k}{2} z(1-z) \\
& C_{gg}^{(1)}(z) = C_{q\bar q}^{(1)}(z) = C_{qq'}^{(1)}(z) = C_{q\bar q'}^{(1)}(z) = 0 \\
& G_{gq}^{(1)}(z) = \frac{k}{2} C_F \frac{1-z}{z} \\
& G_{gg}^{(1)}(z) = \frac{k}{2} C_A \frac{1-z}{z} \; .
\end{align}
Notice that 3 out of 5 of the non-diagonal $C_{ab}$ are actually \emph{vanishing} at first order. Finally, let us give references for the $C_{ab}^{(2)}$ coefficients; since these $2^{nd}$ order coefficients have very long and involved expressions, we will not report them in a completely explicit form, but give them in terms of more fundamental objects whose definitions can be found in the listed references. Also, since they are not needed for the only process currently implemented in {\tt reSolve} (diphoton production), we will not report the $G_{ga}^{(2)}$ coefficients. All the following definitions are extracted from~\cite{Catani:2013tia}, eqs.~(37)-(40) and nearby text.
\begin{align}
\begin{split}
& C_{qq}^{(2)}(z) = \frac{k^2}{2}\left\{ \mathcal{H}_{q\bar q\leftarrow q\bar q}^{DY (2)}(z)|_{\text{no }\delta(1-z)} - \frac{C_F^2}{4}\left[ (2\pi^2 -18)(1-z)
  -(1+z)\log z \right] \right\} , \\
& \qquad \text{with $\mathcal{H}_{q\bar q\leftarrow q\bar q}^{DY (2)}(z)$ defined in eq.~(23) of~\cite{Catani:2012qa};}
\end{split}\\
\begin{split}
& C_{qg}^{(2)}(z) = k^2 \left\{ \mathcal{H}_{q\bar q\leftarrow qg}^{DY (2)}(z) - \frac{C_F}{4}\left[ z\log z
  + \frac{1}{2}(1-z^2) + (\pi^2-8)z(1-z) \right] \right\} , \\
& \qquad \text{with $\mathcal{H}_{q\bar q\leftarrow qg}^{DY (2)}(z)$ defined in eq.~(32) of~\cite{Catani:2012qa};}
\end{split}\\
& C_{q\bar q}^{(2)}(z) = k^2 \, \mathcal{H}_{q\bar q\leftarrow qq}^{DY (2)}(z), \ 
  \text{with $\mathcal{H}_{q\bar q\leftarrow qq}^{DY (2)}(z)|_{\text{no }\delta(1-z)}$ defined in eq.~(24) of~\cite{Catani:2012qa};}\\
& C_{qq'}^{(2)}(z) = k^2 \, \mathcal{H}_{q\bar q\leftarrow q\bar q'}^{DY (2)}(z), \ 
  \text{with $\mathcal{H}_{q\bar q\leftarrow q\bar q'}^{DY (2)}(z)|_{\text{no }\delta(1-z)}$ defined in eq.~(25) of~\cite{Catani:2012qa};}\\
& C_{q\bar q'}^{(2)}(z) = C_{qq'}^{(2)}(z), \text{ by eqs.~(24),~(26) and~(35) of~\cite{Catani:2012qa};}\\
\begin{split}
& C_{gg}^{(2)}(z) = \frac{k^2}{2}\left\{ \mathcal{H}_{gg\leftarrow gg}^{H (2)}(z)|_{\text{no }\delta(1-z)} + C_A^2 \left[ \frac{1+z}{z}\log z
  +2\frac{1-z}{z} \right] \right\} , \\
& \qquad \text{with $\mathcal{H}_{gg\leftarrow g g}^{H (2)}(z)$ defined in eq.~(24) of~\cite{Catani:2011kr};}
\end{split}\\
\begin{split}
& C_{gq}^{(2)}(z) = k^2\left\{ \mathcal{H}_{gg\leftarrow gq}^{H (2)}(z) +C_F^2 \frac{3z}{4} - \frac{C_A C_F}{z} \mspace{-3mu} \left[ (1+z)\log z
  +2(1-z) -\frac{5+\pi^2}{4} z^2 \mspace{-2mu}  \right] \mspace{-2mu} \right\} , \\
& \qquad \text{with $\mathcal{H}_{gg\leftarrow gq}^{H (2)}(z)$ defined in eq.~(23) of~\cite{Catani:2011kr}.}
\end{split}
\end{align}
All $C_{ab}$ coefficients given here are defined in $z$-space. In {\tt reSolve}, as commented in section~\ref{sec:FourierAndMellin}, the corresponding Mellin space expressions are actually used.

\subsection{Examples of process-dependent coefficients}
\label{sec:Hcoeffs}

Here we will list the process-dependent $H_c^{F(n)}$ coefficients for selected processes, again in \emph{hard scheme}. The processes we list them for are Drell-Yan, Higgs and diphoton production, with, remember, only the first and last of these actually implemented in the current version of $\tt reSolve$. The main ref. here is~\cite{Catani:2013tia}: eqs.~(82-83) for DY and eqs.~(85),~(87) for Higgs (obtained in the large $m_t$ limit):
\begin{align}
\label{H1DY}
H_q^{DY(1)} & = k\, C_F\left( \frac{\pi^2}{2} - 4 \right)\\
\label{H2DY}
\begin{split}
H_q^{DY(2)} & = k^2 \left[ C_F C_A \left( \frac{59}{18}\zeta_3 - \frac{1535}{192} + \frac{215}{216}\pi^2 - \frac{\pi^4}{240} \right) \right.\\
& + \frac{1}{4}C_F^2 \left( -15\zeta_3 +\frac{511}{16} - \frac{67}{12}\pi^2 + \frac{17}{45}\pi^4 \right)\\
& + \left. \frac{1}{864} C_F T_R N_f (192\zeta_3 + 1143 - 152\pi^2) \right]
\end{split}\\
\label{H1H}
H_g^{H(1)} & = k \left[ C_A \frac{\pi^2}{2} + \frac{5 C_A - 3 C_F}{2} \right] \\
\label{H2H}
H_g^{H(2)} & = k^2 \left[ C_A^2 \left( \frac{3187}{288} + \frac{7}{8} L_Q + \frac{157}{72}\pi^2 + \frac{13}{144}\pi^4 - \frac{55}{18} \zeta_3 \right) \right. \\
& + C_A C_F \left(-\frac{145}{24} - \frac{11}{8}L_Q - \frac{3}{4}\pi^2 \right) + \frac{9}{4} C_F^2 - \frac{5}{96} C_A - \frac{1}{12} C_F \\
& \left.- C_A T_R N_f\left( \frac{287}{144} + \frac{5}{36}\pi^2 + \frac{4}{9}\zeta_3 \right) + C_F T_R N_f \left( -\frac{41}{24} + \frac{1}{2}L_Q + \zeta_3\right) \right] \; ,
\end{align}
where $L_Q = \log(m_H^2/m_t^2)$. For Higgs at NLO, also the contribution for finite top mass is known: see for instance \cite{Aglietti:2006tp}, eqs.~(21)-(30).

Finally, for the diphoton ($\gamma\gamma$) process, from~\cite{Catani:2013tia} eqs.~(88-91):
\begin{align}
\label{H1yy}
\begin{split}
H_q^{\gamma\gamma(1)} & = k \, \frac{C_F}{2} \bigg[ \pi^2 - 7 + \frac{1}{(1-v)^2 + v^2} \Big[ ((1-v)^2+1)\log^2(1-v) \\
&\ + v(v+2)\log(1-v) + (v^2+1)\log^2 v + (1-v)(3-v) \log v \Big] \bigg]
\end{split}\\
\label{H2yy}
\begin{split}
H_q^{\gamma\gamma(2)} & = \frac{k^2}{4 \mathcal{A}_{LO}} [\mathcal{F}_{inite,q\bar q\gamma\gamma;s}^{0\times2} + \mathcal{F}_{inite,q\bar q\gamma\gamma;s}^{1\times1}]
  + 3\zeta_2 C_F H_q^{\gamma\gamma(1)} - \frac{45}{4} C_F^2 \\
+ \, C_F & C_A\left( \frac{607}{324} + \frac{1181}{144}\zeta_2 - \frac{187}{144}\zeta_3 - \frac{105}{32}\zeta_4\right) + C_F T_R N_f \left(-\frac{41}{162}
  -\frac{97}{72}\zeta_2 + \frac{17}{72}\zeta_3 \right) \; ,
\end{split}
\end{align}
where $v = - u /s$ (partonic variables), $\mathcal{A}_{LO}$ is just proportional to the LO matrix element $u/t + t/u$ (and to KinFac in our code),
\begin{equation}
\mathcal{A}_{LO} = -8N_c \left(\frac{u}{t} + \frac{t}{u}\right) = 8N_c \frac{1-2v+2v^2}{v(1-v)} \; ,
\end{equation}
and $\mathcal{F}_{inite,q\bar q\gamma\gamma;s}^{0\times2}$, $\mathcal{F}_{inite,q\bar q\gamma\gamma;s}^{1\times1}$ are defined in~\cite{Anastasiou:2002zn}, eqs.~(4.6) and~(5.3), see also appendices A and B in the same reference. A subtlety which appears in the $\gamma\gamma$ process is that the $\mathcal{F}_{inite,q\bar q\gamma\gamma;s}^{0\times2}$ factor, and thus $H_q^{\gamma\gamma(2)}$, is actually dependent on the \emph{electric charge} (squared) of the involved quark. So there are actually two different $H_q^{\gamma\gamma(2)}$ coefficients, namely $H_u^{\gamma\gamma(2)}$ and $H_d^{\gamma\gamma(2)}$. This is of no particular consequence for our expressions, but it implies that a modicum of care must be used in setting the resummation scheme for the resummation coefficients: if one was for instance to define a \emph{$\gamma\gamma$ scheme} along the lines of the \emph{DY-H scheme} defined in section~\ref{sec:resscheme}, the charge dependence would creep in the definition of the $C_{ab}^{(2)}$ coefficients, which would be impractical. 
In the $\gamma\gamma$ case, in principle also the $gg$ channel contributes, with the LO term given by the $gg\to\gamma\gamma$ box, so that $H_{gg}^{\gamma\gamma} \neq 0$. However, the $gg\to\gamma\gamma$ partonic process is suppressed by $\as^2$ with respect to the $q\bar q$-initiated one (which is $\mathcal{O}(\as^0)$, a purely electromagnetic process at LO), so a non-trivial $H_{gg}^{\gamma\gamma}$ is only needed for $\text{N}^3$LL onwards, and so is not included in {\tt reSolve}.

\pagebreak

\bibliography{Draft_paper.bib}
\bibliographystyle{elsarticle-num}

\end{document}